\documentclass[aps,prd,twocolumn,nofootinbib,superscriptaddress,floatfix]{revtex4-2}

\usepackage{amsmath,amssymb,bm}
\usepackage{booktabs}
\usepackage{graphicx}
\usepackage{multirow}

\usepackage{xcolor}
\usepackage{tikz}
\usepackage{hyperref}
\hypersetup{hidelinks}
\usepackage{enumitem}
\usetikzlibrary{arrows.meta,positioning,calc,decorations.pathmorphing}

\usepackage{tikz-feynman}
\tikzfeynmanset{compat=1.1.0}

\graphicspath{{figures/}}

\newcommand{\GeV}{\,{\rm GeV}}
\newcommand{\TeV}{\,{\rm TeV}}
\newcommand{\keV}{\,{\rm keV}}

\newcommand{\cm}{\,{\rm cm}}
\newcommand{\s}{\,{\rm s}}
\newcommand{\kms}{\,{\rm km\,s^{-1}}}

\begin{document}

\title{A Comprehensive Interpretation of the LZ High-Energy Recoil Event: A Minimal $U(1)_B$ Dark Sector and Its Experimental Tests}

\author{Mattia Di Mauro}
\affiliation{Istituto Nazionale di Fisica Nucleare, Sezione di Torino, Via P. Giuria 1, 10125 Torino, Italy}

\date{October 8, 2026}

\begin{abstract}
The $248\keV$ nuclear-recoil candidate reported by LUX-ZEPLIN (LZ) motivates an inelastic dark matter (DM) interpretation, in which scattering converts the DM particle into a slightly heavier state. We provide a comprehensive study of this interpretation in an anomaly-free model of gauged baryon number, where a new vector boson, $Z_B$, mediates interactions between DM and quarks. Breaking the baryonic gauge symmetry generates the new particle masses and a small splitting between two DM states, while a residual symmetry keeps the lighter state stable. We identify regions compatible with the LZ candidate, the observed DM abundance, and current ATLAS and CMS dijet constraints. Within the scanned domain, these regions span DM masses $m_\chi\simeq0.19$--$4.5\TeV$, mediator masses $M_{Z_B}\simeq0.05$--$5.0\TeV$, and gauge couplings $g_B\simeq0.004$--$0.74$, with correlated values of these parameters. A representative thermal solution has $m_\chi=1\TeV$, $M_{Z_B}=1.44\TeV$, $g_B\simeq0.228$, and a mass splitting of $300\keV$. A small electroweak mixing allows the heavier DM state to decay rapidly, suppressing unwanted signals from its surviving population. This interpretation can be tested through complementary measurements: additional LZ and XENONnT data can establish whether the high-energy recoil signal persists, ATLAS and CMS Run-3 dijet searches can probe the mediator mass and coupling, and a dedicated IceCube analysis of four-quark final states can test DM capture and annihilation in the Sun. Together, these measurements can test complementary parts of the surviving region and assess whether the LZ candidate is consistent with this interpretation.
\end{abstract}

\maketitle

\section{Introduction}
\label{sec:intro}

A broad set of astrophysical and cosmological observations establishes the existence of non-baryonic dark matter (DM), from galactic and cluster dynamics and gravitational lensing to the cosmic microwave background (CMB) and the growth of large-scale structure~\cite{Bertone:2010zza,Bertone:2016nfn,ArcadiReview,CirelliStrumiaZupan}. These measurements determine with high precision that the cosmological DM abundance is $\Omega_{\chi} h^2=0.120$ \cite{Planck2018}, but they do not identify its microscopic nature.  Despite major advances in direct detection (DD), indirect detection (ID), and collider searches, no unambiguous non-gravitational DM signal has yet been established in astrophysical observations or laboratory experiments.

Weakly interacting massive particles (WIMPs) remain among the most extensively studied DM candidates because thermal freeze-out relates an electroweak-scale annihilation cross section to the observed relic abundance \cite{Lee:1977ua,1978ApJ...223.1015G,1983PhRvL..50.1419G,Ellis:1983ew}. The simplest implementations are, however, increasingly constrained by spin-independent (SI) DD searches from LZ and XENONnT experiments~\cite{LZ:2023,Aprile:2023XENONnT,LZ:2024zvo,XENONnT2025}, gamma rays and charged cosmic rays \cite{Gaskins:2016cha}, and collider measurements~\cite{ArcadiReview,WaningWIMP,ArcadiLHCDark}. Detailed studies of Higgs-portal, vector-mediator, leptophilic, and other WIMP realizations show that viable thermal regions frequently survive through resonances, blind spots, velocity suppression, or non-trivial mediator sectors rather than through the simplest elastic contact interaction~\cite{KongDiMauro,DiMauroArinaFornengo,DiMauroXieResonance,KoechlerDiMauro,ArcadiProfumoBlindSpots,ArcadiLHCDark}. This motivates treating the WIMP paradigm as a broad thermal framework rather than identifying it with one specific SI elastic-scattering limit.

Two especially useful examples are secluded and resonant DM. In secluded models, freeze-out can be controlled by annihilation into lighter hidden-sector particles while the portal connecting the dark sector to the Standard Model (SM) is substantially weaker, thus suppressing laboratory signals~\cite{PospelovSecluded,DiMauroWangSecluded}. 
In resonant models, thermal averaging around an $s$-channel pole can reproduce the relic abundance for couplings much smaller than electroweak ones and can thus evade DD constraints~\cite{GriestSeckel,DiMauroXieResonance,DiMauroArinaFornengo}. These scenarios illustrate how the interaction responsible for setting the thermal relic abundance can be partially decoupled from the processes probed by present-day DD, ID, and collider experiments.

Inelastic DM provides another particularly interesting departure from the
standard elastic WIMP picture, since the scattering changes the DM state,
$\chi_1 N\to\chi_2 N$, with a small mass splitting
$\delta_\chi=m_{\chi_2}-m_{\chi_1}$. The resulting kinematics can
substantially modify the recoil-energy spectrum observed in DD experiments.
This becomes especially relevant when an experimental event occurs in a
recoil-energy region that is difficult to populate efficiently with
conventional elastic scattering.

The extended nuclear-recoil analysis reported by the LUX-ZEPLIN Collaboration (LZ) provides precisely such a situation~\cite{LZ2026}. Using an exposure of $2.84$ tonne-year, LZ identified one event compatible with a nuclear recoil (NR) at
\begin{equation}
E_R=248\pm23_{\rm stat}\pm23_{\rm sys}\keV.
\end{equation}
Among the interaction hypotheses considered by LZ, the largest local significance is $3.4\sigma$, which is reduced to a global significance of $2.6\sigma$ after accounting for the look-elsewhere effect~\cite{LZ2026}. Clearly, a single event cannot be regarded as evidence for DM, and additional exposure together with confirmation by other experiments will be essential to determine its origin. Nevertheless, its unusually large recoil energy raises an interesting and well-defined phenomenological question: a viable DM interpretation must enhance the event rate in the high-energy part of the xenon recoil spectrum without simultaneously predicting an excessive number of events at lower recoil energies.

Endothermic DM provides a direct kinematic mechanism for doing so~\cite{TuckerSmithWeiner,ChangInelastic,DiMauroLZ}.  
If the DM halo state $\chi_1$ scatters into a heavier state $\chi_2$, with mass splitting $\delta_\chi>0$, the minimum incident speed is
\begin{equation}
 v_{\rm min}(E_R)=
 \frac{m_AE_R/\mu_{\chi A}+\delta_\chi}{\sqrt{2m_AE_R}},
 \label{eq:intro_vmin}
\end{equation}
where $m_A$ is the nuclear mass and $\mu_{\chi A}$ the reduced mass.  The minimum of $v_{\rm min}$ occurs at a non-zero recoil energy, so a splitting of a few hundred keV can remove the low-energy part of the recoil spectrum and move the accepted events toward the LZ candidate.  For TeV-scale DM and xenon, the required splitting is naturally of order a few $10^2\keV$~\cite{DiMauroLZ}.

The unusual properties of the event immediately triggered a broad theoretical effort to identify viable DM interpretations. Early studies focused particularly on endothermic and inelastic scattering~\cite{SuYangYang,DiMauroLZ,YamashitaLZ,McCabeSeasonal}, as well as on the particularly predictive case of TeV-scale Higgsino DM~\cite{FreeseHiggsino,DuWangHiggsino,LanghoffHiggsino}. These were rapidly followed by realizations involving exothermic DM~\cite{BaerExothermic}, the inert-doublet model~\cite{WangXiaoIDM}, singlet--doublet fermion DM~\cite{BorahSingletDoublet,LeeYounMixing}, gauged $B-L$ constructions with inelastic scalar DM~\cite{OkadaSetoBL,KumarPrajapatiBL}, magnetic inelastic dark baryons~\cite{AsadiMagneticLZ}, and more general inelastic scenarios~\cite{FanInelastic}. A realization of inelastic DM within a warped extra-dimensional framework has also been proposed~\cite{LeeRandallLZ}. Alternative mechanisms capable of producing a high-energy recoil have been explored, including pseudoscalar-mediated elastic scattering~\cite{UnwinAxion}, boosted DM~\cite{AlhazmiBoostedLZ,HeikinheimoBoosted}, scotogenic and radiative-neutrino-mass constructions~\cite{OkadaLZ,RadiativeDiracLZ}, and non-thermal freeze-in production~\cite{FreezeInLZ}. More recently, the event has also been investigated in the 2HD+$a$ framework~\cite{Arcadi2HDa}.
Several studies have also stressed that the DD signal cannot be considered as the only experimental probe: the same interaction may be constrained by DM Solar capture and neutrino production \cite{IceCubeSolar2025}, the thermal relic density, collider searches, ID, and the high-velocity tail of the Galactic halo~\cite{PospelovRamaniSolar,SolarCapture,HiggsCoupledMDM,OHareHighVelocity}.

In Ref.~\cite{DiMauroLZ}, we showed that a thermal pseudo-Dirac fermion with an off-diagonal vector interaction gives an economical endothermic interpretation.  The transition $\chi_1 N\to\chi_2 N$ controls the LZ signal, while the crossed interaction $\chi_1\chi_2\to q\bar q$ controls thermal freeze-out.  Around $m_\chi\sim1\TeV$, the thermal relic condition selects a DD normalization close to that required by the LZ candidate.  Ref.~\cite{SolarCapture} then demonstrated an important property of the two-state system: Solar capture can remain efficient while the annihilation rate is strongly suppressed.  Once the captured population de-excites to $\chi_1$, the leading coannihilation channel lacks the real $\chi_2$ required in the initial state, and capture--annihilation equilibrium need not be reached.  This mechanism substantially weakens the IceCube constraint \cite{IceCubeSolar2025} that is powerful for other inelastic candidates~\cite{PospelovRamaniSolar,HiggsCoupledMDM,SolarCapture}.

The simplified pseudo-Dirac description introduced in Ref.~\cite{DiMauroLZ} nevertheless leaves the origin of its defining ingredients unspecified.  A complete theory must explain why the vector current is dominantly off diagonal, why the lightest state is stable, how a sub-MeV splitting is generated naturally in a TeV-scale sector, how the mediator acquires its mass, and which additional states are required by gauge consistency.  Gauged baryon number offers an economical answer \cite{FileviezWise2010,DulaneyWise2011,DuerrFileviez2015,Fileviez2019,AnomalyFreeDM,DuHuangXie}.  The SM fermion content is anomalous under a local $U(1)_B$, forcing new fermions to exist.  With the charge assignment used below, the neutral member of this anomaly-cancelling sector is precisely the pseudo-Dirac DM particle.  A single baryonic scalar generates the vector-boson mass, the anomalon masses, the dominant Dirac DM mass, and the small Majorana entry responsible for the endothermic splitting.  

The central purpose of this work is therefore not merely to construct a UV completion of the pseudo-Dirac scenario, but to determine whether imposing gauge consistency leads to a predictive and phenomenologically viable framework. We construct an anomaly-free realization in which the interactions responsible for DM scattering, thermal freeze-out, and collider production arise from the same underlying gauge structure. We then study its phenomenology by computing the LZ recoil spectrum, the DM relic abundance, current collider constraints, ID estimates, and a Solar capture--annihilation diagnostic.
An important consequence of the UV completion is that the parameters controlling these different observables are no longer independent. In particular, requiring the model to simultaneously reproduce the LZ event and the observed relic abundance selects a relatively restricted region of parameter space that can be directly tested through searches for the new mediator at the LHC. We therefore confront the model with existing ATLAS and CMS dijet searches~\cite{ATLASDijet,CMSDijetScouting,CMSLowMassDijet2026} and investigate the sensitivity that can be reached with the completed Run-3 data set, taking advantage of the dedicated trigger-level and scouting strategies developed by the two experiments~\cite{ATLASRun3TLA,CMSRun3Scouting}. This provides a concrete collider target for testing the DM interpretation of the LZ event and illustrates how a possible direct-detection signal can become correlated with complementary searches at the LHC.

The paper is organized as follows. Sec.~\ref{sec:model} presents the anomaly-free $U(1)_B$ completion, symmetry breaking, fermion spectrum, and the origin of the pseudo-Dirac splitting, DM stability, and off-diagonal current. Sec.~\ref{sec:pheno} develops the DD, thermal relic, collider, ID, and Solar capture predictions, including prospective Run-3 sensitivity. Sec.~\ref{sec:results} combines the constraints in a global parameter scan, and Sec.~\ref{sec:conclusions} summarizes the findings. Apps.~\ref{app:anomalies} and~\ref{app:kinmix} detail anomaly cancellation and radiative kinetic mixing. App.~\ref{app:subleadingDD} derives the recoil contributions, their suppressions, and interference; App.~\ref{app:lmc_tail} examines the impact of the LMC on the high-speed halo tail. App.~\ref{app:excitedstate} calculates excited-state decays, cosmological depletion, late-time regeneration, and captured Solar populations, while App.~\ref{app:uvsplitting} discusses possible UV origins of the small Majorana mass.

\section{Description of the model}
\label{sec:model}

We consider a UV-complete realization of the pseudo-Dirac DM scenario used in Ref.~\cite{DiMauroLZ}. The basic idea is to promote baryon number $B$ to a local gauge symmetry, so that the mediator responsible for the interaction between the dark and visible sectors arises as the gauge boson associated with a new $U(1)_B$ factor. Gauged baryon-number extensions of the SM have been extensively studied as theoretically consistent frameworks for leptophobic interactions and DM phenomenology~\cite{FileviezWise2010,DulaneyWise2011,DuerrFileviez2015,Fileviez2019,AnomalyFreeDM}. A crucial requirement of such constructions is the cancellation of the gauge anomalies induced by the SM quark sector. This necessarily introduces additional fermionic states, while the spontaneous breaking of $U(1)_B$ generates the mass of the new gauge boson and can simultaneously provide masses and mass splittings in the dark sector.

In the realization considered here, these ingredients are combined in a minimal way such that the low-energy spectrum contains two nearly degenerate Majorana states forming a pseudo-Dirac DM pair. Their interaction with the baryonic gauge boson is predominantly off diagonal, naturally realizing the endothermic transition required to reproduce the high-energy LZ recoil event. At the same time, the gauge structure fixes the couplings of the mediator to SM quarks and correlates DD, thermal freeze-out, collider production, and Solar capture. The purpose of this section is to define the complete field content and interactions of the model, discuss anomaly cancellation and spontaneous symmetry breaking, and derive the low-energy pseudo-Dirac theory used in the phenomenological analysis.

\subsection{Gauge symmetry and field content}
\label{subsec:gaugefield}

We extend the SM gauge group to
\begin{equation}
\mathcal G=SU(3)_c\times SU(2)_L\times U(1)_Y\times U(1)_B.
\label{eq:gaugegroup}
\end{equation}
The SM baryon-number assignment is conventional, with $B(Q_L)=B(u_R)=B(d_R)=1/3$, while leptons and the SM Higgs are neutral under $U(1)_B$.

Once baryon number is promoted from an accidental global symmetry of the renormalizable SM Lagrangian to a local gauge symmetry, the SM fermion content is no longer consistent by itself. In particular, the mixed $[SU(2)_L]^2U(1)_B$ and $[U(1)_Y]^2U(1)_B$ anomalies do not vanish. Additional chiral fermions are therefore required by gauge consistency rather than introduced solely for phenomenological purposes. A minimal and well-studied possibility is to add fermions that are vectorlike under the SM gauge group but chiral under $U(1)_B$~\cite{Fileviez2019,AnomalyFreeDM}. In the construction considered here, the same fermion sector that restores anomaly cancellation also contains the DM candidate.

The new fermionic and scalar degrees of freedom are summarized in Tab.~\ref{tab:fields}. The fermion sector contains an electroweak doublet $\Psi$, a hypercharged singlet $\eta$, and a neutral SM-singlet fermion $\chi$. The SM-charged states $\Psi$ and $\eta$ form the anomalon sector required for anomaly cancellation, while $\chi$ belongs to the same anomaly-free fermion content and provides the DM candidate. The scalar sector contains, in addition to the SM Higgs doublet, a single complex scalar $S_B$, neutral under the SM gauge group but charged under $U(1)_B$. Its vacuum expectation value spontaneously breaks the baryonic gauge symmetry, generates the mass of the corresponding gauge boson $Z_B$, and enters the Yukawa interactions responsible for the masses of the new fermions and for the splitting of the neutral state.

We denote by $B_1$ and $B_2$ the two $U(1)_B$ charges carried by the chiral components of the new fermions. In the convention adopted here,
\begin{align}
B(\Psi_L)=B(\eta_R)=B(\chi_R)&=B_1,\\
B(\Psi_R)=B(\eta_L)=B(\chi_L)&=B_2.
\end{align}
These charges are constrained simultaneously by anomaly cancellation and by gauge invariance of the fermion mass sector. Cancellation of the mixed gauge anomalies requires
\begin{equation}
B_1-B_2=-3,
\label{eq:Bdiff}
\end{equation}
while the requirement that a single scalar $S_B$ generate the anomalon Dirac masses and the Majorana term in the neutral sector further restricts the allowed charge assignments. The scalar-minimal solution adopted in this work is
\begin{equation}
B_1=-\frac{9}{2},
\,\,
B_2=-\frac{3}{2}.
\label{eq:Bcharges}
\end{equation}
The half-integer charges follow from anomaly cancellation together with the chosen left-handed Majorana operator and one baryonic scalar. Anomaly cancellation alone fixes only the charge difference; alternative one-sided assignments and scalar sectors are possible. The complete anomaly-cancellation conditions are presented in App.~\ref{app:anomalies}.

\begin{table}[htbp] 
\centering 
\caption{Field content beyond the SM. The new fermions are vectorlike under $SU(3)_c\times SU(2)_L\times U(1)_Y$ but chiral under $U(1)_B$. The charge assignment in Eq.~\eqref{eq:Bcharges} cancels all gauge and mixed gravitational anomalies, as shown in App.~\ref{app:anomalies}.} \label{tab:fields} 
\begin{ruledtabular} 
\begin{tabular}{ccccc} 
field & $SU(3)_c$ & $SU(2)_L$ & $Y$ & $B$\\ 
\hline 
$\Psi_L$ & $1$ & $2$ & $-1/2$ & $-9/2$\\ 
$\Psi_R$ & $1$ & $2$ & $-1/2$ & $-3/2$\\ 
$\eta_R$ & $1$ & $1$ & $-1$ & $-9/2$\\ 
$\eta_L$ & $1$ & $1$ & $-1$ & $-3/2$\\ 
$\chi_R$ & $1$ & $1$ & $0$ & $-9/2$\\ 
$\chi_L$ & $1$ & $1$ & $0$ & $-3/2$\\ 
$S_B$ & $1$ & $1$ & $0$ & $-3$ 
\end{tabular} 
\end{ruledtabular} 
\end{table}

For some purposes it is useful to adopt the integer normalization $Q_B=6B$. In this convention, SM quarks carry charge $2$, the two new chiral sectors carry charges $-27$ and $-9$, and $S_B$ carries charge $-18$. This normalization is particularly convenient when discussing the residual discrete symmetry that remains after spontaneous $U(1)_B$ breaking.

\subsection{Renormalizable Lagrangian and spontaneous symmetry breaking}
\label{subsec:lagrangian}

Having specified the field content and charge assignments, we now discuss the renormalizable interactions that determine the spectrum and phenomenology of the model. The Lagrangian can be written schematically as
\begin{equation}
\mathcal L=
\mathcal L_{\rm SM}
+\mathcal L_{\rm gauge}
+\mathcal L_{\rm kin}^{\rm new}
+\mathcal L_{\rm Yuk}
-V(H, S_B),
\label{eq:Ldecomp}
\end{equation}
where $\mathcal L_{\rm SM}$ is the SM Lagrangian, $\mathcal L_{\rm kin}^{\rm new}$ contains the kinetic terms of the anomalons, the neutral fermion $\chi$, and the baryonic scalar $S_B$.
The gauge sector associated with $U(1)_B$ is
\begin{equation}
\mathcal L_{\rm gauge}=
-\frac14 Z_{B\mu\nu}Z_B^{\mu\nu}
-\frac{\epsilon_{YB}}{2}B_{\mu\nu}Z_B^{\mu\nu},
\label{eq:gaugekin}
\end{equation}
and the covariant derivative acting on a field with hypercharge $Y$ and baryon number $B$ is
\begin{equation}
D_\mu=
\partial_\mu
-i g_s T^a G_\mu^a
-i g T^a W_\mu^a
-i g_Y Y B_\mu
-i g_B B Z_{B\mu}.
\label{eq:covder}
\end{equation}
Here $B_\mu$ denotes the SM hypercharge gauge field. The second term in Eq.~\eqref{eq:gaugekin} describes gauge kinetic mixing between $U(1)_Y$ and $U(1)_B$. Even if $\epsilon_{YB}$ is set to zero at some ultraviolet (UV) scale, it is regenerated radiatively because fields carrying both hypercharge and baryon number propagate in loops. Its value therefore depends on the UV boundary condition and on the thresholds of the new fermions. Since these ingredients introduce additional model dependence that is not required for the core LZ analysis, we neglect kinetic mixing in the main numerical study and discuss its renormalization-group evolution and phenomenological consequences separately in App.~\ref{app:kinmix}.

For the charge assignment introduced in Sec.~\ref{subsec:gaugefield}, the relevant renormalizable Yukawa interactions involving the new fermions are
\begin{align}
-\mathcal L_{\rm Yuk}={}&
y_1\overline\Psi_L H\eta_R
+y_2\overline\Psi_L\widetilde H\chi_R \nonumber\\
&+y_3\overline\Psi_R H\eta_L
+y_4\overline\Psi_R\widetilde H\chi_L \nonumber\\
&+\lambda_\Psi S_B\overline\Psi_L\Psi_R
+\lambda_\eta S_B\overline\eta_R\eta_L \nonumber\\
&+\lambda_\chi S_B\overline\chi_R\chi_L
+\frac{y_\chi}{2}S_B^\dagger
\overline{\chi_L^c}\chi_L
+{\rm h.c.},
\label{eq:Yukawa}
\end{align}
where $H$ is the SM Higgs and $\widetilde H=i\sigma_2 H^*$. These interactions play two distinct roles. The couplings $\lambda_\Psi$, $\lambda_\eta$, and $\lambda_\chi$ generate the dominant masses of the new fermions once $U(1)_B$ is spontaneously broken. The couplings $y_1$ and $y_3$ induce mixing between the charged electroweak fermions after electroweak symmetry breaking, while $y_2$ and $y_4$ connect the neutral component of $\Psi$ to the singlet sector. These electroweak Yukawa interactions are also important phenomenologically because they allow the heavier anomalons to decay, preventing the appearance of stable electrically charged states.

The last term in Eq.~\eqref{eq:Yukawa} is qualitatively different. It violates the global fermion-number symmetry present in the Dirac limit and generates a Majorana mass for the neutral sector. As discussed in Sec.~\ref{subsec:pseudodirac}, this term splits the otherwise Dirac fermion into two nearby Majorana mass eigenstates and thereby provides the small inelastic splitting required for endothermic scattering at LZ.

The most general renormalizable scalar potential involving the SM Higgs and the baryonic scalar is
\begin{align}
V(H,S_B)={}&
-\mu_H^2 H^\dagger H
+\lambda_H(H^\dagger H)^2 \nonumber\\
&-\mu_B^2 S_B^\dagger S_B
+\lambda_B(S_B^\dagger S_B)^2 \nonumber\\
&+\lambda_{HB}(H^\dagger H)(S_B^\dagger S_B).
\label{eq:scalarpotential}
\end{align}
After electroweak and baryonic symmetry breaking, the scalar fields can be
parametrized as
\begin{equation}
H=
\begin{pmatrix}
G^+\\
(v+h_0+iG^0)/\sqrt{2}
\end{pmatrix},
\,\,
S_B=\frac{v_B+s_0+iG_B}{\sqrt{2}},
\label{eq:vevs}
\end{equation}
where $v\simeq246\GeV$ and $v_B$ denotes the scale of spontaneous
$U(1)_B$ breaking. The Goldstone bosons $G^\pm$ and $G^0$ are eaten by the
$W^\pm$ and $Z$ bosons, respectively, while $G_B$ is absorbed by the baryonic
gauge boson $Z_B$. Neglecting gauge kinetic mixing, the tree-level $Z_B$
mass is
\begin{equation}
M_{Z_B}=3g_Bv_B,
\label{eq:MZB}
\end{equation}
where the factor of $3$ follows from the magnitude of the baryon charge
$B(S_B)=-3$.

The same symmetry-breaking scale also generates the dominant Dirac masses of the new fermions,
\begin{equation}
M_\Psi=\frac{\lambda_\Psi v_B}{\sqrt2},
\,\,\,
M_\eta=\frac{\lambda_\eta v_B}{\sqrt2},
\,\,\,
M_\chi=\frac{\lambda_\chi v_B}{\sqrt2}.
\label{eq:Diracmasses}
\end{equation}
The Majorana interaction proportional to $y_\chi$ instead generates the characteristic scale
\begin{equation}
m_M=\frac{y_\chi v_B}{\sqrt2},
\label{eq:Majoranamass}
\end{equation}
which controls the splitting of the pseudo-Dirac neutral state as shown in Sec.~\ref{subsec:pseudodirac}.

Eqs.~\eqref{eq:MZB}--\eqref{eq:Majoranamass} illustrate an important consequence of the UV completion. The mediator mass, the DM mass, the anomalon masses, and the inelastic splitting all originate from the same symmetry-breaking scale $v_B$ and are therefore not completely independent parameters. Once $M_{Z_B}$ and $g_B$ are specified, $v_B$ is fixed through Eq.~\eqref{eq:MZB}, while the remaining masses determine the corresponding Yukawa couplings. For Yukawa couplings of natural size, $\lambda_{\Psi,\eta,\chi}\sim\mathcal O(1)$, the new fermion masses are consequently expected to lie parametrically close to the $U(1)_B$ breaking scale. Considerably larger masses can be obtained only by increasing the corresponding Yukawa couplings. For our numerical domain we impose a conventional perturbativity bound, typically $|y_F|\lesssim\sqrt{4\pi}\simeq3.5$ under the usual perturbativity criterion. Since $M_F\simeq y_F v_B/\sqrt{2}$, this prevents the anomalon
masses from being taken arbitrarily above the $U(1)_B$-breaking scale $v_B$. Conversely, a spectrum containing states much lighter than $v_B$ requires correspondingly small Yukawa couplings. Large separations among the mediator, DM, and anomalon mass scales are therefore possible, but they reflect genuine hierarchies in the underlying couplings rather than independent mass parameters of the low-energy theory.

The CP-even neutral scalars $h_0$ and $s_0$ mix after symmetry breaking through
the Higgs-portal interaction proportional to $\lambda_{HB}$, which generates
off-diagonal terms in the CP-even scalar mass matrix. In the $(h_0,s_0)$ basis, their mass matrix is
\begin{equation}
\mathcal M_h^2=
\begin{pmatrix}
2\lambda_Hv^2 & \lambda_{HB}vv_B\\
\lambda_{HB}vv_B & 2\lambda_Bv_B^2
\end{pmatrix}.
\label{eq:scalarmatrix}
\end{equation}
We define the physical CP-even states $h_1$ and $h_2$ through the rotation
\begin{equation}
\begin{pmatrix}
h_1\\
h_2
\end{pmatrix}
=
\begin{pmatrix}
\cos\theta_B & \sin\theta_B\\
-\sin\theta_B & \cos\theta_B
\end{pmatrix}
\begin{pmatrix}
h_0\\
s_0
\end{pmatrix},
\label{eq:scalarrotation}
\end{equation}
We identify $h_1$ with the observed $125\GeV$ Higgs boson, while $h_2$ is predominantly associated with the baryonic scalar. In the small-mixing limit relevant for our phenomenological analysis, $|\sin\theta_B|\ll1$, one therefore has $h_1\simeq h_0$ and $h_2\simeq s_0$.
The mixing angle is determined by
\begin{equation}
\tan2\theta_B=
\frac{\lambda_{HB}vv_B}
{\lambda_Hv^2-\lambda_Bv_B^2}.
\label{eq:thetaB}
\end{equation}

Scalar exchange also induces elastic SI scattering. The baryonic Yukawa interactions in Eq.~\eqref{eq:Yukawa} couple the neutral fermion to the scalar component $s_0$, whereas SM quarks couple to the Higgs component $h_0$. Scalar mixing allows both mass eigenstates, $h_1$ and $h_2$, to couple to these sectors and mediate elastic DM--nucleon scattering. The resulting portal amplitude is proportional to $\sin\theta_B\cos\theta_B(1/m_{h_1}^2-1/m_{h_2}^2)$. At fixed masses and fermion couplings, the portal contribution to the cross section therefore scales as $\sin^2\theta_B\cos^2\theta_B$ and vanishes as $\theta_B\to0$.

There is also a contribution from neutral-fermion mixing. The electroweak Yukawa interactions give the light neutral state a direct coupling to $h_0$, suppressed by $\sin^2\theta_N$ in the symmetric benchmark. Consequently, the full scalar amplitude need not vanish at $\theta_B=0$ when $\theta_N\neq0$. The two contributions can interfere, making the rate sensitive to the sign of the scalar mixing angle. Their suppression is important because elastic scattering predominantly populates lower recoil energies and could otherwise compete with the endothermic interpretation of the high-energy LZ candidate. App.~\ref{app:subleadingDD} derives the complete scalar amplitude and its recoil contribution.

We take $m_{h_2}=3\TeV$ and $\sin\theta_B=+3\times10^{-4}$ throughout the numerical analysis. The sign is part of the benchmark because it affects scalar interference. Elastic scalar scattering is small at the reference point. Channels containing $h_2$ are closed there, but open at larger DM masses and are included in the implemented freeze-out kernel.

\subsection{Anomalon masses and the physical new-fermion spectrum}
\label{subsec:anomalons}

After spontaneous $U(1)_B$ breaking, the Yukawa couplings $\lambda_\Psi$, $\lambda_\eta$, and $\lambda_\chi$ generate the vectorlike mass parameters introduced in Eq.~\eqref{eq:Diracmasses}. These masses originate from the baryonic symmetry-breaking scale $v_B$ and are therefore already present before electroweak symmetry breaking. In the neutral sector we denote the singlet Dirac mass by
\begin{equation}
M_D\equiv M_\chi=\frac{\lambda_\chi v_B}{\sqrt2},
\label{eq:MDdefinition}
\end{equation}
in order to distinguish this mass parameter from the physical Majorana eigenstates $\chi_1$ and $\chi_2$ introduced in Sec.~\ref{subsec:pseudodirac}.

Throughout this work, we assume a CP-conserving new-fermion sector and take its Yukawa couplings to be real. The mass matrices, mixing angles, and numerical analysis below use this convention. The Yukawa couplings $y_i$ to the SM Higgs generate additional mass terms after electroweak symmetry breaking. We define
\begin{equation}
m_i\equiv\frac{y_i v}{\sqrt2},
\,\,
i=1,\ldots,4,
\label{eq:EWmassinsertions}
\end{equation}
where $v\simeq246\GeV$. The quantities $m_i$ are not physical fermion masses, but electroweak-induced mass insertions connecting new fermions carrying the same electric charge.

In the charged sector, the negatively charged component of the electroweak doublet $\Psi$ mixes with the charged singlet $\eta$ through $y_1\overline{\Psi}_L H\eta_R
+
y_3\overline{\Psi}_R H\eta_L
+
{\rm h.c.}$,
which generate the mass insertions $m_1$ and $m_3$. Together with the $U(1)_B$-induced masses $M_\Psi$ and $M_\eta$, these terms lead to
\begin{equation}
\mathcal M_-=
\begin{pmatrix}
M_\Psi & m_1\\
m_3 & M_\eta
\end{pmatrix}.
\label{eq:chargedmatrix}
\end{equation}
Since $\mathcal M_-$ is in general not symmetric, independent rotations of the left- and right-handed fields are required,
\begin{equation}
(U_-^L)^\dagger\mathcal M_-U_-^R
=
{\rm diag}
\left(
m_{X_1},
m_{X_2}
\right),
\label{eq:chargeddiag}
\end{equation}
where $X_1^\pm$ and $X_2^\pm$ denote the physical charged anomalons. In the small charged-mixing regime,
\begin{equation}
|m_1|,|m_3|
\ll
|M_\Psi-M_\eta|,
\label{eq:smallchargedmixing}
\end{equation}
their masses remain close to the corresponding vectorlike mass parameters.

The neutral sector is particularly important for the DM phenomenology. The neutral component $\Psi^0$ of the electroweak doublet mixes with the SM-singlet fermion $\chi$ through $y_2\overline{\Psi}_L\widetilde H\chi_R
+
y_4\overline{\Psi}_R\widetilde H\chi_L
+
{\rm h.c.}$,
which generate $m_2$ and $m_4$.
Before including the small Majorana interaction proportional to $y_\chi$, the neutral Dirac mass matrix is therefore
\begin{equation}
\mathcal M_0^{\rm Dirac}
=
\begin{pmatrix}
M_\Psi & m_2\\
m_4 & M_D
\end{pmatrix}.
\label{eq:neutralmatrix}
\end{equation}

The off-diagonal entries in Eq.~\eqref{eq:neutralmatrix} provide the electroweak bridge between the singlet and doublet neutral sectors. Since the matrix is not symmetric for generic $y_2\neq y_4$, independent rotations are required for its left- and right-handed components. We define
\begin{align}
\begin{pmatrix}
\Psi_{L/R}^0\\
\chi_{L/R}
\end{pmatrix}
&=
\begin{pmatrix}
\cos\theta_{L/R} & -\sin\theta_{L/R}\\
\sin\theta_{L/R} & \cos\theta_{L/R}
\end{pmatrix}
\begin{pmatrix}
N_{HL/R}\\
N_{LL/R}
\end{pmatrix}.
\label{eq:neutralrotationsmain}
\end{align}
The fields $N_H$ and $N_L$ denote the heavy and light neutral Dirac eigenstates before the Majorana splitting is introduced. In the parameter region considered below, $N_H$ is predominantly the neutral component of the electroweak doublet, whereas $N_L$ is predominantly the SM-singlet state.

The corresponding bi-unitary diagonalization is
\begin{equation}
(U_0^L)^\dagger\mathcal M_0^{\rm Dirac}U_0^R
=
{\rm diag}
\left(
m_{N_H},
m_{N_L}
\right).
\label{eq:neutraldiracdiag}
\end{equation}
For real parameters, the two mixing angles satisfy
\begin{align}
\tan2\theta_L
&=
\frac{
2\left(
M_\Psi m_4+M_Dm_2
\right)
}{
M_\Psi^2+m_2^2-M_D^2-m_4^2
},
\nonumber\\
\tan2\theta_R
&=
\frac{
2\left(
M_\Psi m_2+M_Dm_4
\right)
}{
M_\Psi^2+m_4^2-M_D^2-m_2^2
}.
\label{eq:thetaLRmain}
\end{align}
In the limit in which the electroweak-induced entries are much smaller than the vectorlike masses, i.e.~in the small-mixing regime,
\begin{equation}
|m_2|,|m_4|\ll M_\Psi,M_D,
\end{equation}
and the mixing angles reduce to
\begin{align}
\theta_L
&\simeq
\frac{
M_\Psi m_4+M_Dm_2
}{
M_\Psi^2-M_D^2
},
\nonumber\\
\theta_R
&\simeq
\frac{
M_\Psi m_2+M_Dm_4
}{
M_\Psi^2-M_D^2
}.
\label{eq:thetaLRsmallmain}
\end{align}
Consequently, the light neutral state can be written approximately as
\begin{align}
N_{LL}
&\simeq
\chi_L-\theta_L\Psi_L^0,
\nonumber\\
N_{LR}
&\simeq
\chi_R-\theta_R\Psi_R^0.
\label{eq:lightneutralcomposition}
\end{align}
The quantities $\sin^2\theta_L$ and $\sin^2\theta_R$ therefore measure the electroweak-doublet fractions carried by the two chiral components of the light state.

This distinction is important because the analysis performed here does not assume the exact singlet limit that would correspond to
\begin{equation}
y_2=y_4=0
\, \Longleftrightarrow \,
m_2=m_4=0
\,\Longleftrightarrow\,
\theta_L=\theta_R=0.
\label{eq:exactsingletlimit}
\end{equation}
In this limit the light neutral fermion $N_L$ carries no electroweak-doublet component and therefore has no tree-level coupling inherited from $\Psi^0$.

Instead, throughout our phenomenological analysis we adopt a \emph{singlet-dominated}, but not exactly singlet, configuration with $\sin^2\theta_L\ll1$, $\sin^2\theta_R\ll1$.
The distinction is phenomenologically relevant. A very small doublet component has a negligible impact on the dominant baryonic interactions responsible for the LZ signal and on the leading relic-density and collider phenomenology, but it provides a coupling of the light pseudo-Dirac sector to the electroweak gauge bosons. As discussed in detail in App.~\ref{app:excitedstate}, the resulting decay channels make the heavier state $\chi_2$ unstable on a timescale much shorter than the age of the Universe. Consequently, the primordial $\chi_2$ population is efficiently depleted and its present-day cosmological abundance is negligible. The Galactic halo can therefore be treated to excellent accuracy as being composed entirely of the lighter state $\chi_1$. This removes the potentially large exothermic DD contribution from $\chi_2 N\to\chi_1 N$ and suppresses present-day ID signals associated with the otherwise efficient coannihilation process $\chi_1\chi_2\to q\bar q$. The nonzero mixing therefore allows the model to retain the endothermic $\chi_1 N\to\chi_2 N$ transition relevant for the LZ event while avoiding phenomenological constraints associated with a relic population of the excited state. Possible regeneration of $\chi_2$ in dense environments, in particular inside the Sun through inelastic scattering, is treated separately in the corresponding Solar analysis.

The reference benchmark uses real and equal neutral Yukawa couplings,
\begin{equation}
y_2=y_4\equiv y_N.
\label{eq:symmetricneutralYukawa}
\end{equation}
In this limit, $m_2=m_4\equiv t$,
the neutral Dirac mass matrix is symmetric and the two rotations coincide, $\theta_L=\theta_R\equiv\theta_N$.
The exact relation between the mixing angle and the mass parameters then becomes
\begin{equation}
\tan2\theta_N
=
\frac{2t}{M_\Psi-M_D}.
\label{eq:symmetricneutralmixing}
\end{equation}
For small $\theta_N$ this reduces to the intuitive expression
\begin{equation}
\theta_N
\simeq
\frac{t}{M_\Psi-M_D}
=
\frac{y_Nv}{
\sqrt2\left(
M_\Psi-M_D
\right)
}.
\label{eq:thetaNyN}
\end{equation}

Our reference choice for $m_\chi=1\TeV$ and $M_\Psi=1.5\TeV$ is
\begin{equation}
\theta_L=\theta_R=\theta_N=0.005,
\label{eq:mixingbenchmarkangle}
\end{equation}
which corresponds to
\begin{equation}
m_2=m_4\simeq2.50\GeV,
\,\,
y_2=y_4\simeq1.437\times10^{-2}.
\label{eq:mixingbenchmarkyukawa}
\end{equation}
The doublet probability of the light neutral state is therefore only
\begin{equation}
\sin^2\theta_N
\simeq
2.50\times10^{-5},
\label{eq:mixingbenchmarkfraction}
\end{equation}
so that the state remains more than $99.997\%$ singlet at the wave-function probability level. The benchmark is consequently extremely close to the singlet limit for the observables controlled by the dominant $U(1)_B$ interaction, while being physically distinct from the exact $y_2=y_4=0$ configuration.

For numerical work it is convenient to keep the physical light Dirac eigenmass fixed when the mixing is switched on. Defining
\begin{equation}
\Delta_\Psi
\equiv
M_\Psi-m_\chi,
\end{equation}
the symmetric benchmark can be parametrized exactly as
\begin{align}
t
&=
\Delta_\Psi\tan\theta_N,
\nonumber\\
M_D
&=
m_\chi
+
\frac{t^2}{\Delta_\Psi}.
\label{eq:MDmixingadjustment}
\end{align}
For $m_\chi=1\TeV$, $M_\Psi=1.5\TeV$, and $\theta_N=0.005$, this gives $t\simeq2.50002\GeV$, $M_D\simeq1000.0125\GeV$.
We retain this second-order mass shift. The implemented charged sector also has $y_1=y_3=y_2=y_4$, with $M_\Psi=1.5m_\chi$ and $M_\eta=2m_\chi$; thus the charged anomalons are mixed mass eigenstates.
Throughout the scan, we hold $\theta_N=0.005$ fixed and adjust the common electroweak Yukawa coupling according to Eq.~\eqref{eq:MDmixingadjustment}, $y_1=y_2=y_3=y_4=\sqrt2\,\Delta_\Psi\tan\theta_N/v$, so that these couplings vary with $m_\chi$ and take the value $0.01437$ only at $m_\chi=1\TeV$.

\subsection{Pseudo-Dirac dark matter mass splitting}
\label{subsec:pseudodirac}

We now focus on the light neutral DM state. As discussed in Sec.~\ref{subsec:anomalons}, diagonalization of the neutral Dirac mass matrix defines the predominantly singlet light state $N_L$ and the predominantly doublet heavy state $N_H$, with their chiral components related to the interaction basis through the mixing angles $\theta_L$ and $\theta_R$, see Eq.~\eqref{eq:neutralrotationsmain}. Throughout our phenomenological analysis we work in the small-mixing regime, $\theta_L,\theta_R\ll1$, so that $N_L$ remains overwhelmingly composed of $\chi_L$ and $\chi_R$, while retaining a small but phenomenologically relevant electroweak-doublet admixture.

An important consequence of the $U(1)_B$ charge assignment is that the singlet fermion admits both a Dirac mass and a Majorana mass, while at the renormalizable level the latter is generated only for the left-handed singlet component. The relevant interactions are
\begin{equation}
-\mathcal L_\chi\supset\lambda_\chi S_B\overline{\chi_R}\chi_L+\frac{y_\chi}{2}S_B^\dagger\overline{\chi_L^c}\chi_L+{\rm h.c.}
\label{eq:chimasslag}
\end{equation}
The first interaction generates the dominant Dirac mass of the singlet fermion. The second interaction is gauge invariant because $B(\chi_L)=-3/2$ and $B(S_B^\dagger)=+3$, and generates a small Majorana mass for $\chi_L$. By contrast, with the field content and baryon-number assignments of Tab.~\ref{tab:fields}, no analogous renormalizable Majorana operator exists for $\chi_R$. The asymmetric Majorana structure is therefore a consequence of the minimal scalar realization rather than an additional phenomenological assumption.
After $U(1)_B$ breaking, $M_D$ is the singlet Dirac mass parameter defined in Eq.~\eqref{eq:MDdefinition}, while $m_M$ is the Majorana mass defined in Eq.~\eqref{eq:Majoranamass}.

Because the Majorana operator acts on the interaction-basis field $\chi_L$, its projection onto the light Dirac eigenstate is controlled specifically by the left-handed mixing angle. Substituting Eq.~\eqref{eq:neutralrotationsmain} for $\chi_L$ and $\chi_R$ into the second term of Eq.~\eqref{eq:chimasslag} gives
\begin{align}
-\mathcal L_M\supset\frac{m_M}{2}\bigg[&\cos^2\theta_L\,\overline{N_{LL}^c}N_{LL}
 +\sin^2\theta_L\,\overline{N_{HL}^c}N_{HL}\nonumber\\
&+\sin\theta_L\cos\theta_L\,\overline{N_{HL}^c}N_{LL}\nonumber\\
&+\sin\theta_L\cos\theta_L\,\overline{N_{LL}^c}N_{HL}\bigg]
 +{\rm h.c.}
\label{eq:MajoranaProjected}
\end{align}
This expression makes explicit that the original Majorana mass is distributed among the light and heavy neutral eigenstates once the electroweak mixing is nonzero.

For the light state, the relevant effective Majorana mass is therefore
\begin{equation}
m_M^{\rm eff}=m_M\cos^2\theta_L.
\label{eq:effectiveMajorana}
\end{equation}
The terms proportional to $\sin^2\theta_L$ and $\sin\theta_L\cos\theta_L$ in Eq.~\eqref{eq:MajoranaProjected} affect the heavy neutral anomalon and generate an extremely small Majorana mixing between the heavy and light neutral sectors. Since $m_M\ll m_{N_H}-m_{N_L}$ and $\theta_L\ll1$, their effect on the light pseudo-Dirac spectrum is completely negligible for the benchmark considered below.

The pseudo-Dirac system should therefore be written in terms of the chiral components of the light Dirac eigenstate rather than directly in terms of the unmixed interaction-basis fields. In the left-handed basis
\begin{equation}
\mathcal N_L=\begin{pmatrix}N_{LL}\\N_{LR}^c\end{pmatrix},
\end{equation}
the effective mass matrix is
\begin{equation}
\mathcal M_{\rm light}=\begin{pmatrix}m_M^{\rm eff}&m_{N_L}\\m_{N_L}&0\end{pmatrix},
\label{eq:lightpseudodiracmatrix}
\end{equation}
where $m_{N_L}$ denotes the light Dirac eigenmass obtained after diagonalizing the neutral Dirac mass matrix of Eq.~\eqref{eq:neutralmatrix}.

The corresponding physical masses are
\begin{align}
m_{\chi_2/\chi_1}&=\frac{\sqrt{\left(m_M^{\rm eff}\right)^2+4m_{N_L}^2} \pm m_M^{\rm eff}}{2},
\label{eq:chimasses}
\end{align}
where $\chi_1$ denotes the lighter state and $\chi_2$ the heavier state. Within the effective light-state theory, their mass splitting is therefore
\begin{equation}
\delta_\chi\equiv m_{\chi_2}-m_{\chi_1}=m_M^{\rm eff}=m_M\cos^2\theta_L.
\label{eq:deltachi}
\end{equation}
Using Eq.~\eqref{eq:Majoranamass}, this relation becomes
\begin{equation}
\delta_\chi=\frac{y_\chi v_B}{\sqrt2}\cos^2\theta_L.
\label{eq:deltachiyukawa}
\end{equation}
Corrections to Eq.~\eqref{eq:deltachi} from the last term of Eq.~\eqref{eq:MajoranaProjected} arise through mixing of the light pseudo-Dirac pair with the much heavier neutral anomalon and are parametrically suppressed by additional powers of
\begin{equation}
\frac{m_M\sin\theta_L}{m_{N_H}-m_{N_L}}.
\end{equation}
They are entirely negligible for $m_M$ at the sub-MeV scale and a heavy-light mass separation of several hundred GeV.

For $m_M^{\rm eff}\ll m_{N_L}$, the two physical masses reduce to
\begin{equation}
m_{\chi_{1,2}}\simeq m_{N_L}\mp\frac{m_M^{\rm eff}}{2}=m_{N_L}\mp\frac{\delta_\chi}{2},
\label{eq:pseudodiracmasses}
\end{equation}
and the two Majorana states form a pseudo-Dirac pair. In the positive-mass Takagi convention, their left-handed components approach
\begin{align}
\chi_{1L}&\simeq\frac{i}{\sqrt2}\left(N_{LL}-N_{LR}^c\right),\\
\chi_{2L}&\simeq\frac{1}{\sqrt2}\left(N_{LL}+N_{LR}^c\right).
\label{eq:pseudodiracstates}
\end{align}
Thus, in the limit $y_\chi\rightarrow0$, the Majorana splitting vanishes and the two states recombine into the single Dirac fermion $N_L$. In this limit the renormalizable fermion sector acquires an additional conserved fermion-number symmetry, while the electroweak Yukawa couplings $y_2$ and $y_4$ preserve the Dirac character of the neutral states. The small hierarchy $m_M\ll m_{N_L}$ is therefore technically natural in the 't~Hooft sense: setting $y_\chi$ to zero increases the symmetry of the theory. As a consequence, radiative corrections to the Majorana mass remain proportional to the same symmetry-breaking parameter $y_\chi$, and the presence of small but nonzero $\theta_L$ and $\theta_R$ does not destabilize the sub-MeV splitting.

The physical significance of the two neutral mixing angles is somewhat different. Since the renormalizable Majorana operator in Eq.~\eqref{eq:chimasslag} acts only on $\chi_L$, the leading correction to the pseudo-Dirac splitting depends directly on $\theta_L$, as shown in Eq.~\eqref{eq:deltachi}. Both $\theta_L$ and $\theta_R$, however, determine the electroweak neutral current inherited by the light Dirac state. Before the Majorana splitting one obtains
\begin{equation}
\mathcal L_Z\supset\frac{g}{2c_W}Z_\mu\overline{N_L}\gamma^\mu\left(\sin^2\theta_L P_L+\sin^2\theta_R P_R\right)N_L,
\label{eq:lightZcurrent}
\end{equation}
which contains both a vector component proportional to $\sin^2\theta_L+\sin^2\theta_R$ and an axial component proportional to $\sin^2\theta_R-\sin^2\theta_L$,
\begin{align}
\mathcal L_Z\supset{} \frac{g}{4c_W} Z_\mu\overline{N_L}\gamma^\mu
&\bigg[\sin^2\theta_L+\sin^2\theta_R
\nonumber\\
&+\left(\sin^2\theta_R-\sin^2\theta_L\right)\gamma^5\bigg]N_L.
\label{eq:lightZcurrentVA}
\end{align}

For $\theta_L=\theta_R$, the axial component vanishes and the neutral current is purely vectorlike at this order. After the Majorana splitting, this vector current becomes predominantly an off-diagonal $\chi_1$--$\chi_2$ transition current, providing an electroweak connection between the two members of the pseudo-Dirac pair without generating a comparable diagonal axial interaction. This will be explained in Sec.~\ref{subsec:current}. Conversely, $\theta_L\neq\theta_R$ produces a nonzero axial component and therefore additional diagonal electroweak interactions, requiring a dedicated reassessment of DD and Solar phenomenology. We therefore adopt the symmetric reference configuration
\begin{equation}
y_2=y_4,\,\,\theta_L=\theta_R\equiv\theta_N.
\label{eq:symmetricmixingpseudo}
\end{equation}
The consequences of the resulting transition current for the decay and cosmological depletion of $\chi_2$ are discussed in App.~\ref{app:excitedstate}.
The exact singlet case, $y_2=y_4=0,\,\,\theta_L=\theta_R=0$,
would remove the electroweak transition current in Eq.~\eqref{eq:lightZcurrent}.

For the reference physical splitting $\delta_\chi=300\keV$, Eq.~\eqref{eq:deltachi} requires
\begin{equation}
m_M=\frac{\delta_\chi}{\cos^2\theta_L}.
\label{eq:mMfromdelta}
\end{equation}
For $\theta_L=0.005$, this gives
\begin{equation}
m_M\simeq300.008\keV.
\label{eq:mMbenchmark}
\end{equation}
The difference from $300\keV$ is only at the level of a few eV, but we include it consistently in the definition of the benchmark. Using $v_B\simeq2.11\TeV$, the Majorana Yukawa coupling is
\begin{equation}
y_\chi=\frac{\sqrt2\,m_M}{v_B}\simeq2.0\times10^{-7}.
\label{eq:ychibenchmark}
\end{equation}
The Dirac Yukawa remains of order unity,
\begin{equation}
\lambda_\chi=\frac{\sqrt2M_D}{v_B}\simeq0.67,
\label{eq:lambdachibenchmark}
\end{equation}
where the tiny difference between $M_D$ and the physical light Dirac eigenmass induced by the neutral mixing has been retained in the numerical implementation.

The complete neutral-sector reference benchmark used in the phenomenological analysis is therefore
\begin{align}
m_\chi&=1\TeV,& \delta_\chi&=300\keV,& M_\Psi&=1.5\TeV,\nonumber\\
\theta_L&=0.005,& \theta_R&=0.005,& y_2=y_4&=0.01437,\nonumber\\
M_{Z_B}&=1.44\TeV,& g_B&=0.22782,& v_B&\simeq2.11\TeV.
\label{eq:neutralFullBenchmarkPseudo}
\end{align}
Together with $M_\eta=2.0\TeV$, this benchmark gives
\begin{equation}
\lambda_\Psi\simeq1.01,\,\,\lambda_\eta\simeq1.34,\,\,\lambda_\chi\simeq0.67,\,\, y_\chi\simeq2.0\times10^{-7}.
\label{eq:fullYukawaBenchmarkPseudo}
\end{equation}
The hierarchy among these couplings has a simple physical interpretation. The order-one couplings $\lambda_\Psi$, $\lambda_\eta$, and $\lambda_\chi$ generate the TeV-scale masses associated with $U(1)_B$ breaking; the smaller electroweak Yukawas $y_2$ and $y_4$ generate the small singlet-doublet admixture; and the much smaller $y_\chi$ controls the technically natural sub-MeV Majorana splitting.

App.~\ref{app:uvsplitting} discusses two illustrative UV origins of the small Majorana Yukawa coupling---a symmetry-protected higher-dimensional operator and a schematic dark-sector seesaw---neither of which is required as an additional assumption in the scan.

\subsection{Off-diagonal baryonic current and DD matching}
\label{subsec:current}

A central feature of the model is that the interaction responsible for the LZ signal is not imposed as an effective off-diagonal coupling. Rather, it follows directly from the $U(1)_B$ charges of the light pseudo-Dirac fermion after transforming first from the interaction basis to the neutral Dirac mass basis and subsequently to the Majorana mass basis.

As discussed in Sec.~\ref{subsec:anomalons}, before the small Majorana mass is introduced the light neutral Dirac state $N_L$ is predominantly composed of the singlet fields $\chi_L$ and $\chi_R$, with a small admixture of the neutral doublet components $\Psi_L^0$ and $\Psi_R^0$. 
Using the rotations defined in Eq.~\eqref{eq:neutralrotationsmain}, the diagonal $Z_B$ coupling of the light Dirac state can be written as
\begin{equation}
\mathcal L_{N_LZ_B}=g_BZ_{B\mu}\overline{N_L}\gamma^\mu\left(B_L^{\rm eff}P_L+B_R^{\rm eff}P_R\right)N_L+\ldots,
\label{eq:NLZBchiral}
\end{equation}
where the ellipsis denotes interactions involving the heavy neutral eigenstate and
\begin{align}
B_L^{\rm eff}&=B_1\sin^2\theta_L+B_2\cos^2\theta_L=-\frac32-3\sin^2\theta_L,\\
B_R^{\rm eff}&=B_2\sin^2\theta_R+B_1\cos^2\theta_R=-\frac92+3\sin^2\theta_R.
\label{eq:effectiveBcharges}
\end{align}
It is useful to introduce the corresponding effective vector and axial charges,
\begin{equation}
B_V^{\rm eff}=\frac{B_R^{\rm eff}+B_L^{\rm eff}}{2}=-3+\frac32\left(\sin^2\theta_R-\sin^2\theta_L\right),
\label{eq:BVeff}
\end{equation}
and
\begin{equation}
B_A^{\rm eff}=\frac{B_R^{\rm eff}-B_L^{\rm eff}}{2}=-\frac32+\frac32\left(\sin^2\theta_L+\sin^2\theta_R\right).
\label{eq:BAeff}
\end{equation}
The baryonic current of the light Dirac state is therefore
\begin{equation}
\mathcal L_{N_LZ_B}=g_BZ_{B\mu}\overline{N_L}\gamma^\mu\left(B_V^{\rm eff}+B_A^{\rm eff}\gamma^5\right)N_L+\ldots.
\label{eq:DiracZBcurrent}
\end{equation}

For the symmetric neutral-mixing configuration adopted throughout the numerical analysis, $\theta_L=\theta_R\equiv\theta_N$,
the vector charge simplifies to $B_V^{\rm eff}=-3$,
independently of the value of $\theta_N$, while the axial charge becomes $B_A^{\rm eff}=-3/2\cos\left(2\theta_N\right)$.
The exact cancellation in the vector part follows from the particular $U(1)_B$ charge assignments: the left-handed doublet component carries the same baryon charge as the right-handed singlet component, while the right-handed doublet component carries the same charge as the left-handed singlet component. Consequently, an equal left- and right-handed singlet--doublet admixture leaves the vector baryonic charge of the light state unchanged.

The effect of the small Majorana mass becomes transparent by decomposing the light Dirac state into the two Majorana eigenstates. In the pseudo-Dirac limit, consistently with Eq.~\eqref{eq:pseudodiracstates}, we write
\begin{equation}
N_L=\frac{1}{\sqrt2}\left(\chi_2-i\chi_1\right),\,\, N_L^c=\frac{1}{\sqrt2}\left(\chi_2+i\chi_1\right),
\label{eq:DiracMajoranaDecomp}
\end{equation}
with $\chi_i^c=\chi_i$. The corresponding vector and axial currents satisfy
\begin{align}
\overline{N_L}\gamma^\mu N_L&=i\overline{\chi_1}\gamma^\mu\chi_2,\\
\overline{N_L}\gamma^\mu\gamma^5N_L&=\frac12\left(\overline{\chi_1}\gamma^\mu\gamma^5\chi_1+\overline{\chi_2}\gamma^\mu\gamma^5\chi_2\right).
\label{eq:MajoranaCurrentRelations}
\end{align}
The diagonal vector current of each individual Majorana state vanishes identically, whereas the vector current connecting the two members of the pseudo-Dirac pair remains nonzero.

Substituting Eq.~\eqref{eq:MajoranaCurrentRelations} into Eq.~\eqref{eq:DiracZBcurrent} gives
\begin{align}
\mathcal L_{\chi Z_B}\simeq{}&-3ig_BZ_{B\mu}\overline{\chi_1}\gamma^\mu\chi_2\nonumber\\
&-\frac{3g_B}{4}\cos\left(2\theta_N\right)Z_{B\mu}\left(\overline{\chi_1}\gamma^\mu\gamma^5\chi_1+\overline{\chi_2}\gamma^\mu\gamma^5\chi_2\right).
\label{eq:MajoranaZBcurrent}
\end{align}
We retain the Hermitian transition current $J_{12}^\mu=i\overline{\chi_1}\gamma^\mu\chi_2$. With $\chi_i^c=\chi_i$, its factor of $i$ cannot simply be dropped while keeping real coefficients: the off-diagonal vector bilinear is antisymmetric. The diagonal term in Eq.~\eqref{eq:MajoranaZBcurrent} carries the standard Majorana factor $1/2$, so its physical vertex is $g_B B_A^{\rm eff}$, rather than $g_B B_A^{\rm eff}/2$. The physically relevant result is that, for the symmetric benchmark,
\begin{equation}
\left|g_{\chi_1\chi_2Z_B}\right|=3g_B, \,\, \cos\left(2\theta_N\right)\simeq1.
\label{eq:transitionZBcoupling}
\end{equation}

Corrections associated with the departure from the exact pseudo-Dirac limit remain suppressed by $\delta_\chi/m_\chi\sim3\times10^{-7}$ for the reference benchmark. The neutral mixing also generates $Z_B$ interactions connecting the light and heavy neutral Dirac eigenstates, proportional to $\sin\theta_N\cos\theta_N$. These couplings are relevant for the phenomenology of the heavy anomalons but do not contribute to low-energy DD involving the light pseudo-Dirac pair.

For a generic Dirac fermion $f$ with left- and right-handed $U(1)_B$ charges $B_L^f$ and $B_R^f$, we write the interaction with the baryonic gauge boson as
\begin{equation}
\mathcal L_{Z_B f}
=
Z_{B\mu}\overline f\gamma^\mu
\left(
g_V^f+g_A^f\gamma^5
\right)f,
\label{eq:ZBfermionVA}
\end{equation}
with
\begin{equation}
g_V^f=\frac{g_B}{2}\left(B_L^f+B_R^f\right),
\,\,
g_A^f=\frac{g_B}{2}\left(B_R^f-B_L^f\right).
\label{eq:ZBVAcharges}
\end{equation}

For SM quarks, the left- and right-handed fields carry the same baryon charge, $B_L^q=B_R^q=1/3$,
so the interaction is purely vectorial,
\begin{equation}
g_V^q=\frac{g_B}{3},
\,\,
g_A^q=0,
\end{equation}
and can be written as
\begin{equation}
\mathcal L_{qZ_B}
=
\frac{g_B}{3}
Z_{B\mu}
\sum_q\overline q\gamma^\mu q.
\label{eq:quarkZBcurrent}
\end{equation}
This coupling controls both resonant production of $Z_B$ at hadron colliders and its visible decay into dijets.

By contrast, SM leptons carry no baryon number, $B_L^\ell=B_R^\ell=0$,
and therefore have no direct tree-level coupling to $Z_B$,
\begin{equation}
g_V^\ell=g_A^\ell=0.
\label{eq:ZBleptonTree}
\end{equation}
The model is thus leptophobic at the renormalizable level. A small coupling to charged leptons and neutrinos is nevertheless induced through hypercharge--baryon kinetic mixing. After electroweak symmetry breaking and diagonalization of the neutral gauge sector, the physical $Z_B$ acquires suppressed couplings to the SM electromagnetic and neutral currents,
\begin{equation}
\mathcal L_{Z_B\ell}^{\rm kin}
=
Z_{B\mu}\,
\overline\ell\gamma^\mu
\left(
g_{V,\rm kin}^{\ell}
+
g_{A,\rm kin}^{\ell}\gamma^5
\right)\ell,
\label{eq:ZBleptonKinetic}
\end{equation}
with $g_{V,\rm kin}^{\ell},g_{A,\rm kin}^{\ell}\propto\epsilon_{YB}$. In our numerical analysis, we set the renormalized kinetic mixing to $\epsilon_{YB}=0$ and neglect its loop-induced effects that are negligible. In this approximation, $Z_B$ has no induced couplings to SM leptons, while its direct coupling to quarks is $g_B/3$. Consequently, leptonic decay modes are neglected, and dijet searches provide the collider constraints on the mediator used in our analysis.

\subsection{Residual gauge symmetry and DM stability}
\label{subsec:residual}

An important consequence of the spontaneous breaking of $U(1)_B$ is the emergence of a residual discrete gauge symmetry that stabilizes the DM candidate. The stability of $\chi_1$ therefore does not require imposing an additional ad hoc $Z_2$ symmetry on the Lagrangian.

To make the residual symmetry transparent, it is convenient to use the integer normalization $Q_B=6B$.
In this normalization, the SM quarks have $Q_B(q)=2$,
while the new fermionic chiralities have charges $Q_B=-27$, $Q_B=-9$,
and the scalar responsible for baryonic symmetry breaking carries $Q_B(S_B)=-18$.
When $S_B$ develops a vacuum expectation value, the continuous $U(1)_B$ symmetry is broken, but transformations satisfying
\begin{equation}
e^{iQ_B(S_B)\alpha}=1
\end{equation}
leave the vacuum invariant. The spontaneous breaking therefore leaves a residual discrete $Z_{18}$ subgroup of the original gauge symmetry.

Of particular relevance is the $Z_2$ subgroup generated by
\begin{equation}
P_B=(-1)^{Q_B}=(-1)^{6B}.
\label{eq:PB}
\end{equation}
Under this transformation, all SM fields are even. Indeed, SM quarks have the even integer charge $Q_B=2$, while leptons and the Higgs have $Q_B=0$. By contrast, the new fermions have odd charges, $Q_B=-27$ or $-9$, and are therefore odd under $P_B$. The scalar $S_B$ is even because $Q_B(S_B)=-18$.
Consequently, the new fermion sector is separated from the SM by an exact parity inherited from the broken gauge symmetry. If the lightest parity-odd particle is the neutral state $\chi_1$, it cannot decay exclusively into SM particles, since such a decay would violate $P_B$. Requiring the physical charged and heavier neutral anomalons to lie above $\chi_1$ therefore guarantees the stability of the DM candidate.
This mechanism is conceptually important because DM stability follows directly from the gauge structure and the symmetry-breaking pattern. The stabilizing parity is a residual discrete gauge symmetry rather than a global symmetry imposed by hand, and is consequently tied to the same $U(1)_B$ structure responsible for anomaly cancellation, the mediator interactions, and the pseudo-Dirac mass splitting.

\subsection{Representative processes and connection to the phenomenology}
\label{subsec:processesmodel}

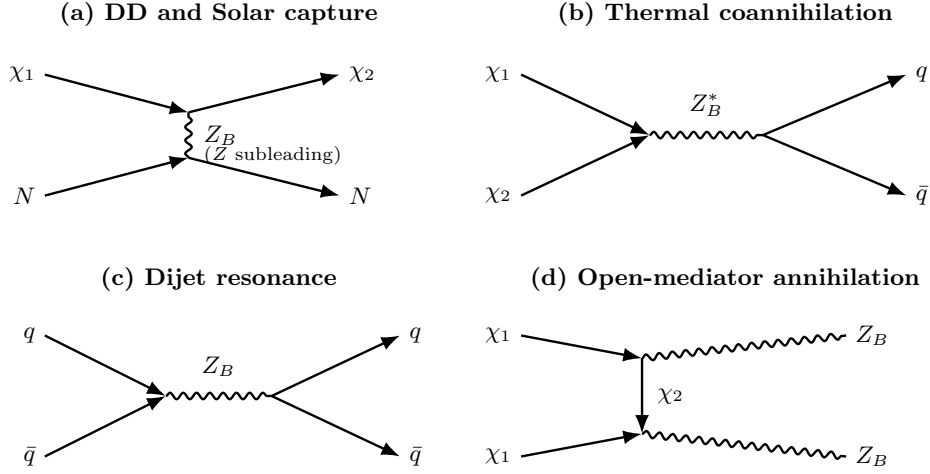
\begin{figure*}[tbp]
\centering
\begin{tikzpicture}[>=Latex,line width=0.9pt,font=\small]

\node[font=\bfseries] at (2.4,4.35) {(a) DD and Solar capture};
\draw[->] (0.1,3.55) -- (2.0,3.05) node[pos=0,left] {$\chi_1$};
\draw[->] (2.0,3.05) -- (4.0,3.55) node[pos=1,right] {$\chi_2$};
\draw[->] (0.1,1.95) -- (2.0,2.45) node[pos=0,left] {$N$};
\draw[->] (2.0,2.45) -- (4.0,1.95) node[pos=1,right] {$N$};
\draw[decorate,decoration={snake,amplitude=1.2pt,segment length=5pt}] (2.0,2.45) -- (2.0,3.05);
\node[right] at (2.08,2.75) {$Z_B$};
\node[right,font=\scriptsize] at (2.08,2.48) {$(Z\ {\rm subleading})$};

\node[font=\bfseries] at (9.1,4.35) {(b) Thermal coannihilation};
\draw[->] (6.4,3.55) -- (8.1,2.75) node[pos=0,left] {$\chi_1$};
\draw[->] (6.4,1.95) -- (8.1,2.75) node[pos=0,left] {$\chi_2$};
\draw[decorate,decoration={snake,amplitude=1.2pt,segment length=5pt}] (8.1,2.75) -- (9.6,2.75);
\node[above] at (8.85,2.88) {$Z_B^\ast$};
\draw[->] (9.6,2.75) -- (11.5,3.55) node[pos=1,right] {$q$};
\draw[->] (9.6,2.75) -- (11.5,1.95) node[pos=1,right] {$\bar q$};

\node[font=\bfseries] at (2.4,0.85) {(c) Dijet resonance};
\draw[->] (0.1,0.1) -- (1.7,-0.7) node[pos=0,left] {$q$};
\draw[->] (0.1,-1.5) -- (1.7,-0.7) node[pos=0,left] {$\bar q$};
\draw[decorate,decoration={snake,amplitude=1.2pt,segment length=5pt}] (1.7,-0.7) -- (3.1,-0.7);
\node[above] at (2.4,-0.56) {$Z_B$};
\draw[->] (3.1,-0.7) -- (4.8,0.1) node[pos=1,right] {$q$};
\draw[->] (3.1,-0.7) -- (4.8,-1.5) node[pos=1,right] {$\bar q$};

\node[font=\bfseries] at (9.1,0.85) {(d) Open-mediator annihilation};
\draw[->] (6.4,0.1) -- (8.0,-0.2) node[pos=0,left] {$\chi_1$};
\draw[->] (6.4,-1.5) -- (8.0,-1.2) node[pos=0,left] {$\chi_1$};
\draw[->] (8.0,-0.2) -- (8.0,-1.2);
\node[right] at (8.08,-0.70) {$\chi_2$};
\draw[decorate,decoration={snake,amplitude=1.2pt,segment length=5pt}] (8.0,-0.2) -- (10.7,0.1) node[pos=1,right] {$Z_B$};
\draw[decorate,decoration={snake,amplitude=1.2pt,segment length=5pt}] (8.0,-1.2) -- (10.7,-1.5) node[pos=1,right] {$Z_B$};

\end{tikzpicture}
\caption{Representative tree-level processes of the $U(1)_B$ completion. Panel (a) shows the endothermic transition relevant for DD and Solar capture. The dominant mediator is $Z_B$, while the nonzero singlet--doublet mixing generates an additional mixing-suppressed contribution mediated by the SM $Z$ boson. Panel (b) shows the $\chi_1\chi_2$ coannihilation channel that controls thermal freeze-out over the heavy-mediator branch; electroweak channels induced by the small doublet admixture are subleading and are not shown. Panel (c) shows resonant production and decay of $Z_B$ into dijets at the LHC. Panel (d) shows the $Z_BZ_B$ annihilation channel that becomes kinematically accessible for $M_{Z_B}<m_{\chi_1}$.}
\label{fig:feynman}
\end{figure*}

The structure developed above implies that several apparently different observables are controlled by the same small set of underlying parameters. Fig.~\ref{fig:feynman} summarizes the representative processes that will be studied in detail in Sec.~\ref{sec:pheno}. In addition to the dominant interactions mediated by the baryonic gauge boson $Z_B$, the nonzero singlet--doublet mixing introduced in Sec.~\ref{subsec:anomalons} generates additional electroweak interactions of the light pseudo-Dirac states. For the reference choice $\theta_L=\theta_R=0.005$, these contributions are strongly suppressed and do not modify the dominant phenomenological picture, but they are important for the decay and depletion of the excited state $\chi_2$, as discussed in App.~\ref{app:excitedstate}.

The first process is the endothermic transition
\begin{equation}
\chi_1 N\rightarrow\chi_2 N,
\end{equation}
shown in Fig.~\ref{fig:feynman}a. The dominant contribution is mediated by $Z_B$ through the off-diagonal vector current derived in Sec.~\ref{subsec:current}. The same interaction governs DD experiments on Earth and the inelastic component of DM capture in the Sun. The positive splitting $\delta_\chi=m_{\chi_2}-m_{\chi_1}>0$ modifies the scattering kinematics and selects the high-velocity component of the incoming DM distribution, which is the essential ingredient allowing the model to populate the high-recoil region observed by LZ.

The nonzero doublet admixture also induces an additional off-diagonal interaction mediated by the SM $Z$ boson. For the symmetric benchmark $\theta_L=\theta_R\equiv\theta_N$, this contribution is proportional to $\sin^2\theta_N$ at the amplitude level. For $\theta_N=0.005$, the corresponding effective nucleon amplitude is at the few-$10^{-3}$ level relative to the baryonic one and therefore has a negligible impact on the LZ recoil spectrum and on the dominant Solar-capture rate. We nevertheless retain the electroweak mixing in the model because the same interaction plays an essential role in the decay of $\chi_2$.

The second representative process, shown in Fig.~\ref{fig:feynman}b, is
\begin{equation}
\chi_1\chi_2\rightarrow Z_B^\ast\rightarrow q\bar q.
\end{equation}
Because the dominant baryonic vector interaction is off diagonal in the pseudo-Dirac basis, coannihilation between $\chi_1$ and $\chi_2$ plays an important role during thermal freeze-out. Since $\delta_\chi\ll T_f$, where $T_f$ denotes the freeze-out temperature, both states are populated with essentially equal abundances during chemical decoupling. The sub-MeV splitting relevant for DD therefore has a negligible effect on their relative thermal abundance at freeze-out.

The nonzero values of $\theta_L$ and $\theta_R$ also generate electroweak annihilation and coannihilation processes involving the $Z$, the SM Higgs, and the heavier electroweak anomalons. Their rates are suppressed by the small doublet admixture and, for the benchmark $\theta_N=0.005$, remain subleading with respect to the baryonic channels considered in the relic-density analysis. The freeze-out phenomenology is therefore still dominantly controlled by $M_{Z_B}$, $g_B$, and the masses of the pseudo-Dirac states. The small electroweak mixing can nevertheless be retained straightforwardly in the complete numerical implementation.

The third process, shown in Fig.~\ref{fig:feynman}c, is resonant production of the baryonic gauge boson at hadron colliders,
\begin{equation}
q\bar q\rightarrow Z_B\rightarrow q\bar q.
\end{equation}
Since $Z_B$ couples directly to the SM baryon current, the model predicts a dijet resonance whose production rate and width depend on the same gauge coupling $g_B$ that enters DD and freeze-out. Existing ATLAS and CMS dijet searches therefore provide a direct constraint on the LZ-motivated parameter space, while the larger Run-3 data set offers an important future test of the model. This process is essentially unaffected by the small neutral singlet--doublet mixing because the production and decay of $Z_B$ into SM quarks are controlled directly by the gauged baryon current.

The nonzero electroweak mixing additionally implies electroweak interactions of the heavy neutral and charged anomalons and therefore opens complementary collider signatures involving their production and decay. These signatures depend on the anomalon masses and on the Yukawa couplings $y_i$ and are conceptually distinct from the dijet constraint on $Z_B$. A complete collider analysis of the anomalon sector is beyond the scope of the present work.

Finally, when $M_{Z_B}<m_{\chi_1}$, the annihilation channel
\begin{equation}
\chi_1\chi_1\rightarrow Z_BZ_B
\end{equation}
becomes kinematically accessible, as illustrated in Fig.~\ref{fig:feynman}d. Each mediator subsequently decays predominantly into SM quarks, producing a four-fermion final state. This channel is relevant for assessing the open-mediator region of parameter space and its possible ID and Solar-annihilation signatures. Its leading rate is controlled by the baryonic gauge interaction and is modified only at a negligible level by the small singlet--doublet mixing of the benchmark.

The important feature of Fig.~\ref{fig:feynman} is therefore not simply that these processes coexist in the same model, but that they are correlated. The same mediator mass $M_{Z_B}$ and gauge coupling $g_B$ enter the dominant LZ scattering rate, the thermal relic abundance, Solar capture, dijet production, and, where kinematically allowed, present-day annihilation. The neutral-sector parameters $\theta_L$, $\theta_R$, and $\delta_\chi$ provide an additional connection between these observables and the cosmological evolution of the excited state. In particular, the very small electroweak admixture has only a negligible impact on the dominant $Z_B$ phenomenology while being sufficient to provide efficient decay channels for $\chi_2$. The UV completion consequently turns the LZ interpretation into a multi-observable framework in which the DD signal, thermal history, present-day state composition, collider signatures, and Solar phenomenology are simultaneously correlated.

The gauge structure and anomalon charge assignment are closely related to other anomaly-free gauged-baryon-number constructions~\cite{Fileviez2019,AnomalyFreeDM,DuHuangXie}. The realization considered here is distinguished by its scalar-minimal structure: the same field $S_B$ that breaks $U(1)_B$ generates the vectorlike masses of the new fermions and provides the single renormalizable Majorana mass term responsible for the pseudo-Dirac splitting. The gauge-boson mass is
\begin{equation}
M_{Z_B}=3g_Bv_B,
\end{equation}
while in the presence of the small singlet--doublet mixing the physical pseudo-Dirac splitting is
\begin{equation}
\delta_\chi=\frac{y_\chi v_B}{\sqrt2}\cos^2\theta_L,
\label{eq:deltaSummaryProcesses}
\end{equation}
up to the completely negligible corrections associated with mixing of the light pseudo-Dirac pair with the heavy neutral anomalon. In the exact singlet limit $\theta_L\rightarrow0$, Eq.~\eqref{eq:deltaSummaryProcesses} reduces to $\delta_\chi=y_\chi v_B/\sqrt2$. In constructions containing independent scalar fields for the two chiral Majorana masses, the splitting instead depends on additional independent symmetry-breaking parameters.

The analysis below exploits this reduced parameter freedom by solving the thermal relic-density problem and consistently confronting the same parameter space with DD, collider searches, ID, and Solar-capture constraints. The small but nonzero neutral mixing is retained throughout where relevant, and its most important consequence, the decay and depletion of the excited state $\chi_2$, is discussed in detail in App.~\ref{app:excitedstate}.

\section{Phenomenological signatures and constraints}
\label{sec:pheno}

The usefulness of the UV completion is that the leading phenomenological observables are correlated.  On the central branch of parameter space, the same pair $(M_{Z_B},g_B)$ controls the coherent endothermic DD transition, the coannihilation rate at freeze-out, and the visible dijet resonance rate.  ID and Solar capture then test the late-time state composition and the channels that remain open once the excited state has been depleted.  In this section we discuss each ingredient separately before combining them in Sec.~\ref{sec:results}.

Unless stated otherwise, throughout this section we use as a reference benchmark $m_\chi=1\TeV$, $\delta_\chi=300\keV$, $M_{Z_B}=1.44\TeV$, and $g_B=0.2278$, corresponding through $M_{Z_B}=3g_Bv_B$ to $v_B\simeq2.11\TeV$. The neutral sector is taken in the singlet-dominated but nonzero-mixing configuration $\theta_L=\theta_R=0.005$, corresponding to $y_2=y_4\simeq0.01437$ for $M_\Psi=1.5\TeV$. The remaining parameters are fixed to $m_{h_2}=3\TeV$, $|\sin\theta_B|=3\times10^{-4}$, $M_\Psi=1.5\TeV$, and $M_\eta=2.0\TeV$.

\subsection{Direct detection and the LZ high-recoil candidate}
\label{subsec:LZ}

The momentum transfers relevant for xenon DD are at most of order a few hundred MeV, whereas the assessed scan starts at $M_{Z_B}=50\GeV$. Since $q^2\ll M_{Z_B}^2$, the baryonic mediator can be integrated out. As shown in Sec.~\ref{subsec:current}, for the symmetric neutral-mixing configuration $\theta_L=\theta_R=\theta_N$ the off-diagonal baryonic vector coupling remains $|g_{\chi_1\chi_2Z_B}|=3g_B$, so the leading $Z_B$-mediated interaction relevant for the high-energy LZ signal is unchanged with respect to the exact-singlet limit. At the quark level one obtains
\begin{equation}
\mathcal L_{\rm eff}^{q}=\frac{g_B^2}{M_{Z_B}^2}J_{12\mu}\sum_q\overline q\gamma^\mu q,
\label{eq:effquark}
\end{equation}
with the Hermitian current defined in Sec.~\ref{subsec:current} ($J_{12}^\mu=i\overline{\chi_1}\gamma^\mu\chi_2$); an additional Hermitian-conjugate term is unnecessary.

Introducing the conserved baryon current $J_B^\mu=1/3\sum_q\overline q\gamma^\mu q$, Eq.~\eqref{eq:effquark} becomes
\begin{equation}
\mathcal L_{\rm eff}^{B}=\frac{3g_B^2}{M_{Z_B}^2}J_{12\mu}J_B^\mu.
\label{eq:effbaryon}
\end{equation}
The nucleon matrix element of the baryon current is normalized as
\begin{equation}
\langle N|J_B^\mu|N\rangle=\overline N\gamma^\mu N,
\end{equation}
because both the proton and neutron carry baryon number one. The effective transition coupling to either nucleon is therefore
\begin{equation}
G_N^B=\frac{3g_B^2}{M_{Z_B}^2},
\label{eq:GN}
\end{equation}
so the baryonic vector interaction is isoscalar, $f_p^B=f_n^B=G_N^B$.

For a nucleus containing $Z$ protons and $A-Z$ neutrons, the vector amplitudes add coherently,
\begin{equation}
f_A^B=Zf_p^B+(A-Z)f_n^B=AG_N^B.
\label{eq:nuclearcoupling}
\end{equation}
The conventional zero-momentum per-nucleon normalization is then
\begin{equation}
\sigma_N^0=\frac{\mu_{\chi N}^2}{\pi}\left(G_N^B\right)^2=\frac{9\mu_{\chi N}^2g_B^4}{\pi M_{Z_B}^4},
\label{eq:sigmaN}
\end{equation}
where $\mu_{\chi N}$ denotes the DM--nucleon reduced mass.

We stress that $\sigma_N^0$ is only a convenient normalization of the interaction strength. For endothermic scattering, the physical process
\begin{equation}
\chi_1+A\rightarrow\chi_2+A
\end{equation}
cannot occur at zero momentum transfer because the incoming DM particle must provide the energy required to produce the heavier state $\chi_2$. The LZ event rate must therefore be calculated from the complete recoil spectrum using the full inelastic kinematics as done later in this section (see Eqs.~\ref{eq:rate} and \ref{eq:dsigmadER}).

The baryonic interaction also contains a diagonal axial current (see Eq.~\eqref{eq:MajoranaZBcurrent}). In the symmetric benchmark the effective axial charge of the light Dirac state is
\begin{equation}
B_A^{\rm eff}=-\frac{3}{2}\cos(2\theta_N),
\end{equation}
and, after the pseudo-Dirac state is decomposed into $\chi_1$ and $\chi_2$, the axial current becomes diagonal. The coupling of the lighter state can be written as
\begin{equation}
\mathcal L_{\chi_1\chi_1Z_B}^{A}
=
\frac{g_B B_A^{\rm eff}}{2}\,
Z_{B\mu}\,
\overline{\chi_1}\gamma^\mu\gamma^5\chi_1.
\label{eq:axialZBchi1}
\end{equation}
Integrating out $Z_B$ therefore generates the elastic axial-DM--vector-nucleon operator
\begin{equation}
\mathcal L_{\rm eff}^{A,B}
=
-\frac12G_{A,N}^{B}
\left(\overline{\chi_1}\gamma^\mu\gamma^5\chi_1\right)
\left(\overline N\gamma_\mu N\right),
\label{eq:axialZBeffective}
\end{equation}
where $G_{A,N}^{B}$ denotes the signed physical vertex product used by the recoil code,
\begin{equation}
G_{A,N}^{B}
=
\frac{g_B^2B_A^{\rm eff}}{M_{Z_B}^2}
=
-\frac{3g_B^2}{2M_{Z_B}^2}\cos(2\theta_N)
\end{equation}
Equivalently,
\begin{equation}
G_{A,N}^{B}=-\frac{\cos(2\theta_N)}{2}\,G_N^B.
\label{eq:axialVectorRatio}
\end{equation}
For $\theta_N=0.005$, $\cos(2\theta_N)\simeq0.99995$, so the axial coupling itself is not strongly suppressed by neutral-fermion mixing. Its scattering amplitude is nevertheless nonrelativistically suppressed. The axial-DM--vector-nucleon interaction generates the operators $\mathcal O_8=\mathbf S_\chi\cdot\mathbf v^\perp$ and $\mathcal O_9=i\mathbf S_\chi\cdot(\mathbf S_N\times\mathbf q/m_N)$, where $\mathbf S_\chi$ and $\mathbf S_N$ are the DM and nucleon spins, $\mathbf q$ is the momentum transfer, and $\mathbf v^\perp$ is the transverse relative velocity. Thus the $\mathcal O_8$ contribution is velocity suppressed but retains a coherent nuclear response while the $\mathcal O_9$ contribution instead carries an explicit factor $q^2/m_N^2$ in the squared amplitude and probes nuclear spin without the coherent $A^2$ enhancement.
These suppressions make the axial interaction negligible for the normalization of the high-energy endothermic signal around the LZ candidate. It nevertheless mediates elastic scattering, $\chi_1A\to\chi_1A$, without an endothermic threshold. It therefore becomes comparatively important at low recoil energies, where the off-diagonal vector transition is kinematically forbidden, and generates the residual low-energy tail discussed below.

A further contribution arises from the nonzero singlet--doublet mixing and is absent in the exact-singlet limit. As discussed in Sec.~\ref{subsec:pseudodirac}, for $\theta_L=\theta_R=\theta_N$ the light Dirac state inherits a purely vectorial coupling to the SM $Z$ boson,
\begin{equation}
\mathcal L_{N_LZ}=\frac{g}{2c_W}\sin^2\theta_N\,Z_\mu\overline{N_L}\gamma^\mu N_L,
\label{eq:lightZvectorCurrent}
\end{equation}
which becomes an off-diagonal transition current after transforming to the Majorana basis,
\begin{equation}
\mathcal L_{\chi Z}\simeq i\frac{g}{2c_W}\sin^2\theta_N\,Z_\mu\overline{\chi_1}\gamma^\mu\chi_2.
\label{eq:MajoranaZtransition}
\end{equation}
The SM $Z$ therefore mediates the same endothermic process $\chi_1N\rightarrow\chi_2N$ as the baryonic vector interaction, but its amplitude is suppressed by the small doublet probability $\sin^2\theta_N$.

Integrating out the $Z$ gives the effective vector couplings
\begin{equation}
G_p^Z=-\frac{\sin^2\theta_N}{v^2}\left(\frac12-2s_W^2\right),\,\,
G_n^Z=+\frac{\sin^2\theta_N}{2v^2},
\label{eq:GNZ}
\end{equation}
so, including both vector mediators, the coherent transition coupling to a nucleus is
\begin{equation}
f_A^{\rm vec}=AG_N^B+ZG_p^Z+(A-Z)G_n^Z.
\label{eq:fullNuclearVectorCoupling}
\end{equation}
These signs include the relative sign of the two mediator exchanges for $J_{12}^\mu$. We use $C_p=G_N^B+G_p^Z$ and $C_n=G_N^B+G_n^Z$ in the numerical modules. The weak contribution is not isoscalar.

For the benchmark $\theta_N=0.005$ one has $\sin^2\theta_N\simeq2.5\times10^{-5}$ and
\begin{equation}
|G_n^Z|\simeq\frac{\sin^2\theta_N}{2v^2}\simeq2.1\times10^{-10}\GeV^{-2},
\label{eq:GnZestimate}
\end{equation}
whereas for $g_B=0.22782$ and $M_{Z_B}=1.44\TeV$ the baryonic vector interaction gives
\begin{equation}
G_N^B=\frac{3g_B^2}{M_{Z_B}^2}\simeq7.5\times10^{-8}\GeV^{-2}.
\label{eq:GNbenchmark}
\end{equation}
The corresponding diagonal baryonic axial coupling is
\begin{equation}
|G_{A,N}^{B}|\simeq3.75\times10^{-8}\GeV^{-2},
\end{equation}
although, as emphasized above, its contribution to the scattering rate is strongly reduced by its nonrelativistic velocity and momentum dependence. By contrast, the $Z$-mediated vector amplitude is only at the few-$10^{-3}$ level relative to the dominant baryonic vector amplitude even for neutrons and becomes still smaller relative to the coherent baryonic xenon amplitude once the proton and neutron weak charges are combined. Its effect on the recoil spectrum is therefore negligible for the reference benchmark.

The DD phenomenology is controlled by three qualitatively different contributions. The off-diagonal vector $Z_B$ interaction dominates the endothermic transition $\chi_1N\to\chi_2N$ and fixes the normalization required to reproduce the high-energy LZ candidate, while the reverse process $\chi_2N\to\chi_1N$ is included according to the excited-state halo fraction. The diagonal axial $Z_B$ interaction instead gives an elastic, nonrelativistically suppressed contribution that survives below the endothermic threshold and controls the residual low-energy tail; in our calculation it is treated in the coherent $\mathcal O_8$ nuclear-response approximation. We also include scalar-mediated SI elastic scattering. Finally, the mixing-induced SM $Z$ contribution is off diagonal and endothermic but numerically negligible for $\theta_N=0.005$, so the one-event determination of $g_B$ and the high-energy recoil spectrum are essentially unchanged by the neutral mixing.

The same electroweak admixture is nevertheless phenomenologically important for another reason: nonzero $\theta_L$ and $\theta_R$ generate the decay and efficient depletion of the excited state $\chi_2$, as discussed in App.~\ref{app:excitedstate}. Mixing between the SM Higgs and the baryonic scalar also induces an elastic SI interaction. For the small scalar and neutral-fermion mixings adopted in the benchmark this contribution is subleading; its derivation, together with the detailed nonrelativistic treatment of the axial $Z_B$ interaction and the other suppressed DD terms, is presented in App.~\ref{app:subleadingDD}.

For a nucleus of mass $m_A$, the minimum incoming DM speed required to produce a recoil energy $E_R$ is
\begin{equation}
v_{\rm min}^{A}(E_R)=\frac{1}{\sqrt{2m_AE_R}}\left(\frac{m_AE_R}{\mu_{\chi A}}+\delta_\chi\right),
\label{eq:vmininelastic}
\end{equation}
where $\mu_{\chi A}$ is the DM--nucleus reduced mass. The function $v_{\rm min}(E_R)$ has a minimum at
\begin{equation}
E_R^\star=\frac{\mu_{\chi A}}{m_A}\delta_\chi,\,\, v_{\rm min}^\star=\sqrt{\frac{2\delta_\chi}{\mu_{\chi A}}},
\label{eq:ridge}
\end{equation}
while the largest splitting that can be accessed for a given DM mass and a maximum laboratory-frame speed $v_{\rm max}$ is
\begin{equation}
\delta_\chi^{\rm max}=\frac12\mu_{\chi A}v_{\rm max}^2.
\label{eq:ceiling}
\end{equation}
These two quantities encode different information. The first identifies the recoil energy at which the inelastic transition is kinematically easiest, whereas the second defines the hard kinematic boundary above which no halo particle can induce the transition.

\begin{figure}[htbp]
\centering
\includegraphics[width=0.98\columnwidth]{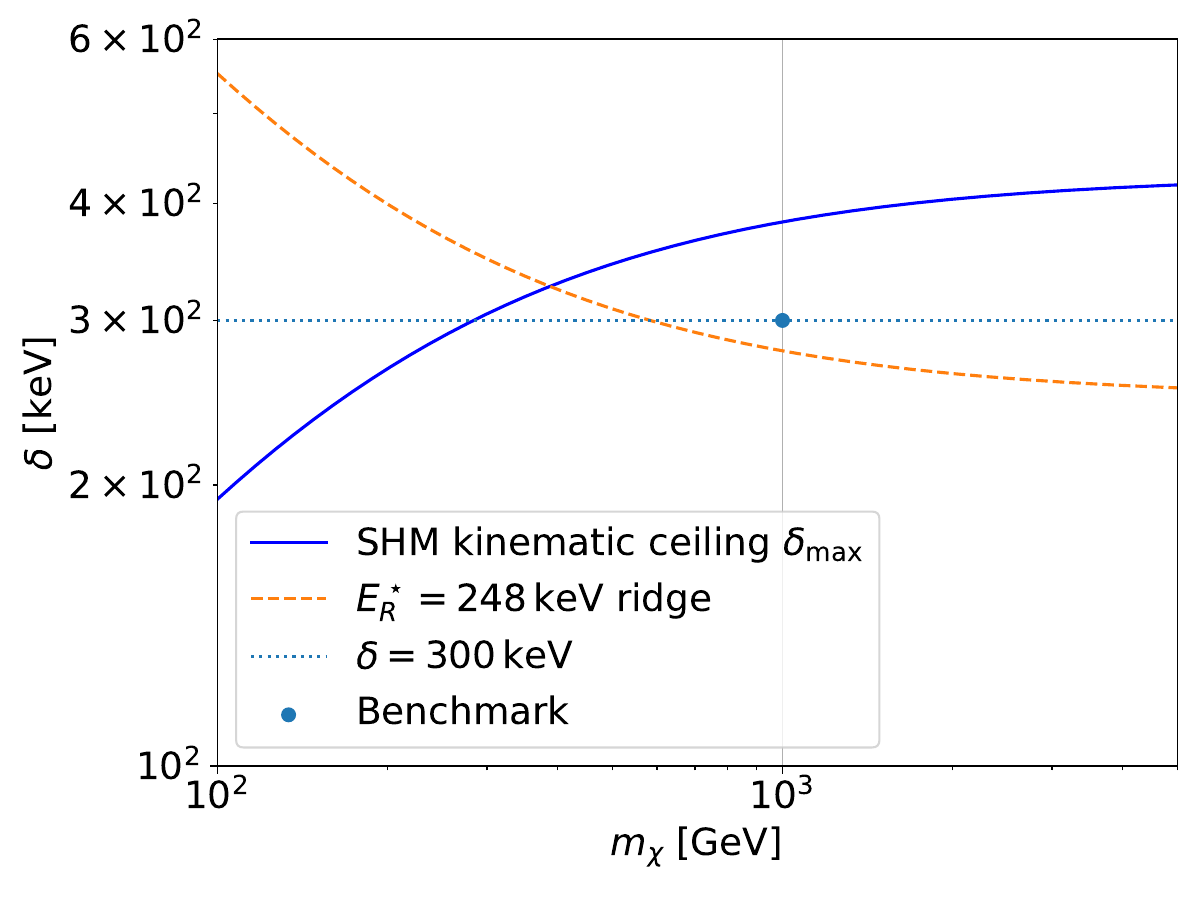}
\caption{Endothermic xenon kinematics in the $(m_\chi,\delta_\chi)$ plane for the SHM parameters adopted in this work. The blue solid curve shows the maximum kinematically accessible splitting for the effective isotope $A=131.3$, $\delta_\chi^{\rm max}=\mu_{\chi A}v_{\rm max}^2/2$, while the orange dashed curve gives the locus for which the minimum of $v_{\rm min}(E_R)$ occurs at $E_R=248\keV$. The horizontal dotted line denotes the reference splitting $\delta_\chi=300\keV$, and the point marks the $m_\chi=1\TeV$ benchmark.}
\label{fig:kinematic}
\end{figure}

For $m_\chi=1\TeV$, requiring the minimum of $v_{\rm min}(E_R)$ to occur exactly at the LZ candidate energy, $E_R^\star=248\keV$,
gives approximately $\delta_\chi\simeq278\keV$.
The reference value $\delta_\chi=300\keV$ therefore places the minimum of the velocity requirement at a recoil energy somewhat above the observed candidate.
Fig.~\ref{fig:kinematic} isolates this purely kinematic information in the $(m_\chi,\delta_\chi)$ plane. The blue solid curve shows the largest splitting that can be accessed for the adopted halo model from Eq.~\eqref{eq:ceiling}, whereas the dashed curve identifies the combinations of $(m_\chi,\delta_\chi)$ for which the minimum of $v_{\rm min}$ occurs exactly at $E_R=248\keV$. The solid curve is therefore a genuine kinematic boundary, while the dashed curve is not an exclusion limit but simply the locus where the recoil energy of minimum velocity requirement coincides with the observed event.
For $\delta_\chi=300\keV$, scattering becomes inaccessible below approximately $m_\chi\simeq284\GeV$
for the effective isotope $A=131.3$ used in the kinematic diagnostic and the SHM parameters adopted below. At the $m_\chi=1\TeV$ benchmark, the maximum accessible splitting is approximately $\delta_\chi^{\rm max}\simeq382\keV$,
while the $248\keV$ ridge corresponds to $\delta_\chi\simeq278\keV$. The benchmark therefore lies above the ridge but safely below the SHM kinematic ceiling. In the limit $m_\chi\gg m_A$, the reduced mass approaches the nuclear mass, $\mu_{\chi A}\rightarrow m_A$, so the kinematic benefit of increasing $m_\chi$ eventually saturates. This is why a single high-energy recoil has relatively little ability to discriminate between TeV-scale and multi-TeV DM once the heavy-mass regime is reached.

For a xenon isotope with mass number $A$, the coherent transition cross section in the baryonic limit is
\begin{equation}
\frac{d\sigma_A}{dE_R}
=
\frac{m_A}{2\pi v^2}
A^2\left[G_N^B\right]^2F_A^2(q),
\label{eq:dsigmadER}
\end{equation}
where $q=\sqrt{2m_AE_R}$ and $F_A(q)$ is the Helm form factor. The weak correction replaces $AG_N^B$ by $f_A^{\rm vec}$ in the nuclear amplitude.
Summing over the natural xenon isotopic composition, the differential recoil rate is
\begin{equation}
\frac{dR}{dE_R}
=
\frac{\rho_\chi}{m_\chi}
\sum_A\xi_A
\frac{A^2\left[G_N^B\right]^2}{2\pi}
F_A^2(q)
\eta\!\left[
v_{\rm min}^{A}(E_R)
\right],
\label{eq:rate}
\end{equation}
where $\rho_\chi$ is the local DM density taken to be $0.3$ GeV/cm$^3$, $\xi_A$ denotes the isotope mass fraction in the rate per detector mass. The halo dependence is
\begin{equation}
\eta(v_{\rm min})
=
\int_{|\mathbf v|>v_{\rm min}}
d^3v\,
\frac{f_{\rm lab}(\mathbf v)}{v}
\end{equation}
This is the usual inverse-speed integral.

For the baseline analysis we adopt the Standard Halo Model (SHM) with
\begin{equation}
v_0=238\kms,
\,\,
v_{\rm esc}=544\kms,
\,\,
v_E=250.2\kms.
\label{eq:SHMparameters}
\end{equation}
The Galactic-frame velocity distribution is taken to be an isotropic Maxwell--Boltzmann distribution with characteristic speed $v_0$, truncated at the Galactic escape speed $v_{\rm esc}$ and boosted into the laboratory frame by the Earth speed $v_E$. These parameters are commonly used in recent DD studies and coincide, for example, with those adopted in the DarkSide-50 signal modeling~\cite{DarkSide50Bayesian}. With this prescription, the maximum laboratory-frame speed is
\begin{equation}
v_{\rm max}
=
v_{\rm esc}+v_E
\simeq794.2\kms.
\label{eq:vmaxSHM}
\end{equation}

The precise choice of the velocity distribution is particularly important in the present case. For splittings of several hundred keV, Eq.~\eqref{eq:vmininelastic} selects particles in the extreme high-velocity tail of the local DM distribution. For the reference values $m_\chi=1\TeV$ and $\delta_\chi=300\keV$, one finds $v_{\rm min}\simeq704\kms$ around the recoil region of the LZ candidate, to be compared with the SHM maximum speed $v_{\rm max}\simeq794.2\kms$. The predicted spectrum is consequently much more sensitive to the halo model than in conventional low-energy elastic scattering. Changes in the high-speed tail can alter both the maximum accessible splitting and the relative normalization of the different recoil-energy regions~\cite{OHareHighVelocity,McCabeSeasonal}.

An especially relevant modification is induced by the recent infall of the Large Magellanic Cloud (LMC). Numerical simulations indicate two physically distinct effects. First, DM particles originally associated with the LMC can populate the fastest part of the local velocity distribution. Second, the gravitational response of the MW halo to the LMC can accelerate native Galactic DM particles to larger velocities~\cite{BeslaPeterGaravitoLMC,SmithOrlikLMC}. Both effects enhance the population at velocities near and above the canonical SHM cutoff. Since the endothermic process considered here probes precisely this region, the LMC can increase the kinematically accessible splitting, change the relative population of the recoil lobes, and modify the interaction strength required to reproduce the LZ candidate. We study this effect explicitly in App.~\ref{app:lmc_tail}, where the SHM result is compared with the present-day MW-only and MW+LMC distributions extracted from cosmological simulations.
In particular, the enhanced high-speed population in the adopted MW--LMC model reduces the baryonic coupling required to reproduce a fixed signal-window count by approximately $30\%$ for $\delta_\chi=300\keV$ and $73\%$ for $\delta_\chi=350\keV$ relative to its SHM comparison, as quantified in App.~\ref{app:lmc_tail}. These illustrative reductions do not directly rescale the combined allowed model parameter region, since the modified halo velocity profile also changes the recoil spectrum and would require a new DD fit.

For the nuclear response we use the Helm form factor~\cite{LewinSmith1996,Helm1956},
\begin{equation}
F_A(q)
=
3\frac{j_1(qR_1)}{qR_1}
\exp\left(
-\frac{q^2s^2}{2}
\right),
\label{eq:HelmFF}
\end{equation}
with
\begin{equation}
R_1=\sqrt{c_A^2+\frac73\pi^2 a^2-5s^2},
\end{equation}

Here $c_A=\left[1.23A^{1/3}-0.60\right]\,{\rm fm}$, $a=0.52\,{\rm fm}$, and $s=0.9\,{\rm fm}$, matching the production calculation. At the recoil energies relevant for the LZ candidate, the detailed form of $F_A(q)$ is important. The spherical Bessel function $j_1(qR_1)$ possesses a sequence of diffraction zeros, so the coherent nuclear response does not decrease monotonically with recoil energy. These oscillations are a consequence of the finite spatial extent of the nucleus: the nuclear form factor is essentially the Fourier transform of the nucleon density distribution, and at sufficiently large momentum transfer amplitudes from different regions of the nucleus interfere destructively, producing diffraction minima. The structure therefore reflects finite-size nuclear diffraction and the progressive loss of coherence, rather than detailed nuclear shell effects. Summing over the natural xenon isotopic abundances smooths the exact zeros because the isotopes have slightly different nuclear radii, but the characteristic multi-lobed structure remains visible.

The quantity directly relevant for the comparison with the LZ data is the expected number of events in a specified recoil-energy interval,
\begin{equation}
N(E_R^{\rm min},E_R^{\rm max})
=
\mathcal{E}
\int_{E_R^{\rm min}}^{E_R^{\rm max}}
dE_R\,
\epsilon(E_R)\,
\frac{dR}{dE_R},
\label{eq:NexpLZ}
\end{equation}
where $\mathcal E$ denotes the effective exposure normalization that is $\mathcal E_{\rm LZ}=2.84~{\rm tonne\,yr}$, and $\epsilon(E_R)$ is the smooth acceptance approximation reported in \cite{DiMauroLZ}. Its turn-on and turn-off follow the published high-energy response scales~\cite{LZ2026}; it is not the official detector-space likelihood. In the simplified likelihood adopted here, Eq.~\eqref{eq:NexpLZ} is evaluated in two recoil-energy intervals. The high-energy interval,
\begin{equation}
200\keV\leq E_R\leq300\keV,
\end{equation}
is used as the signal region containing the observed LZ candidate. We therefore define the expected number of events in the signal region as
\begin{equation}
N_{\rm sig}
=
N(200\keV,300\keV),
\label{eq:NsigLZ}
\end{equation}
and, for each $(m_\chi,\delta_\chi,M_{Z_B})$, determine the central LZ coupling by requiring $N_{\rm sig}=1$. The exact central $68.27\%$ Poisson interval associated with one observed event is then propagated into the corresponding allowed interval in $g_B$.
The adjacent lower-energy interval,
\begin{equation}
0\keV\leq E_R<200\keV,
\end{equation}
is instead used as a control region, for which no corresponding event is observed in our simplified two-bin treatment. We define
\begin{equation}
N_{\rm low}
=
N(0\keV,200\keV),
\label{eq:NlowLZ}
\end{equation}
and use the ratio $N_{\rm low}/N_{\rm sig}$ to test whether a given value of $\delta_\chi$ would predict an excessive population of lower-energy recoils. Thus the $200$--$300\keV$ interval determines the one-event normalization in $g_B$, whereas the $0$--$200\keV$ interval provides additional information on the recoil-spectrum shape and is used in the $\delta_\chi$ likelihood selection.
These analysis windows should not be confused with the experimental uncertainty on the reconstructed recoil energy of the candidate. The latter is used separately when discussing the kinematic range of $\delta_\chi$, while the broader $200$--$300\keV$ interval is the signal-counting window of the simplified likelihood.

In the analysis we combine the central $68.27\%$ Poisson interval for one event in the signal window with a separate one-sided $95\%$ upper-count criterion in the low-energy window. Their intersection defines a simplified compatibility selection, without an assigned joint confidence level, and does not reproduce the LZ likelihood. The assumption $n_{\rm low}=0$ is used as a diagnostic requirement rather than as the observed event count over the full low-energy LZ sample.

The event prediction therefore depends on the complete recoil spectrum within each relevant energy interval, including the inelastic kinematics, the halo integral, the xenon isotopic composition, the momentum-dependent nuclear response, and the detector efficiency. It cannot be inferred directly from the nominal zero-momentum quantity $\sigma_N^0$.
In Sec.~\ref{subsec:deltaprofile} the details of this analysis will be further explained.

\begin{figure}[htbp]
\centering
\includegraphics[width=0.98\columnwidth]{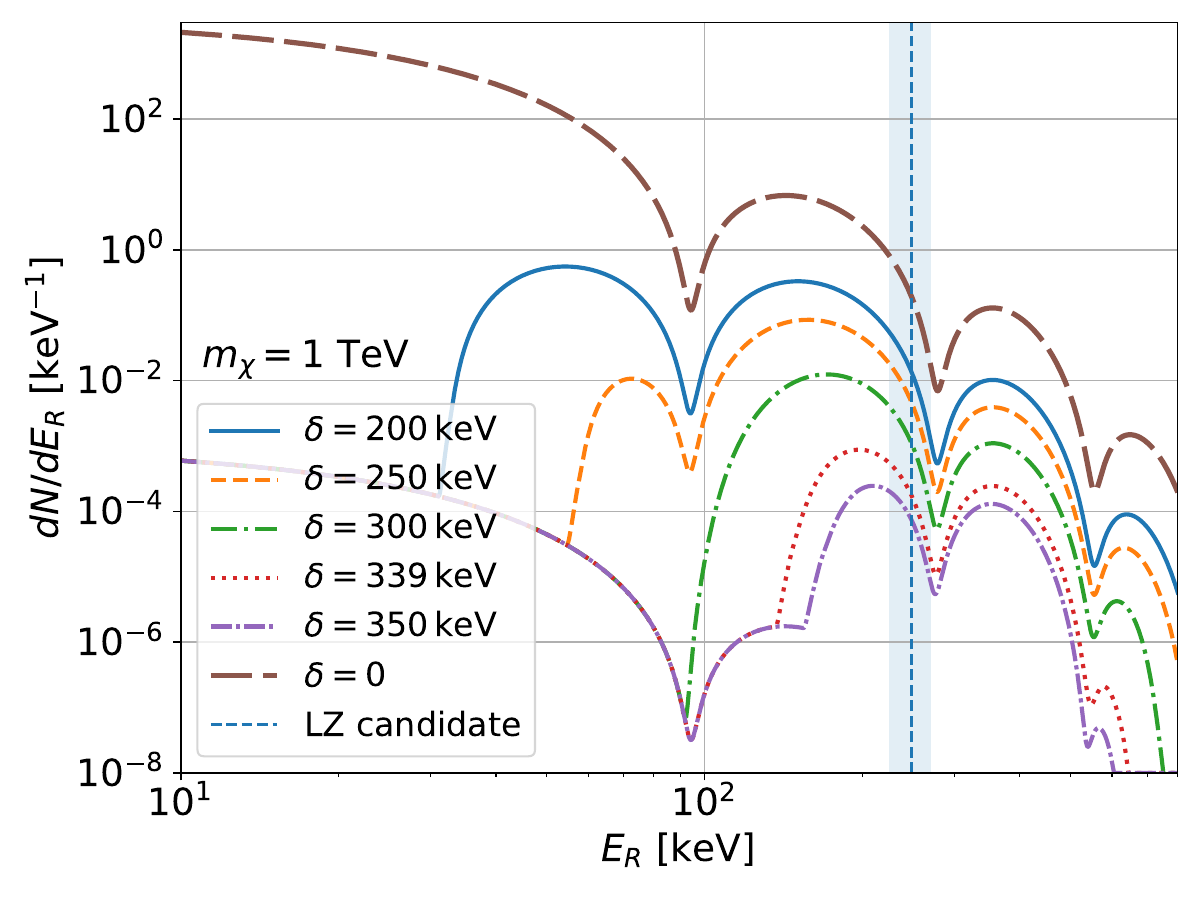}
\includegraphics[width=0.98\columnwidth]{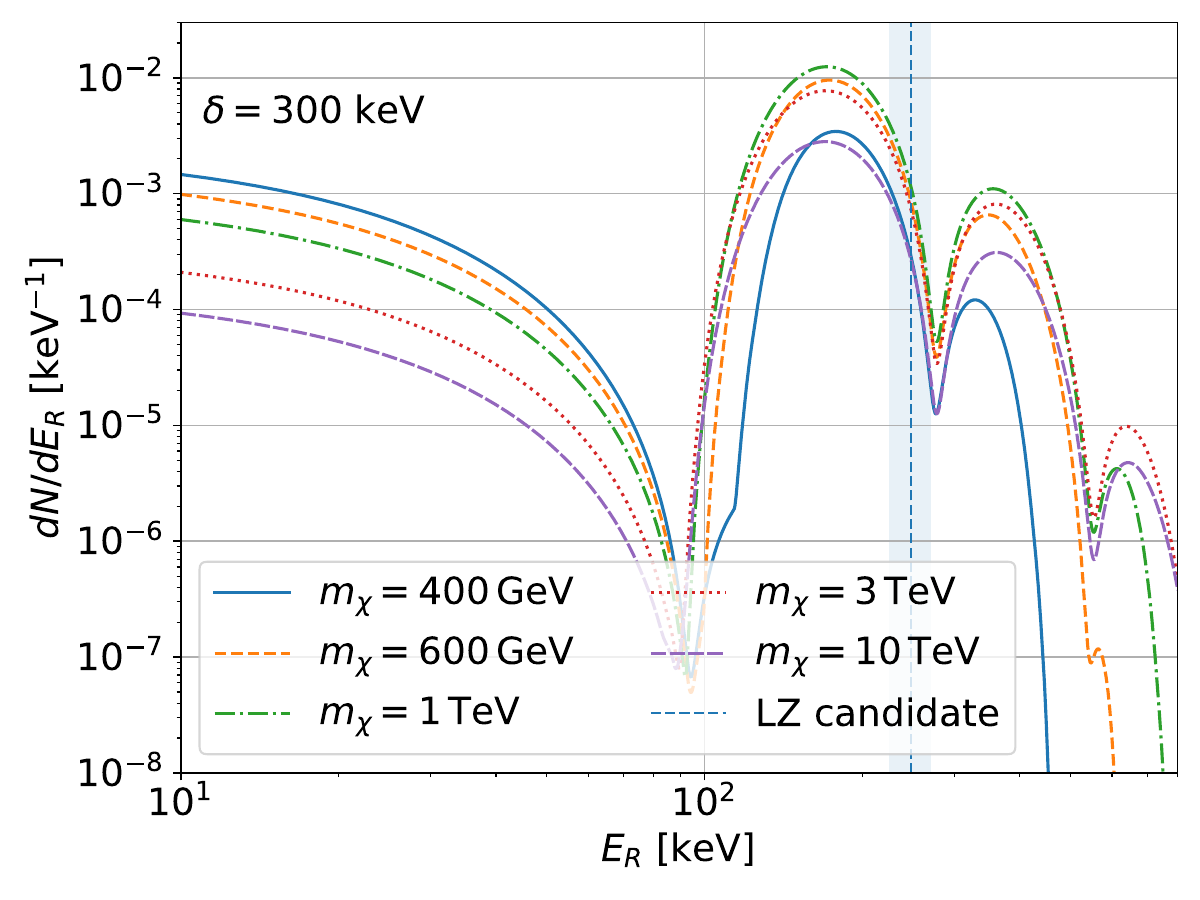}
\caption{Xenon recoil spectra for endothermic DM scattering. Top: spectra for $m_\chi=1\TeV$ and representative mass splittings, together with the elastic SI limit $\delta_\chi=0$ evaluated at the same microscopic normalization. Bottom: spectra for fixed $\delta_\chi=300\keV$ and representative DM masses. The vertical dashed line marks the LZ candidate at $E_R=248\keV$, while the shaded band denotes its $\pm23\keV$ statistical interval.}
\label{fig:recoil}
\end{figure}

The upper panel of Fig.~\ref{fig:recoil} illustrates the xenon recoil spectra for $m_\chi=1\TeV$ and representative values of $\delta_\chi$. All curves are evaluated with the same microscopic interaction normalization, so the differences arise from the inelastic kinematics and nuclear response rather than from an independent rescaling of the coupling. The elastic SI limit, $\delta_\chi=0$, is shown for comparison. The dominant vector-mediated endothermic contribution is strongly suppressed at low recoil energies because the incoming DM particle must provide the excitation energy $\delta_\chi$. Increasing the splitting therefore progressively removes the low-energy support and shifts the surviving inelastic events toward larger recoil energies. A small residual component nevertheless remains at low $E_R$ due to the diagonal axial $Z_B$ current, which mediates elastic $\chi_1A\to\chi_1A$ scattering and is not subject to the endothermic threshold. Although velocity- and momentum-suppressed, this contribution becomes visible once the dominant inelastic channel is kinematically suppressed and produces a nonvanishing low-energy tail.

The lower panel of Fig.~\ref{fig:recoil} shows the complementary dependence on the DM mass for fixed $\delta_\chi=300\keV$. Close to the kinematic threshold, exemplified by the $m_\chi=400\GeV$ curve, only a restricted recoil interval is accessible and the spectrum is sharply truncated. Increasing $m_\chi$ increases the DM--xenon reduced mass, lowers the minimum velocity required for the transition, and opens a broader recoil range. Once $m_\chi$ becomes much larger than the xenon nuclear mass, however, $\mu_{\chi A}$ approaches $m_A$ and the spectral shape changes only slowly. The residual decrease in normalization at fixed interaction strength is then primarily due to the DM number-density factor $\rho_\chi/m_\chi$.

At larger recoil energies, the Helm form factor generates the characteristic multi-lobed structure visible in Fig.~\ref{fig:recoil}. For splittings around $\delta_\chi\simeq300$--$350\keV$, the recoil region containing the LZ candidate can be accompanied by one or more additional lobes at higher energy, whose relative population depends sensitively on $\delta_\chi$ through the inelastic velocity threshold. This motivates considering recoil-energy sidebands rather than only the interval surrounding the candidate. Recent analyses have emphasized precisely this point~\cite{RoddSafdiSlatyerXu,DentNewsteadLZSidebands,DelepineKhalilLZSidebands}. In particular, Ref.~\cite{RoddSafdiSlatyerXu} showed, in the context of inelastic Higgsino DM, that normalizing the signal to approximately one event around $E_R\simeq250\keV$ can imply approximately $3$--$10$ events in the high-energy $S1c$ sideband, corresponding roughly to $350\keV\lesssim E_R\lesssim590\keV$, depending on the assumed halo model and nuclear response. The absence of observed events in this region can therefore provide an important consistency test of an inelastic interpretation.

A quantitative constraint from this high-energy region is presently limited by the fact that the LZ acceptance for WIMP-like nuclear recoils in the sideband has not been published; Ref.~\cite{RoddSafdiSlatyerXu} assumes unit acceptance for this estimate. More generally, the spectra shown here are recoil-level predictions, and a precise comparison with events at several hundred keV requires the detector acceptance, reconstruction efficiency, and energy response in this region. We therefore do not impose the high-energy sideband as a hard exclusion in the numerical analysis, but regard it as an experimentally testable prediction of the model.

\begin{figure}[htbp]
\centering
\includegraphics[width=0.98\columnwidth]{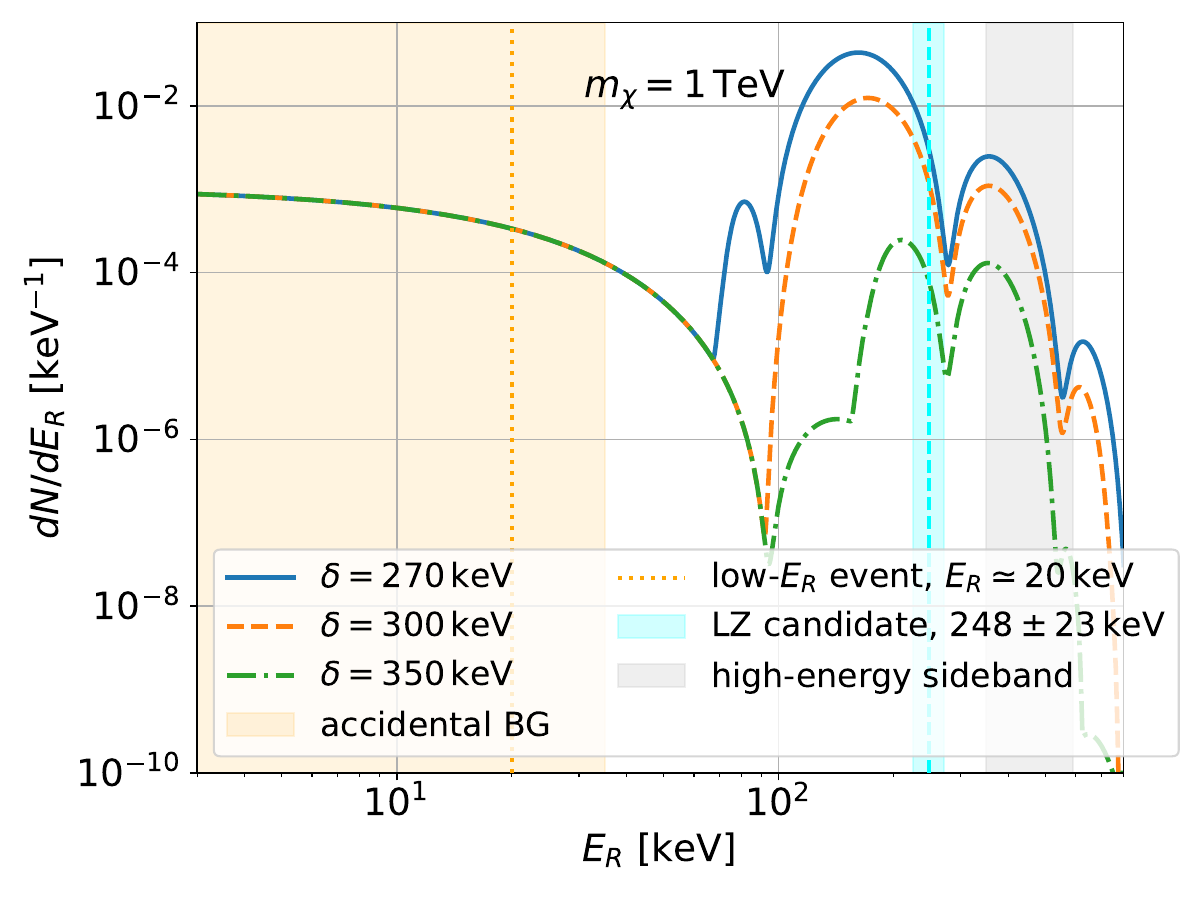}
\caption{Predicted xenon recoil spectra for $m_\chi=1\TeV$ and $\delta_\chi=270$, $300$, and $350\keV$, evaluated at the same microscopic interaction normalization. The cyan band marks the reconstructed-energy interval of the $248\keV$ LZ candidate. The orange-shaded region denotes $E_R\lesssim35\keV$, where accidental events are particularly relevant, and the orange vertical line indicates schematically the additional low-energy event around $E_R\simeq20\keV$. The gray band shows the high-energy sideband, $350\keV\lesssim E_R\lesssim590\keV$.}
\label{fig:recoil_regions}
\end{figure}

Fig.~\ref{fig:recoil_regions} brings together the three recoil-energy regions that are particularly relevant for testing this scenario. The cyan band denotes the reconstructed-energy interval of the $248\keV$ LZ candidate, while the gray band shows the high-energy sideband discussed above. At low energy, the orange-shaded region $E_R\lesssim35\keV$ highlights the range in which accidental events are particularly relevant in the LZ analysis and where the residual elastic axial contribution becomes comparatively more important.

The LZ data contain an additional low-energy event, indicated schematically in Fig.~\ref{fig:recoil_regions} by the orange vertical line around $E_R\simeq20\keV$. This event lies outside the calibrated NR band and was interpreted by the LZ Collaboration as compatible with the accidental-coincidence background~\cite{LZ2026}, and therefore cannot presently be identified with a DM-induced nuclear recoil. Nevertheless, its location overlaps the recoil-energy range in which the residual axial component predicted by the model is present. Future exposure and an improved characterization of the low-energy event population could therefore test this contribution, while the high-energy sideband independently probes the additional recoil lobes generated by the endothermic interaction.

The model consequently predicts a characteristic recoil structure extending over widely separated energies: a small elastic component at low $E_R$, the dominant endothermic signal around the LZ candidate, and additional form-factor lobes at higher recoil energies. These regions provide complementary tests of the same microscopic interaction.

\subsection{Complementary high- and low-recoil normalizations}
\label{subsec:LZhighlow}

The recoil spectra in Fig.~\ref{fig:recoil_regions} suggest two complementary recoil-energy regions in which the model can be tested. The high-energy LZ candidate probes primarily the endothermic transition $\chi_1A\to\chi_2A$ generated by the off-diagonal vector $Z_B$ interaction, whereas at low recoil energies the endothermic process is strongly suppressed and the residual spectrum is generated by the elastic interactions discussed in App.~\ref{app:subleadingDD}. It is therefore useful to compare the values of $g_B$ required to obtain one event in these two regions.

To illustrate this complementarity, we consider $m_\chi=1\TeV$ and $\delta_\chi=350\keV$, and define
\begin{equation}
N_{\rm high}\equiv N(200\keV,300\keV),
\,\,
N_{\rm low}^{100}\equiv N(0,100\keV),
\label{eq:highlowwindows}
\end{equation}
where in both cases the recoil spectrum is folded with the LZ efficiency. The high-energy prediction contains the dominant coherent endothermic $Z_B$ contribution together with the small mixing-induced SM-$Z$ amplitude and their interference. In the low-energy interval, the rate is instead dominated by the elastic diagonal axial $Z_B$ interaction, with a smaller scalar-mediated elastic contribution.

\begin{figure}[htbp]
\centering
\includegraphics[width=0.98\columnwidth]{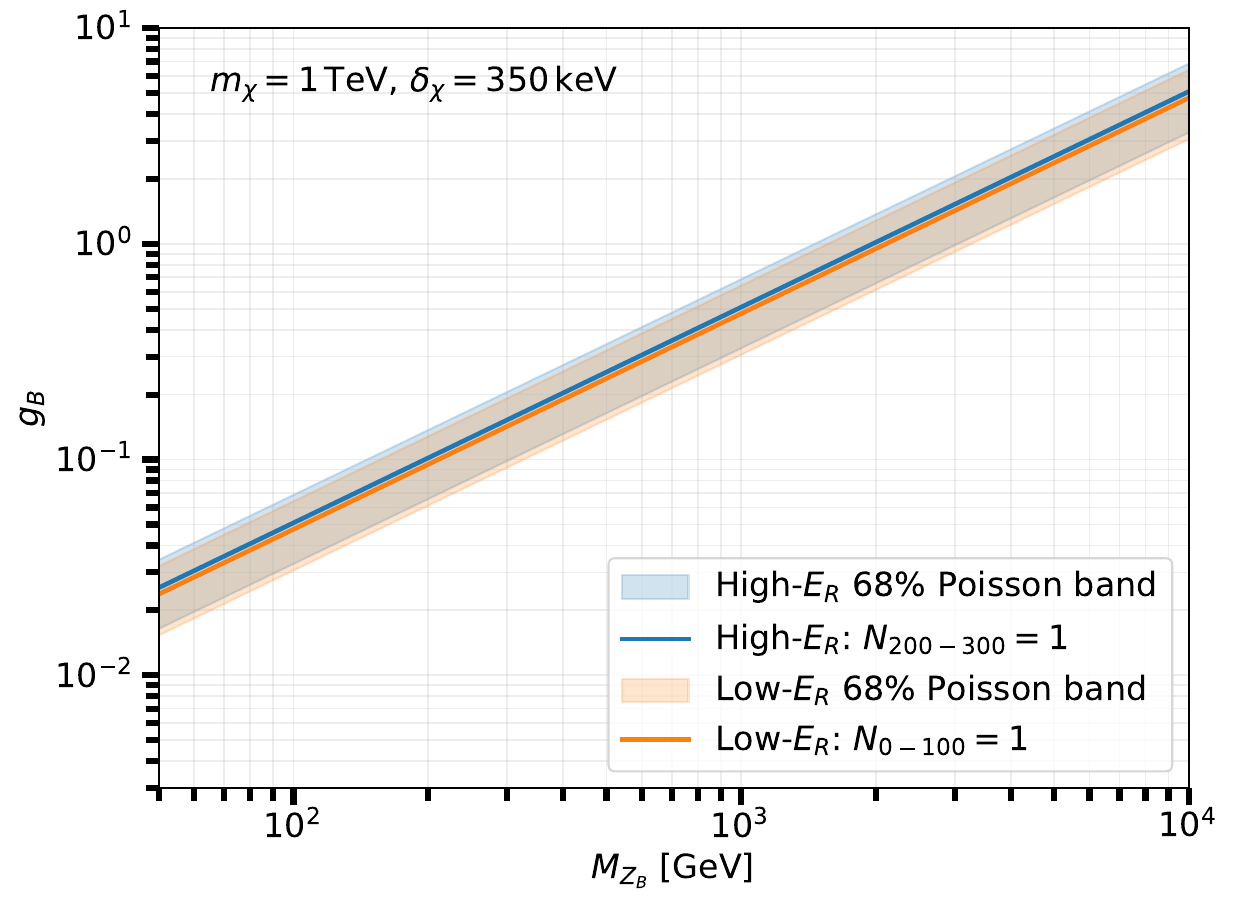}
\caption{Baryonic gauge coupling $g_B$ required to obtain one event in the high- and low-recoil LZ regions as a function of $M_{Z_B}$, for $m_\chi=1\TeV$ and $\delta_\chi=350\keV$. The blue curve and band correspond to $N(200\keV,300\keV)=1$ and its central $68.27\%$ Poisson interval, while the orange curve and band show the analogous result for $N(0,100\keV)=1$, including the LZ efficiency.}
\label{fig:LZhighlowcouplings}
\end{figure}

For each value of $M_{Z_B}$ we determine separately the coupling satisfying $N_{\rm high}=1$ and $N_{\rm low}^{100}=1$. For one observed event and negligible background, the exact central $68.27\%$ Poisson interval for the expected mean is $0.173\lesssim\mu\lesssim3.30$, which we propagate to the corresponding intervals in $g_B$.
Fig.~\ref{fig:LZhighlowcouplings} shows the resulting coupling bands as a function of $M_{Z_B}$. The blue curve and band correspond to the high-energy LZ candidate, while the orange curve and band show the coupling that would yield one event after integrating the low-energy recoil spectrum over $0<E_R<100\keV$. Both curves increase approximately linearly with $M_{Z_B}$, reflecting the dominant dependence of the effective interactions on the ratio $g_B/M_{Z_B}$.

An interesting feature appears for $\delta_\chi=350\keV$: the blue and orange $68.27\%$ Poisson bands overlap. There is therefore a range of couplings for which the same microscopic model can accommodate, within the present Poisson uncertainty, both one event in the high-energy region and one event in the low-energy region. The origin of this compatibility can be understood from the different dependence of the two contributions on $\delta_\chi$. Increasing the splitting suppresses the endothermic high-energy rate and therefore increases the value of $g_B$ required to reproduce the $248\keV$ candidate. By contrast, the low-energy spectrum is generated predominantly by the elastic axial $Z_B$ interaction and is not subject to the endothermic threshold. The high-energy normalization consequently moves closer to the coupling preferred by the low-energy recoil population as $\delta_\chi$ is increased.

This coincidence is particularly intriguing because the two events would probe different components of the same underlying baryonic interaction: the high-energy recoil is generated by the off-diagonal vector current, while the low-energy signal originates predominantly from the diagonal axial current. The overlap of the two bands therefore shows that, for this choice of $(m_\chi,\delta_\chi)$, the model can simultaneously produce recoil rates of the appropriate order in two widely separated energy regions. Given the extremely small statistics, this should of course be regarded only as a suggestive consistency rather than evidence for a common DM origin.

Indeed, the additional low-energy event observed by LZ at $E_R\sim20\keV$ lies outside the calibrated NR band and was classified by the Collaboration as compatible with the accidental-coincidence population~\cite{LZ2026}. We therefore do not include it as a second DM signal in the combined likelihood. The orange curve in Fig.~\ref{fig:LZhighlowcouplings} should instead be interpreted as a diagnostic result, quantifying the interaction strength that would be required if one event in the $0$--$100\keV$ interval were attributed to the elastic component of the model. Nevertheless, the fact that this normalization overlaps with that independently selected by the high-energy event for $\delta_\chi=350\keV$ is a tantalizing feature of the recoil phenomenology and provides additional motivation for studying the low-energy event population with future exposure.

Moreover, the present low-energy calculation retains the leading coherent $\mathcal O_8$ response of the diagonal axial interaction. A precision interpretation of this region would require the complete xenon NREFT response, including the $\mathcal O_9$ contribution and the corresponding nuclear-response interference terms.

Fig.~\ref{fig:LZhighlowcouplings} therefore provides an interesting example in which the high-energy endothermic candidate and the low-energy elastic recoil can be compatible with the same underlying interaction strength. Although the low-energy event can naturally arise from the accidental background and cannot presently be identified as a nuclear recoil signal, the compatibility of the two recoil regions for $\delta_\chi=350\keV$ represents a nontrivial and potentially testable feature of the model. Future LZ exposure and an improved characterization of the low-energy background will determine whether this apparent consistency persists.

\subsection{Thermal relic density}
\label{subsec:relic}

The relic-density calculation in this model is intrinsically a two-state freeze-out problem. 
For the splittings relevant to the LZ interpretation, $\delta_\chi\sim {\cal O}(100)\keV$, and for DM masses in the TeV range, freeze-out occurs at temperatures of order
\begin{equation}
T_f \sim \frac{m_\chi}{20-30}\sim {\cal O}(10)\GeV,
\end{equation}
so that $\frac{\delta_\chi}{T_f}\ll1$.
The two Majorana states are therefore thermally indistinguishable during chemical decoupling. 
Defining
\begin{equation}
\Delta_i\equiv\frac{m_{\chi_i}-m_{\chi_1}}{m_{\chi_1}},
\end{equation}
their equilibrium weights are
\begin{align}
r_i(x)&=
\frac{
g_i(1+\Delta_i)^{3/2}e^{-x\Delta_i}
}{
g_{\rm eff}(x)
},
\\
g_{\rm eff}(x)&=
\sum_i g_i(1+\Delta_i)^{3/2}e^{-x\Delta_i},
\label{eq:ri}
\end{align}
with $g_1=g_2=2$. Since $\Delta_i\sim\delta_\chi/m_\chi\ll1$, one has to an excellent approximation $r_1\simeq r_2\simeq 1/2$.
The effective annihilation rate entering the Boltzmann equation is then
\begin{equation}
\langle\sigma_{\rm eff}v\rangle=
\sum_{i,j} r_i r_j \langle \sigma_{ij}v\rangle,
\label{eq:sigmaeff}
\end{equation}
so that the off-diagonal process $\chi_1\chi_2$ enters with the statistical weight $2r_1r_2\simeq 1/2$.

It is useful to distinguish the various annihilation channels according to the mass ordering. 
The phenomenology is not described by a single universal process over the full parameter space; rather, different channels dominate in different kinematic regions.

The most important channel on the heavy-mediator branch is the coannihilation process
\begin{equation}
\chi_1\chi_2 \to Z_B^\ast \to q\bar q.
\label{eq:coann}
\end{equation}
This process is unsuppressed and, on the branch relevant for our most precise results, it is the dominant contribution to the effective annihilation rate. 
Because the initial state involves one $\chi_1$ and one $\chi_2$, this channel is naturally tied to the pseudo-Dirac structure of the model and is the cosmological analogue of the off-diagonal interaction responsible for the LZ signal.

By contrast, the diagonal annihilation process
\begin{equation}
\chi_1\chi_1 \to q\bar q
\end{equation}
is subleading. In the pseudo-Dirac limit the light state couples diagonally to $Z_B$ through the axial current, and the resulting annihilation into SM quarks is velocity suppressed. 
It therefore gives only a small correction to the freeze-out rate in the parameter region of interest and is not the channel that fixes the relic density on the controlled branch.

A qualitatively different regime is entered once the mediator becomes lighter than the DM particle, $M_{Z_B}<m_\chi$.
In this open-mediator branch, the secluded annihilation channel
\begin{equation}
\chi_1\chi_1\to Z_B Z_B
\end{equation}
becomes kinematically allowed. 
This process proceeds through $t$- and $u$-channel exchange of the heavier neutral state (see Fig.~\ref{fig:feynman}), together with the additional contributions required by the gauge-complete model. 
Once open, this channel can rapidly become important and eventually dominate the total annihilation rate. 
It is therefore the leading effect that distinguishes the controlled heavy-mediator branch from the open-mediator branch.

The production kernel includes SM fermion pairs and the bosonic channel groups $Z_BZ_B$, $Z_Bh_1$, $Z_Bh_2$, $h_1h_1$, $h_1h_2$, $h_2h_2$, $W^+W^-$, $ZZ$, $Z_BZ$, $Zh_1$, and $Zh_2$, with the physical mixed-fermion exchange amplitudes. At freeze-out, the nearly degenerate light pair can be represented by the corresponding Dirac ensemble; corrections scale as $\delta_\chi/T_f$. Heavy anomalons are retained as virtual particles, while their initial-state coannihilations and heavy-fermion final states are omitted. The adopted ratios $M_\Psi=1.5m_\chi$ and $M_\eta=2m_\chi$ suppress their equilibrium populations, although this is not a universal argument for arbitrary spectra.

At the reference point, $m_\chi<M_{Z_B}<2m_\chi$, the fermionic coannihilation contribution exceeds $99.999\%$ of the effective rate. The same inequality does not by itself guarantee single-channel dominance throughout the scan: scalar and mixed vector--scalar final states can open elsewhere. We therefore evaluate the included channels at each point.
\begin{equation}
m_\chi<M_{Z_B}<2m_\chi.
\label{eq:controlledbranch}
\end{equation}

\begin{figure}[htbp]
\centering
\includegraphics[width=0.98\columnwidth]{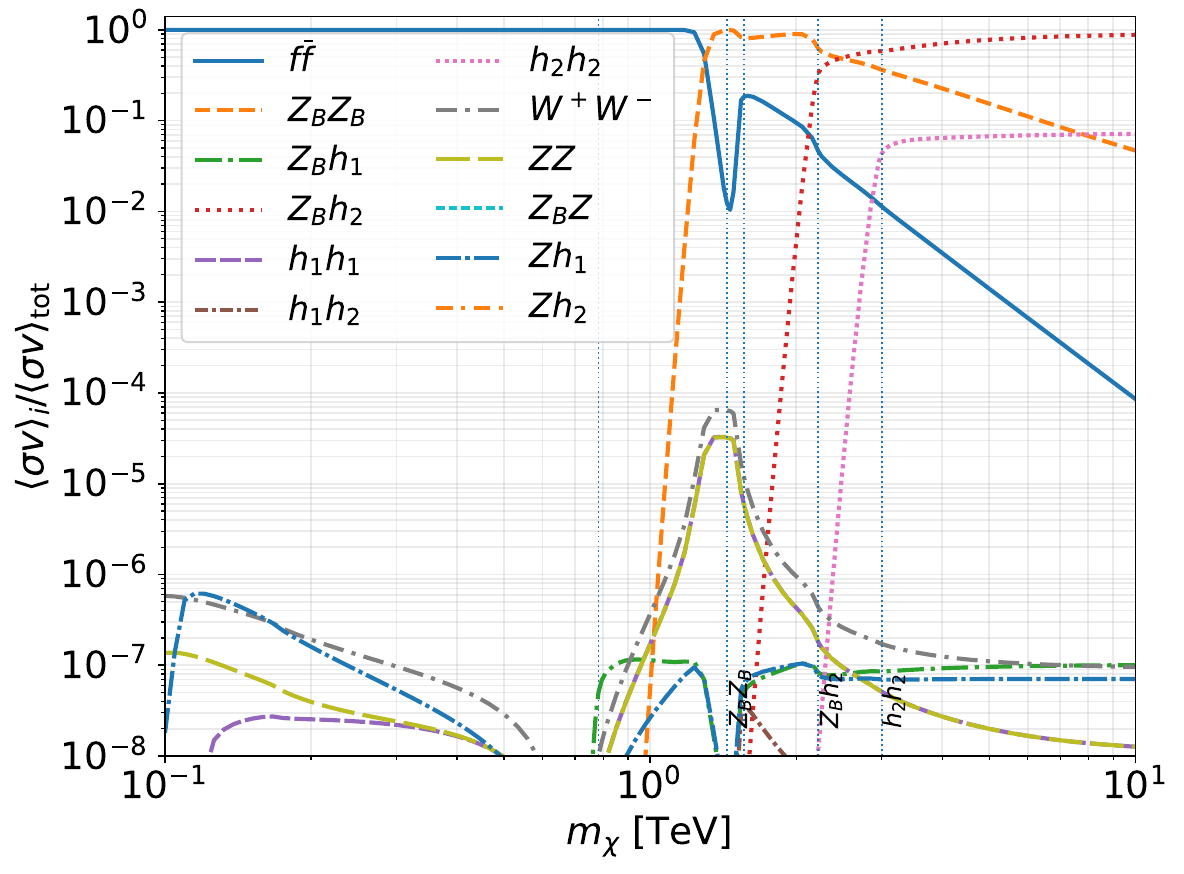}
\caption{Fractions of the included thermal annihilation channels at $x=25$, for $M_{Z_B}=1.44\TeV$, $g_B=0.22782$, $m_{h_2}=3\TeV$, $M_\Psi=1.5m_\chi$, and $M_\eta=2m_\chi$. The curves show channel groups in the full implemented light-fermion kernel; this fixed-coupling diagnostic is not evaluated along the relic-density contour.}
\label{fig:relicBR}
\end{figure}

Near the mediator shoulder, a zero-temperature estimate is not sufficient. 
We therefore retain the physical mediator width in the propagator. For each quark flavor,
\begin{align}
\Gamma(Z_B\to q\bar q)={}&
\frac{N_cg_q^2M_{Z_B}}{12\pi}
\left(1+\frac{2m_q^2}{M_{Z_B}^2}\right)
\sqrt{1-\frac{4m_q^2}{M_{Z_B}^2}},
\label{eq:ZBwidth}
\end{align}
with $g_q=g_B/3$, so that below non-SM thresholds and neglecting quark masses
\begin{equation}
\frac{\Gamma_{Z_B}}{M_{Z_B}}\simeq \frac{n_qg_B^2}{36\pi},\,\,n_q=6\ \text{above the top threshold}.
\end{equation}
The exact $s$-dependent cross section is then thermally averaged with the Gondolo--Gelmini expression~\cite{GondoloGelmini},
\begin{align}
&\langle\sigma_{12}v_{\rm rel}\rangle=
\frac{1}{8m_{\chi_1}^2m_{\chi_2}^2T}
\frac{1}{K_2(m_{\chi_1}/T)K_2(m_{\chi_2}/T)}
\nonumber\\
&\times
\int_{(m_{\chi_1}+m_{\chi_2})^2}^{\infty} ds\,
\frac{\lambda(s,m_{\chi_1}^2,m_{\chi_2}^2)}{\sqrt{s}}
\sigma_{12}(s) K_1(\sqrt{s}/T).
\label{eq:thermalaverage}
\end{align}
Here $K_n$ are modified Bessel functions, and $\lambda$ is the K\"all\'en function, $\lambda(a,b,c)=a^2+b^2+c^2-2ab-2ac-2bc$. We then solve the Boltzmann equation directly,
\begin{equation}
\frac{dY}{dx}
=
-\frac{s}{Hx}
\left(
1+\frac13\frac{d\ln g_{\ast s}}{d\ln T}
\right)
\langle\sigma_{\rm eff}v\rangle
\left(
Y^2-Y_{\rm eq}^2
\right),
\label{eq:Boltzmann}
\end{equation}
using a stiff implicit solver up to $x=100$, followed by integration of the inverse yield to $x=10^6$ after the equilibrium source is negligible, together with a smooth SM interpolation of $g_{\ast s}(T)$. 
This procedure retains the finite-width and finite-temperature effects that shift the relic contour in the vicinity of the resonance. 
The scan uses a tabulated positive interpolation of the finite-temperature rate and direct Boltzmann integration.
The recoil-count construction is specified in Sec.~\ref{subsec:LZ}. Eq.~\eqref{eq:MDmixingadjustment} fixes the physical light eigenmass, and App.~\ref{app:subleadingDD} gives the corresponding recoil coefficients.
At the benchmark point $m_\chi=1\TeV, M_{Z_B}=1.44\TeV$,
the numerical solution of Eq.~\eqref{eq:Boltzmann} gives
\begin{equation}
g_B^{\rm relic}=0.2278.
\end{equation}
The freeze-out point occurs near $x_f\simeq23.9$,
and the effective annihilation rate at $x=25$ is
\begin{equation}
\langle\sigma_{\rm eff}v\rangle_{x=25}
\simeq1.91\times10^{-26}\cm^3\s^{-1}.
\end{equation}
The corresponding threshold value is approximately
\begin{equation}
\langle\sigma_{\rm eff}v\rangle_{\rm thr}
\simeq2.50\times10^{-26}\cm^3\s^{-1}.
\end{equation}
Their difference quantifies the importance of performing the true thermal average rather than replacing the relic calculation by a zero-temperature estimate, especially close to the mediator shoulder.

\begin{figure}[htbp]
\centering
\includegraphics[width=0.98\columnwidth]{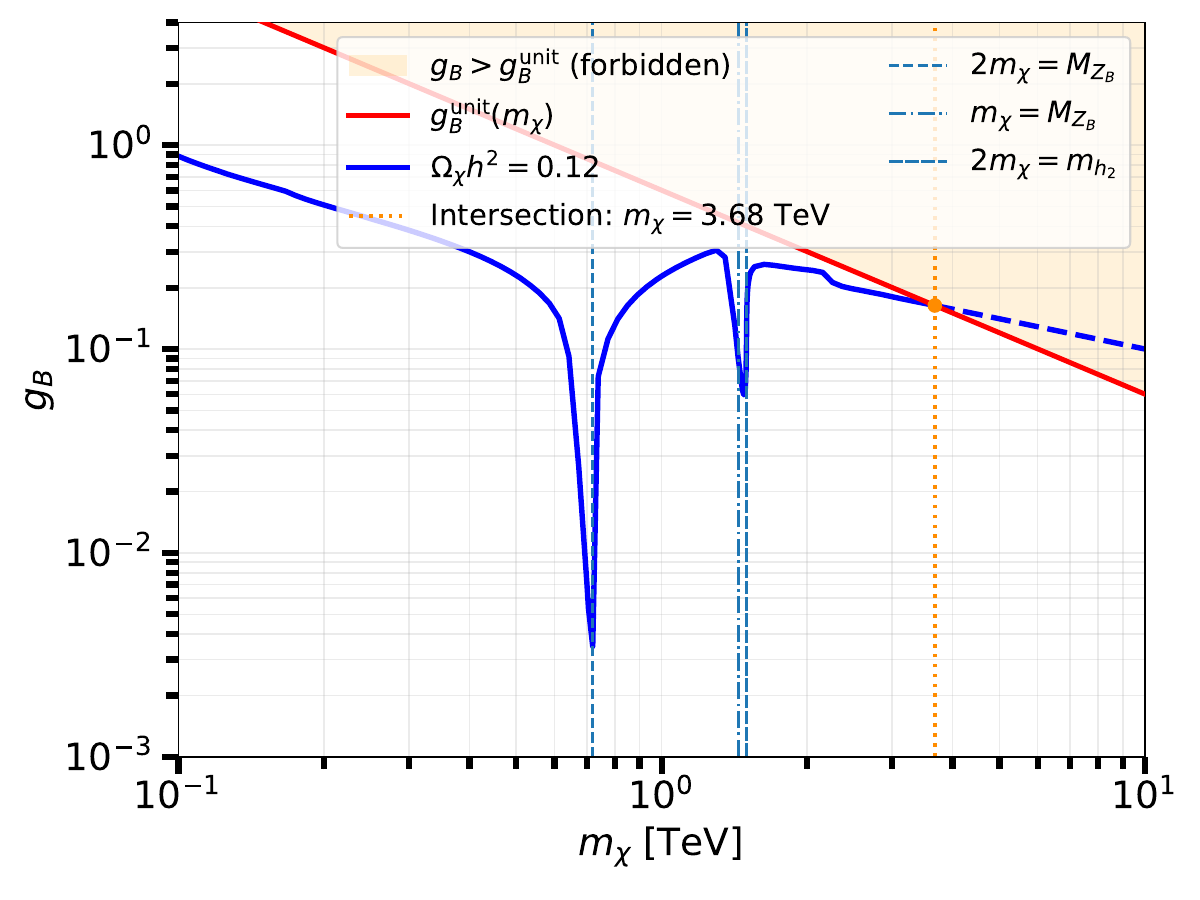}
\caption{Baryonic coupling $g_B^{\rm relic}$ required to reproduce $\Omega_\chi h^2\simeq0.12$ as a function of the DM mass, for $M_{Z_B}=1.44\TeV$ and $m_{h_2}=3\TeV$. The anomalon masses scale as $M_\Psi=1.5m_\chi$ and $M_\eta=2m_\chi$. The blue curve shows the numerical relic-density solution, with solid and dashed segments satisfying and violating the adopted Yukawa perturbativity criterion, respectively. The red curve gives the bound in Eq.~\eqref{eq:gBUnitarity}; the light orange region above it is excluded by this criterion. The orange dotted vertical line marks the intersection at $m_\chi\simeq3.68\TeV$ and $g_B\simeq0.164$. The other vertical lines identify the labeled resonance and kinematic thresholds.}
\label{fig:relic_mass_scan}
\end{figure}

Fig.~\ref{fig:relic_mass_scan} shows the value of the baryonic gauge coupling $g_B^{\rm relic}$ required to reproduce the observed relic abundance as a function of $m_\chi$, keeping $M_{Z_B}=1.44\TeV$ and $m_{h_2}=3\TeV$ fixed. For the mass scan shown here we scale the anomalon masses with the DM mass according to
\begin{equation}
M_\Psi=1.5m_\chi,
\,\,
M_\eta=2m_\chi.
\label{eq:anomalonMassScaling}
\end{equation}
The non-monotonic structure of the relic curve reflects the interplay between resonant annihilation and the opening of additional final states discussed in Fig.~\ref{fig:relicBR}. In particular, the pronounced minimum around
\begin{equation}
2m_\chi\simeq M_{Z_B}
\end{equation}
is produced by the resonant enhancement of the off-diagonal process $\chi_1\chi_2\rightarrow Z_B^\ast\rightarrow q\bar q$. Away from the resonance, a larger coupling is required to maintain an effective annihilation rate compatible with $\Omega_\chi h^2\simeq0.12$.

A second characteristic structure occurs around $m_\chi\simeq1.4$--$1.5\TeV$. In this narrow interval two physically distinct effects occur close to one another: the $Z_BZ_B$ final state opens at $m_\chi=M_{Z_B}$, while the heavy-scalar resonance is encountered near $2m_\chi=m_{h_2}$. Their combined effect generates the second pronounced feature of the relic-density curve. At still larger masses, the evolution of $g_B^{\rm relic}$ reflects the increasing importance of the additional bosonic channels, in particular $Z_Bh_2$ and $h_2h_2$, as illustrated in Fig.~\ref{fig:relicBR}.

The relic-density equation itself continues to admit numerical solutions toward larger $m_\chi$. However, in the UV completion the anomalon masses originate from the same baryonic symmetry-breaking scale $v_B=M_{Z_B}/3g_B$
so their masses and Yukawa couplings cannot be varied independently. With the mass prescription in Eq.~\eqref{eq:anomalonMassScaling}, the largest baryonic Yukawa coupling is associated with $\eta$,
\begin{equation}
\lambda_\eta
=
\frac{\sqrt{2}M_\eta}{v_B}
=
\frac{6\sqrt{2}\,m_\chi g_B}{M_{Z_B}}.
\label{eq:lambdaEtaRelicScan}
\end{equation}
We require this coupling to remain within the perturbative regime, for which we adopt the conventional numerical criterion $\lambda_\eta<\sqrt{4\pi}$.
Equivalently, the relic-density solution must satisfy
\begin{equation}
g_B<
g_B^{\rm unit}(m_\chi)
\equiv
\frac{\sqrt{\pi}\,M_{Z_B}}
{3\sqrt{2}\,m_\chi}.
\label{eq:gBUnitarity}
\end{equation}

This criterion is not a derived coupled-channel partial-wave bound. The upper value allowed by Eq.~\eqref{eq:gBUnitarity} decreases approximately as $1/m_\chi$, whereas the coupling required by the relic-density condition decreases more slowly. The two curves therefore cross at a finite DM mass, around $m_\chi\simeq3.7\TeV$ for the scan shown in Fig.~\ref{fig:relic_mass_scan}. This transition is indicated by the orange dotted vertical line.

The blue relic-density curve is shown solid below this transition and dashed above it. Importantly, the dashed segment does not indicate a failure of the Boltzmann solver or the absence of a formal solution of $\Omega_\chi h^2=0.12$. Rather, it identifies relic-density solutions for which the anomalon Yukawa coupling required by the assumed UV spectrum exceeds the perturbativity criterion for $\lambda_\eta$. For example, at $m_\chi=10\TeV$ a value $g_B^{\rm relic}\simeq0.10$ corresponds to
\begin{equation}
\lambda_\eta\simeq5.9,
\end{equation}
which exceeds $\sqrt{4\pi}\simeq3.54$.

We therefore retain the dashed continuation in Fig.~\ref{fig:relic_mass_scan} to display the formal thermal relic-density solution and its mass dependence, but we do not regard this region as part of the perturbatively controlled UV parameter space. In the combined phenomenological analysis presented below, configurations violating Eq.~\eqref{eq:gBUnitarity} are consequently excluded by the perturbativity requirement, independently of the LZ, collider, and Solar constraints.

\subsection{Collider signatures and present dijet constraints}
\label{subsec:collider}

The same baryonic interaction responsible for DD and thermal freeze-out also determines the collider phenomenology. As discussed in Sec.~\ref{subsec:current}, SM quarks carry equal left- and right-handed baryon charges, $B_L^q=B_R^q=1/3$, so their coupling to the mediator is purely vectorial, $g_V^q=\frac{g_B}{3}$, $g_A^q=0$.
Consequently, the production rate of an $s$-channel $Z_B$ resonance at a hadron collider is directly correlated with the same $g_B$ and $M_{Z_B}$ that enter the LZ and relic-density calculations.

An essential ingredient in the collider reinterpretation is the physical mediator width. For a diagonal Dirac pair with $r_f=m_f^2/M_{Z_B}^2$, the vector and axial contributions are
\begin{align}
\Gamma_f={}&\frac{N_cM_{Z_B}}{12\pi}\sqrt{1-4r_f}\bigg[
\left|g_V^f\right|^2\left[1+2r_f\right]\nonumber\\
&+\left|g_A^f\right|^2\left[1-4r_f\right]\bigg].
\label{eq:ZBwidthVA}
\end{align}
The numerical width sums quark pairs and all kinematically accessible neutral and charged new-fermion pairs in the physical mass basis, including mixed heavy--light channels. It uses the degenerate-Dirac limit for the light pseudo-Dirac pair, retaining both vector and axial pieces; resolving the two Majorana thresholds changes the result only at order $\delta_\chi/m_\chi$ away from threshold. Symbolically,
\begin{equation}
\Gamma_{Z_B}=\sum_q\Gamma_{q\bar q}
             +\sum_{i,j}\Gamma_{N_i\bar N_j}
             +\sum_{i,j}\Gamma_{X_i\bar X_j}.
\label{eq:ZBtotalwidth}
\end{equation}
The latter sums run over allowed physical pairs without double counting. The baseline sets the renormalized kinetic mixing to zero, so no mixing-induced leptonic width enters this calculation.

Below the dark and anomalon thresholds the total width is dominated by quark decays. Neglecting quark masses and assuming that all six flavors are open,
\begin{equation}
\frac{\Gamma(Z_B\to q\bar q)}{M_{Z_B}}
\simeq
\frac{g_B^2}{6\pi}.
\label{eq:ZBnarrowwidth}
\end{equation}
For the reference point $M_{Z_B}=1.44\TeV$ and $g_B=0.2278$ this gives
\begin{equation}
\frac{\Gamma_{Z_B}}{M_{Z_B}}
\simeq
2.75\times10^{-3},
\end{equation}
so the benchmark corresponds to a narrow resonance. Moreover, for
\begin{equation}
M_{Z_B}<m_{\chi_1}+m_{\chi_2}\simeq2m_\chi
\end{equation}
the decay $Z_B\to\chi_1\chi_2$ is closed and, below the anomalon thresholds, the mediator decays almost entirely into quarks. Dijet resonance searches are therefore the leading collider probe in this region.

To compare the baryonic model consistently with experimental interpretations expressed in terms of a universal quark coupling, we define the equivalent coupling
\begin{equation}
g_q^{\rm eq}
=
\frac{g_B}{3}
\left(
\frac{{\rm BR}_{jj}^{\rm model}}
{{\rm BR}_{jj}^{\rm ref}}
\right)^{1/2},
\label{eq:gqequiv}
\end{equation}
where ${\rm BR}_{jj}^{\rm ref}$ is the dijet branching fraction assumed in the corresponding experimental reference model. Within the narrow-resonance approximation this mapping accounts for the reduction of the visible dijet rate whenever dark-sector or anomalon decay modes become accessible. Below all non-quark thresholds and for a reference mediator with the same accessible quark flavors, ${\rm BR}_{jj}^{\rm model}\simeq{\rm BR}_{jj}^{\rm ref}$, and Eq.~\eqref{eq:gqequiv} reduces simply to $g_q^{\rm eq}=g_B/3$.

At very low mediator masses we include the CMS EXO-24-007 search for boosted low-mass resonances decaying into a quark-antiquark pair~\cite{CMSLowMassDijet2026}. The analysis uses $138\,{\rm fb}^{-1}$ of $\sqrt{s}=13\TeV$ data and provides an observed $95\%$ CL constraint on a narrow spin-1 resonance with universal quark coupling over approximately
\begin{equation}
50\GeV\lesssim M_{Z_B}\lesssim300\GeV.
\end{equation}
The published interpretation assumes ${\rm BR}(Z'\to q\bar q)=1$ and constrains approximately $g_q\sim0.03$--$0.13$, depending on the resonance mass. In our recast we use Eq.~\eqref{eq:gqequiv} with ${\rm BR}_{jj}^{\rm ref}=1$ and evaluate the physical branching ratio point by point. Because the published interpretation assumes a narrow resonance, this constraint is applied only within the corresponding narrow-width regime.

At intermediate mediator masses we use the observed ATLAS trigger-level analysis based on $132\,{\rm fb}^{-1}$ and the CMS data-scouting analysis based on $117\,{\rm fb}^{-1}$~\cite{ATLASTLA,CMSDijetScouting}. The ATLAS trigger-level analysis provides Gaussian signal templates with relative reconstructed widths of $5\%$, $10\%$, and $15\%$. For each model point, the physical Breit--Wigner width is mapped onto an effective reconstructed Gaussian width using the Voigt prescription adopted in our recast, and we interpolate between the published ATLAS templates. No extrapolation beyond the largest $15\%$ Gaussian template is performed. This finite-width treatment is distinct from the narrow-resonance coupling interpretations. In particular, the CMS scouting coupling limits and the narrow-width ATLAS high-mass coupling curve are used only for
\begin{equation}
\frac{\Gamma_{Z_B}}{M_{Z_B}}\leq5\%,
\end{equation}
where the narrow-width approximation remains appropriate. For broader resonances in the ATLAS trigger-level mass range we instead use the finite-width template recast.

The resulting low- and intermediate-mass constraints are shown in Fig.~\ref{fig:run3}. The CMS EXO-24-007 result extends the direct dijet sensitivity down to mediator masses of order $50\GeV$, while in the TeV region the observed limits remain at the level of a few $\times10^{-1}$ in $g_B$. At the reference point $M_{Z_B}=1.44\TeV$, the strongest observed boundary is approximately
\begin{equation}
g_B^{95}\simeq0.240,
\label{eq:collidercap}
\end{equation}
only slightly above the relic-density value $g_B=0.2278$. The benchmark therefore lies close to the present collider sensitivity. This proximity is particularly relevant because $g_B$ is not an independent collider parameter: changing it simultaneously modifies the DD normalization and the thermal abundance.

A particularly important threshold occurs at
\begin{equation}
M_{Z_B}\simeq2m_\chi.
\end{equation}
It has two related but distinct consequences. In the early Universe, the condition $M_{Z_B}\simeq2m_\chi$ produces resonant enhancement of $\chi_1\chi_2\to Z_B^\ast\to q\bar q$, leading to the pronounced structure in the relic-density solution. At colliders, the same mass relation marks the opening of
\begin{equation}
Z_B\to\chi_1\chi_2.
\end{equation}
Above threshold, the dark partial width increases $\Gamma_{Z_B}$ and reduces the visible dijet branching fraction,
\begin{equation}
{\rm BR}_{jj}
=
\frac{\Gamma(Z_B\to u,d,s,c,b)}
{\Gamma_{Z_B}},
\label{eq:BRjj}
\end{equation}
where the top-quark width, when open, is included in the denominator but not in the light-dijet numerator. The larger intrinsic width also broadens the resonance and reduces the sensitivity relative to the narrow-width case. Both effects are included point by point in the collider recast.

For the high-mass and potentially broad-resonance region we use the CMS EXO-19-012 dijet search~\cite{CMSHighMassDijet}, based on $137\,{\rm fb}^{-1}$ at $\sqrt{s}=13\TeV$. This analysis provides spin-1 signal templates for intrinsic widths up to
\begin{equation}
\frac{\Gamma}{M}=55\%.
\end{equation}
We use the observed quark-only coupling limit of Fig.~11 of Ref.~\cite{CMSHighMassDijet} as the reference mass-dependent constraint and incorporate the degradation of the sensitivity with increasing intrinsic width using the broad spin-1 limits of Fig.~10(c).

For every model point we calculate
\begin{equation}
w_{\rm model}
=
\frac{\Gamma_{Z_B}}{M_{Z_B}},
\,\,
{\rm BR}_{jj}^{\rm model},
\end{equation}
and obtain the corresponding equivalent quark coupling from Eq.~\eqref{eq:gqequiv}. The width-dependent reference bound is written as
\begin{align}
g_{q,95}^{\rm broad}(M_{Z_B},w_{\rm model})\!
=\!g_{q,95}^{0}(M_{Z_B}) \left[\frac{R_\sigma(M_{Z_B},w_{\rm model})}
{R_\sigma(M_{Z_B},w_{\rm CMS})}\right]^{1/2},
\label{eq:CMSbroadrecast}
\end{align}
where $R_\sigma(M_{Z_B},w)$ parametrizes the degradation of the CMS limit on $\sigma\times{\rm BR}\times A$ as the intrinsic width increases. A point is excluded when
\begin{equation}
g_q^{\rm eq}>g_{q,95}^{\rm broad}.
\end{equation}

The corresponding high-mass exclusion is shown in Fig.~\ref{fig:run3}. The additional dotted curve marks
\begin{equation}
\frac{\Gamma_{Z_B}}{M_{Z_B}}=55\%,
\end{equation}
namely the maximum intrinsic width for which the CMS broad-resonance templates are available. It is therefore a \emph{coverage boundary}, not an experimental exclusion. The same convention is used throughout the collider analysis: whenever a search ceases to be applicable because the mediator lies outside the published mass or width range, that search is simply removed rather than extrapolated into an unvalidated region.

The evolution of the mediator width at large $M_{Z_B}$ also reflects the additional states required by the UV completion. For $m_\chi=1\TeV$, $M_\Psi=1.5\TeV$, and $M_\eta=2\TeV$, the dominant diagonal anomalon thresholds lie approximately at
\begin{equation}
M_{Z_B}=2M_\Psi=3\TeV,
\,\,
M_{Z_B}=2M_\eta=4\TeV.
\end{equation}
Once these channels open, they provide additional contributions to $\Gamma_{Z_B}$ and reduce the visible branching fraction. The full mediator width and ${\rm BR}_{jj}$ are therefore recomputed at every point in the scan with all kinematically accessible states included.

Combining these searches provides collider sensitivity over roughly two orders of magnitude in $M_{Z_B}$: CMS EXO-24-007 probes the $50$--$300\GeV$ range, the ATLAS trigger-level and CMS scouting analyses cover the intermediate region, and the CMS EXO-19-012 broad-resonance search extends the analysis to multi-TeV mediator masses. The different searches are applied only within their validated mass- and width-dependent domains, so the disappearance of a collider curve at large coupling or mass should not be interpreted as an additional exclusion.

The anomalons themselves provide a complementary collider signature. The doublet-like states can be produced electroweakly, while the charged singlet is produced through its hypercharge interactions and mixing. Their decays are controlled primarily by the electroweak Yukawa couplings $y_{1,2,3,4}$ rather than by the three parameters governing the LZ--relic--dijet correlation. Depending on these mixings, prompt multiboson final states, displaced decays, or long-lived charged signatures can arise~\cite{ATLASVLL,ATLASLLP}. We therefore regard direct anomalon searches as complementary tests and do not include them in the primary $(m_\chi,M_{Z_B},g_B)$ exclusion mask. Radiatively generated kinetic mixing also induces suppressed dilepton couplings, as discussed in App.~\ref{app:kinmix}.

\subsubsection{A Run-3 collider target}
\label{subsec:run3}

The completed LHC Run 3 substantially improves the prospects for testing the near-boundary benchmark. ATLAS reports $332\,\mathrm{fb}^{-1}$ of recorded proton--proton data during Run 3, while CMS reports $355\,\mathrm{fb}^{-1}$ delivered with a $92\%$ recording efficiency, corresponding to approximately $327\,\mathrm{fb}^{-1}$ recorded~\cite{ATLASRun3Dataset,CMSRun3Dataset}. Both experiments have maintained the high-rate trigger strategies required for low-mass dijet searches~\cite{ATLASRun3TLA,CMSRun3Scouting}.

Our Run-3 projections use the scaling $g_B^{\rm lim}\propto L^{-1/4}$, appropriate for background-dominated searches with $\sigma\propto g_B^2$, and remain indicative because the analysis-specific usable luminosity, the change in collision energy, and changes in trigger acceptance, efficiencies, and systematic uncertainties are not explicitly modeled.

\begin{figure}
\centering
\includegraphics[width=0.99\columnwidth]{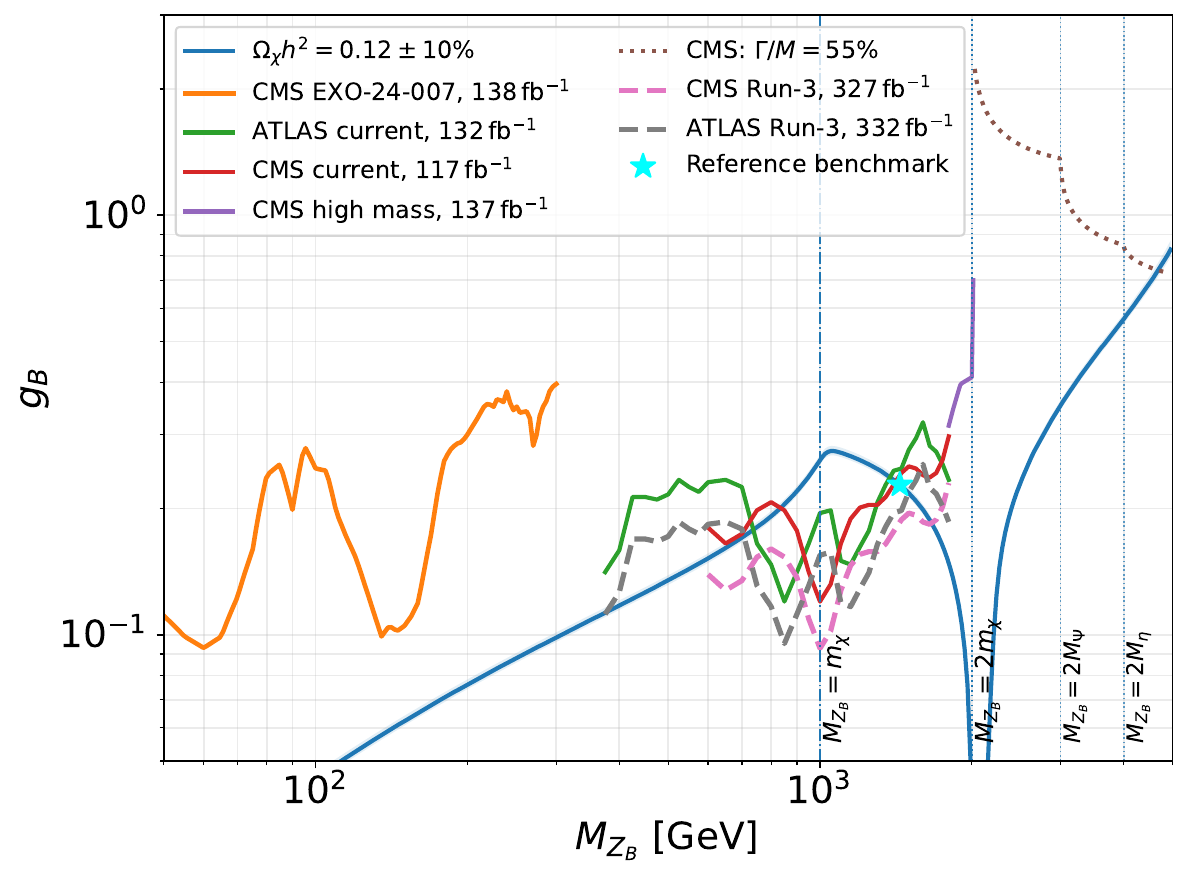}
\caption{Collider constraints and Run-3 sensitivity projections for $m_\chi=1\TeV$, together with the relic-density solution. The blue curve and band show $\Omega_\chi h^2=0.12\pm10\%$. The orange curve is the CMS EXO-24-007 low-mass dijet limit~\cite{CMSLowMassDijet2026}, while the green and red curves are the current ATLAS trigger-level and CMS scouting constraints~\cite{ATLASTLA,CMSDijetScouting}. The purple curve denotes the CMS EXO-19-012 high-mass broad-resonance constraint~\cite{CMSHighMassDijet}, and the brown dotted curve marks the coverage boundary $\Gamma_{Z_B}/M_{Z_B}=55\%$. The magenta and gray dashed curves are the luminosity-scaled CMS and ATLAS Run-3 projections, respectively. The cyan star denotes the reference benchmark. Vertical lines indicate $M_{Z_B}=m_\chi$, $2m_\chi$, $2M_\Psi$, and $2M_\eta$.}
\label{fig:run3}
\end{figure}

Fig.~\ref{fig:run3} includes the present collider constraints used in our analysis: the CMS EXO-24-007 low-mass search, the ATLAS trigger-level and CMS scouting results, and the CMS EXO-19-012 broad-resonance constraint. For the Run-3 sensitivity estimate we extrapolate only the ATLAS trigger-level and CMS scouting coupling limits, for which a direct luminosity scaling can be applied. We do not rescale the EXO-24-007 or EXO-19-012 constraints, since their boosted-jet and finite-width treatments contain analysis-specific information that is not captured by a simple luminosity extrapolation.

\subsection{Present-day annihilation and indirect detection}
\label{subsec:indirect}

The present-day annihilation phenomenology is qualitatively different from thermal freeze-out. At freeze-out the splitting satisfies $\delta_\chi/T_f\ll1$, so both Majorana states are populated and the relic abundance is controlled predominantly by the unsuppressed off-diagonal process $\chi_1\chi_2\to Z_B^\ast\to q\bar q$. In the Galactic halo, instead, the phenomenology depends crucially on the abundance of the excited state $\chi_2$.

If the halo is dominated by the ground state $\chi_1$, the leading annihilation into SM fermions is $\chi_1\chi_1\to Z_B^\ast\to q\bar q$. The diagonal Majorana current is axial whereas the quark current is vectorial, so the annihilation is $p$ wave. Neglecting quark masses,
\begin{equation}
\frac{\langle\sigma v\rangle_{11}}
{\langle\sigma v\rangle_{12}}
\simeq
\frac{v_{\rm rel}^2}{24}.
\label{eq:pwavehalo}
\end{equation}
For $v_{\rm rel}\sim10^{-3}$ and the reference off-diagonal rate $\langle\sigma v\rangle_{12}^{v\to0}\simeq5.0\times10^{-26}\cm^3\s^{-1}$, this gives a present-day rate of order $10^{-33}\cm^3\s^{-1}$. The corresponding continuum gamma-ray, charged-cosmic-ray, and neutrino signals are therefore negligible.

A surviving excited population would restore $s$-wave coannihilation and generate exothermic DD. At the reference point the accepted exothermic yield per incident particle is approximately $530$ times the endothermic yield in $200$--$300\keV$ under the production response. The requirement that the former not exceed the latter gives $f_2\lesssim1.9\times10^{-3}$ as a diagnostic, not an experimental confidence limit. The scan imposes the more conservative primordial-depletion target $f_2^{\rm prim}\leq10^{-6}$.

Since $\delta_\chi<2m_e$, the charged-lepton decay is forbidden. The electroweak admixture nevertheless permits $\chi_2\to\chi_1\nu\bar\nu$ through a virtual SM $Z$. For $\theta_N=0.005$ and $\delta_\chi=300\keV$, the checked three-flavor width gives $\tau_{\nu\bar\nu}=6.24\times10^7$ yr and $\log_{10}f_2^{\rm prim}(t_0)\simeq-96.38$, starting from equal populations. This establishes present-day primordial depletion without assuming a radiative lifetime.

The same admixture generates $\chi_2\to\chi_1\gamma$. Including both physical charged anomalons, the $W$ diagrams, their charged Goldstone partners, and the mixed diagrams required by gauge invariance gives $|\mu_{12}|=5.34\times10^{-10}\GeV^{-1}$ at the reference point. The radiative lifetime is $\tau_\gamma=2.68\times10^5\,\mathrm{s}\simeq3.11$ days, approximately $7.3\times10^9$ times shorter than the neutrino lifetime. App.~\ref{app:radiativedecay} supplies the matching and independent checks.
Representative decay mechanisms are shown in Fig.~\ref{fig:chi2_decay_diagrams} of App.~\ref{app:excitedstate}. This channel depletes the primordial population before recombination. The scan retains the neutrino-only survival criterion as a conservative sufficient condition, so it does not rely on the faster radiative decay to accept points.

Pure hypercharge--baryon kinetic mixing cannot replace this mechanism: the complete neutral-propagator contribution to the neutrino transition cancels at zero momentum, leaving a $q^2$ suppression and an extremely long lifetime. App.~\ref{app:mixing} derives the cancellation. Smooth-halo regeneration and the distinct Solar population are discussed in App.~\ref{app:regeneration}.

A qualitatively different present-day annihilation channel becomes available when $M_{Z_B}<m_\chi$, where the secluded process $\chi_1\chi_1\to Z_BZ_B$ is kinematically open. Because the dominant vector coupling is off diagonal, this process proceeds through $t$- and $u$-channel $\chi_2$ exchange, together with the additional diagrams required by the gauge-complete theory. Unlike direct annihilation into $q\bar q$, the $Z_BZ_B$ channel can contain an $s$-wave contribution and can therefore remain relevant at Galactic velocities.
The subsequent decay of the two mediators produces four-fermion final states. When both $Z_B$ bosons are on shell and narrow, the inclusive cascade rate factorizes as
\begin{equation}
\langle\sigma v\rangle_{\rm incl.\,4f}
\simeq
\langle\sigma v\rangle_{Z_BZ_B},
\end{equation}
whereas a specific exclusive final state carries the corresponding product of branching fractions. Consequently, the very small rates obtained from explicit exclusive $2\to4$ MadDM calculations should not be interpreted as an additional universal suppression of the inclusive on-shell $Z_BZ_B$ cascade. A precision continuum constraint requires an independently checked ground-state amplitude and the actual cascade spectra. We retain the implemented open-channel calculation as a diagnostic. Other bosonic ground-state channels, including $Z_Bh_1$ where open, also require attention.

Finally, the charged anomalons generate monochromatic gamma-ray final states at one loop, $\chi_1\chi_1\to\gamma\gamma$ and $\chi_1\chi_1\to\gamma Z$. The values used here are not obtained from a new complete one-loop calculation of the present mixed pseudo-Dirac model. Instead, we use the charged-anomalon amplitudes derived for anomaly-free $U(1)_B$ models in Ref.~\cite{AnomalyFreeDM} as an order-of-magnitude proxy. For $M_\Psi=1.5\TeV$ and $M_\eta=2.0\TeV$, this gives
\begin{align}
\langle\sigma v\rangle_{\gamma\gamma}
&\sim1.8\times10^{-34}\cm^3\s^{-1},
\\
\langle\sigma v\rangle_{\gamma Z}
&\sim6.4\times10^{-32}\cm^3\s^{-1}.
\label{eq:linebench}
\end{align}
These values should be regarded as benchmark estimates rather than precision predictions, since the exact amplitudes depend on the physical charged-anomalon eigenstates and their mixing. Nevertheless, they lie several orders of magnitude below the current TeV-scale gamma-ray line sensitivity, of order $10^{-28}\cm^3\s^{-1}$~\cite{HESSLines}, and therefore do not presently constrain the benchmark.

\subsection{Solar capture and the IceCube recast}
\label{subsec:solar}

Solar capture provides a particularly sharp consistency test of endothermic DM. DM particles falling into the Solar potential are gravitationally accelerated before scattering, so splittings which select only the extreme high-velocity tail at terrestrial detectors can still permit capture on heavy Solar elements. The absence of a Solar-neutrino excess can then constrain the underlying DD interaction, provided that the annihilation final state and the captured spatial distribution are specified. Translating an equilibrium neutrino limit into a scattering bound additionally requires sufficiently rapid annihilation~\cite{IceCubeSolar2025}. These assumptions require particular care in a pseudo-Dirac model, where capture, state conversion and annihilation can involve different combinations of the two mass eigenstates~\cite{PospelovRamaniSolar,NguyenLindenHooperSolar}.

The dominant interaction responsible for capture in the $U(1)_B$ completion is the coherent inelastic vector interaction which also controls the LZ signal. Its baryonic singlet limit corresponds to the isoscalar $\mathcal O_1^s$ operator, with the zero-momentum per-nucleon normalization given in Eq.~\eqref{eq:sigmaN}. The fermion admixture also introduces ordinary-$Z$ exchange and hence a small difference between the proton and neutron transition amplitudes, denoted by $C_p$ and $C_n$. The present capture diagnostic incorporates this difference through an Fe-56-equivalent per-nucleon normalization, \begin{equation} \sigma_{\rm tr}^{\rm Fe\text{-}eq}=\frac{\mu_N^2}{\pi}\left[\frac{26C_p+30C_n}{56}\right]^2, \end{equation} expressed in physical cross-section units, where $\mu_N$ is the DM--nucleon reduced mass. This prescription reduces to the isoscalar normalization when $C_p=C_n$. It is an approximation for translating the interaction strength into a capture rate; it does not replace a calculation of the separate Solar elements and their nuclear responses.

\begin{figure}[htbp]
\centering
\includegraphics[width=0.98\columnwidth]{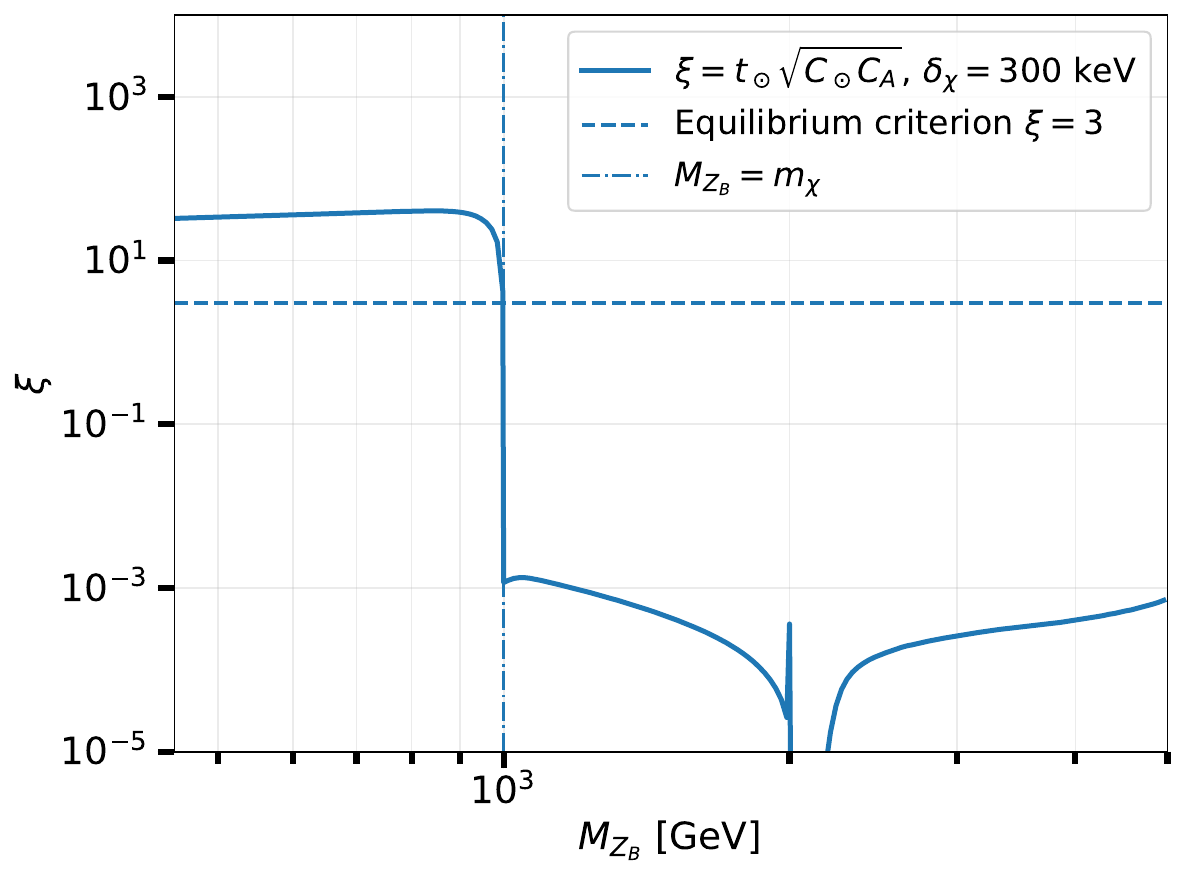}
\caption{Solar capture--annihilation equilibrium diagnostic for $m_\chi=1\TeV$ and $\delta_\chi=300\keV$, along the central relic-density solution $g_B^{\rm relic}$. Capture follows Eq.~\eqref{eq:solarcaptureproxy}, while $C_A=\langle\sigma v\rangle_\odot/V_{\rm eff}$ assumes a thermalized, predominantly ground-state population. The annihilation rate uses $\chi_1\chi_1\to Z_BZ_B$ for $M_{Z_B}<m_\chi$ and Solar-temperature $p$-wave annihilation into quarks above this threshold, rather than a fixed thermal-relic cross section. The horizontal dashed line marks $\xi=3$, and the vertical dash-dotted line marks $M_{Z_B}=m_\chi$. This one-population diagnostic does not evolve captured-state conversion, decay, or coannihilation.}
\label{fig:solar}
\end{figure}

The capture normalization is inherited from the dedicated pseudo-Dirac benchmark of Ref.~\cite{SolarCapture}, with 
\begin{equation} 
C_\odot^{\rm ref}=1.48\times10^{20}\s^{-1},\,\, \sigma_{\rm ref}=6.34346\times10^{-43}\cm^2, 
\end{equation} 
at $m_\chi=1\TeV$ and $\delta_\chi=300\keV$. 
In the optically thin approximation, capture scales linearly with the transition cross section. For the fixed mass and splitting used in Fig.~\ref{fig:solar}, we therefore adopt \begin{equation} C_\odot=C_\odot^{\rm ref}\frac{\sigma_{\rm tr}^{\rm Fe\text{-}eq}}{\sigma_{\rm ref}}. \label{eq:solarcaptureproxy} \end{equation} The Solar composition, escape-speed profile, nuclear form factors and incident halo distribution are represented by the reference calibration, rather than recomputed at each mediator mass. Away from the reference mass and splitting, the package additionally rescales capture using the inverse of the tabulated inelastic $b\bar b$ cross-section limit~\cite{NguyenLindenHooperSolar}. This extra factor equals unity for the figure considered here and should not be interpreted as an independent Solar capture calculation.

The parameters entering the annihilation estimate follow the same specified UV benchmark as the relic-density analysis: $M_\Psi=1.5m_\chi$, $M_\eta=2m_\chi$, $m_{h_2}=3\TeV$, scalar mixing $\sin\theta_B=3\times10^{-4}$, and equal real fermion mixing angles $\theta_L=\theta_R=0.005$. At each mediator mass, the plotting script takes the central coupling $g_B^{\rm relic}$ which reproduces $\Omega_\chi h^2\simeq0.12$ in the freeze-out calculation. It retains converged relic-density roots, without requiring that every displayed point also pass the LZ, collider or theory-domain selections. The figure is consequently an equilibrium diagnostic along the relic-density branch, rather than a plot restricted to the combined allowed region.

For this diagnostic, we assume that the captured population can be represented by a single, predominantly ground-state population which has thermalized into a compact Solar-core distribution. Neglecting evaporation and self-capture, and taking the capture and annihilation coefficients to be time independent, its evolution is 
\begin{equation} 
\frac{{\rm d}N}{{\rm d}t}=C_\odot-C_A N^2,
\,\, 
C_A=\frac{\langle\sigma v\rangle_\odot}{V_{\rm eff}}. 
\end{equation} 
For an initially negligible population, this gives 
\begin{equation} 
\Gamma_A=\frac{C_\odot}{2}\tanh^2\xi,
\,\, 
\xi=t_\odot\sqrt{C_\odot C_A},\,\, 
t_\odot=4.57\,\mathrm{Gyr}. 
\label{eq:solareq} 
\end{equation} 
We use 
\begin{equation} 
V_{\rm eff}=5.7\times10^{27}\cm^3\left[\frac{100\GeV}{m_\chi}\right]^{3/2}, 
\end{equation} 
so that $V_{\rm eff}=1.80\times10^{26}\cm^3$ at $m_\chi=1\TeV$. This effective volume assumes thermalization; the calculation does not establish that the captured particles actually attain this distribution. The horizontal line $\xi=3$ denotes the practical equilibrium criterion, for which $\tanh^2\xi\simeq0.990$. When applying the Solar proxy, we retain the continuous suppression factor $\tanh^2\xi$, rather than treating this line as a discontinuous exclusion boundary.

The annihilation rate entering $C_A$ is not fixed to a canonical thermal-relic value. The freeze-out abundance involves a thermally populated pseudo-Dirac pair and all included annihilation channels, whereas the Solar diagnostic assumes predominantly $\chi_1$ initial states at much smaller velocities. For $M_{Z_B}<m_\chi$, the code uses the ground-state $\chi_1\chi_1\to Z_BZ_B$ kernel, evaluated near the zero-velocity limit at $v_{\rm rel}=10^{-4}$ in units of the speed of light. The implemented ground-state prescription retains vector--vector and axial--axial contributions, while removing the vector--axial combinations associated with the unlike-state initial pair. This is a low-velocity estimate of the open channel, rather than a full Solar thermal average or a sum over every possible ground-state annihilation channel.

For $M_{Z_B}\geq m_\chi$, the two-mediator channel is closed and the diagnostic instead uses a diagonal $\chi_1\chi_1\to q\bar q$ estimate. The baryonic interaction of an individual Majorana state is axial, and this annihilation into the vector quark current is $p$ wave. The code assumes $T_\odot=1.30\keV$ and evaluates the leading velocity dependence using 
\begin{equation} 
\langle v_{\rm rel}^2\rangle=\frac{6T_\odot}{m_\chi}=7.8\times10^{-9}
\,\,
\text{for }m_\chi=1\TeV, 
\end{equation} 
corresponding to an rms relative speed of approximately $26.5\,\mathrm{km\,s^{-1}}$. It sums the kinematically accessible quark flavours and retains the mediator propagator. Near $M_{Z_B}\simeq2m_\chi$, the Solar estimate uses the approximate width regulator $\Gamma_{\rm reg}=0.02M_{Z_B}g_B^2$, rather than the complete physical width employed in the relic-density calculation. The small resonance feature on this branch is therefore diagnostic and should not be interpreted as a precision prediction. Additional electroweak or scalar-mediated ground-state channels are not included in this branch estimate.

These prescriptions explain the large difference between the two branches of Fig.~\ref{fig:solar}. Near $M_{Z_B}=0.50\TeV$, the relic-density coupling is approximately $g_B^{\rm relic}=0.135$, and the implemented open-channel estimate gives $C_\odot\simeq1.20\times10^{21}\s^{-1}$, $\langle\sigma v\rangle_\odot\simeq8.25\times10^{-27}\cm^3\s^{-1}$ and $\xi\simeq34$. Near $M_{Z_B}=1.44\TeV$, the corresponding coupling is approximately $0.228$, but the diagonal quark-annihilation estimate gives only $\langle\sigma v\rangle_\odot\simeq1.58\times10^{-35}\cm^3\s^{-1}$. With $C_\odot\simeq1.44\times10^{20}\s^{-1}$, this yields $\xi\simeq5.1\times10^{-4}$. Thus the displayed closed branch is far from equilibrium within the assumed one-population treatment. The sharp change near $M_{Z_B}=m_\chi$ reflects the closing of the channel used in the diagnostic and the switch to the strongly suppressed quark-annihilation estimate.

A separate issue is whether the predominantly ground-state assumption is justified inside the Sun. Inelastic capture continuously produces $\chi_2$, while scattering can exchange the states. The matched three-day decay time is now short compared with the age of the Sun. In a thermalized one-zone treatment with suppressed upward conversion, a coupled population calculation gives $f_2^\odot\simeq1.9\times10^{-12}$ at the reference point; the mixed-pair annihilation contribution increases the displayed ground-state rate by approximately $1.1\%$. App.~\ref{app:solarpopulation} gives the equations and assumptions. This checks the population limit of the diagnostic, but it does not establish thermalization of the actual captured orbits or compute the elemental capture and neutrino response.

On the open branch, we estimate the IceCube sensitivity to the hadronic cascade using the published direct $b\bar b$ response, since no dedicated $Z_BZ_B\to4q$ template is supplied in the recast used here. Neglecting quark masses and adopting an isotropic-decay approximation, non-relativistic annihilation gives quark-energy endpoints \begin{equation} E_q^\pm=\frac{m_\chi}{2}\left[1\pm\beta_B\right],\,\, \beta_B=\sqrt{1-\frac{M_{Z_B}^2}{m_\chi^2}}. \label{eq:qbox} \end{equation} We match the rms primary-quark energy to an effective direct $b\bar b$ mass, \begin{equation} m_{\rm eff}^{\rm RMS}=\frac{m_\chi}{2}\sqrt{1+\frac13\left[1-\frac{M_{Z_B}^2}{m_\chi^2}\right]}. \label{eq:meff} \end{equation} Treating the entire cascade as a $4b$ yield proxy, the doubled number of primary quarks motivates \begin{equation} \sigma_{N,{\rm IC}}^{4b,\,\rm proxy}\simeq\frac12\,\sigma_{N,{\rm IC}}^{b\bar b}\left[m_{\rm eff}^{\rm RMS},\delta_\chi\right]. \label{eq:icecubeproxy} \end{equation} This factor of two is a primary-particle yield approximation, not a calculated ratio of IceCube event rates. A generic $Z_B$ decays into all accessible quark flavours; the proxy does not include the actual flavour branching fractions, mediator polarization, or a dedicated neutrino transport and detector-response calculation for the $4q$ cascade. We do not extrapolate the mapping outside the tabulated mass and splitting range.

The equilibrium response is adjusted continuously by comparing the transition cross section multiplied by $\tanh^2\xi$ with the proxy limit. The effective-mass mapping and doubled-yield prescription affect this approximate IceCube comparison; they do not determine the annihilation coefficient $C_A$ or the equilibrium curve itself. Within the specified capture, annihilation and thermalization assumptions, the open branch can reach equilibrium and can consequently face a substantial Solar-neutrino constraint. A quantitative exclusion of the UV model nevertheless requires an independently validated ground-state annihilation calculation, element-by-element Solar capture, captured-state evolution and thermalization, and neutrino yields for the actual mediator decay channels.

\section{Combined results}
\label{sec:results}

We now combine the LZ recoil information, the thermal relic-density requirement, the present collider constraints, and the Solar-capture diagnostic. The important point is that all these observables depend on the same baryonic coupling $g_B$. For each choice of $(m_\chi,\delta_\chi,M_{Z_B})$, we therefore scan $g_B$ and retain only those values that satisfy the different requirements simultaneously.

The first ingredient is the LZ recoil normalization. For every point in $(m_\chi,\delta_\chi,M_{Z_B},g_B)$ we calculate the full xenon recoil spectrum and integrate it according to Eq.~\eqref{eq:NexpLZ}, including the nuclear form factor, the halo velocity distribution, and the detector acceptance (see Sec.~\ref{subsec:LZ}). Since the high-energy LZ result contains only one candidate event, we do not interpret it as a precise measurement of the interaction strength. Instead, we associate to the observed event the exact $68\%$ Poisson interval for the expectation value. For one observed event this corresponds approximately to
\begin{equation}
0.17\lesssim N_{\rm sig}\lesssim3.30,
\label{eq:LZpoisson}
\end{equation}
which defines the central $68.27\%$ LZ count band used in the results. 
We denote the corresponding allowed coupling interval by ${\cal I}_{\rm LZ}(g_B,m_\chi,\delta_\chi,M_{Z_B})$.
This procedure retains the large statistical uncertainty associated with a single event. It should therefore not be confused with either a Gaussian error on $g_B$ or a direct fit to the zero-momentum cross section $\sigma_N^0$. A complete detector-level LZ likelihood would be required for a more precise statistical interpretation.
Our direct-detection calculation includes the off-diagonal vector transitions $\chi_1N\to\chi_2N$ and $\chi_2N\to\chi_1N$, weighted by the corresponding halo populations, together with diagonal axial-vector elastic scattering and scalar-mediated SI elastic scattering, with the axial contribution evaluated in the coherent $\mathcal O_8$ nuclear-response approximation.

The relic density is evaluated independently at the same model parameter configurations. We compute the finite-temperature annihilation rate including the light-fermion channel groups specified in Sec.~\ref{subsec:relic} and solve the Boltzmann equation numerically (see Sec.~\ref{subsec:relic}). The central calculation assumes the standard thermal history, namely radiation domination, entropy conservation, and thermal freeze-out. Rather than imposing the measured cosmological abundance with its $\sim 1\%$ observational error \cite{Planck2018}, we use the broader requirement
\begin{equation}
0.108\leq\Omega_\chi h^2\leq0.132,
\label{eq:relicbandcombined}
\end{equation}
corresponding to a $10\%$ interval around the observed value $\Omega_\chi h^2=0.12$ \cite{Planck2018}. This band is intended as a theory and cosmology tolerance. It accounts for residual uncertainties in the approximate particle-physics treatment and can also represent modest departures from the standard cosmological history, for example small modifications of the expansion rate or entropy evolution around freeze-out \cite{Arcadi:2024jzv}. It should not, however, be interpreted as covering an arbitrary non-standard cosmology. We denote the corresponding coupling interval by ${\cal I}_{\rm relic}$.

The third ingredient is the collider constraint. At each scan point we calculate the physical $Z_B$ total width and branching fractions and compare the predicted dijet signal with the observed ATLAS and CMS limits described in Sec.~\ref{subsec:collider}. The collider-allowed region is denoted by ${\cal I}_{\rm coll}$. The primary combined condition is then
\begin{equation}
{\cal I}_{\rm primary}
=
{\cal I}_{\rm LZ}
\cap
{\cal I}_{\rm relic}
\cap
{\cal I}_{\rm coll},
\,\,
{\cal I}_{\rm primary}\neq\varnothing.
\label{eq:primaryintersection}
\end{equation}
Only present observed collider limits enter this condition. The Run-3 curves shown in Sec.~\ref{subsec:run3} are luminosity-only sensitivity estimates and are never used as exclusions.

The collider recast is performed with the mediator width evaluated point by point rather than with a fixed narrow-width approximation. For $M_{Z_B}<2m_\chi$, the decay $Z_B\to\chi_1\chi_2$ is closed. Below the anomalon thresholds the mediator is then predominantly hadronic, and the dijet interpretation is particularly clean. This is the situation at the reference point $m_\chi=1\TeV$, $M_{Z_B}=1.44\TeV$, where the relic-density solution gives $g_B=0.2278$ and the present observed dijet boundary is approximately $g_B^{95}\simeq0.240$.

Once $M_{Z_B}>m_{\chi_1}+m_{\chi_2}\simeq2m_\chi$,
the decay $Z_B\to\chi_1\chi_2$ opens. Because the transition coupling to the dark sector is sizeable, this channel can increase the total mediator width substantially while simultaneously reducing ${\rm BR}(Z_B\to jj)$. Both effects weaken the dijet constraint. In this region we therefore use the physical $\Gamma_{Z_B}/M_{Z_B}$ and visible branching fraction in the recast. The CMS high-mass broad-resonance information is included only within the published intrinsic-width range; if the model width exceeds that range before an exclusion is reached, the corresponding line is interpreted as the boundary of the validated collider recast rather than as an experimental exclusion.

After identifying an LZ--relic--collider overlap, we impose the remaining theory-validity conditions at the actual compatible value of $g_B$. These include the stated gauge, Yukawa, and scalar perturbativity criteria and the explicit numerical validation of the relic solution. No universal cut on $\Gamma_{Z_B}/M_{Z_B}$ is imposed. The long-range indicator is recorded as a diagnostic, while the physical width remains in the thermal rate and applicable collider templates. Interpolation of the relic-density surface is used only as a numerical guide and as an initial condition for the root finder; it is never sufficient by itself to classify a point as viable.

The final primary region shown in the combined plots therefore corresponds schematically to
\begin{equation}
{\rm LZ}
+
{\rm relic}
+
{\rm collider}
+
{\rm theory\ validity}.
\label{eq:primarycombined}
\end{equation}

The Solar diagnostic described in Sec.~\ref{subsec:solar} is applied only as an additional diagnostic layer, rather than as a hard exclusion of the parameter space, because its interpretation requires further assumptions. In particular, the dominant Solar annihilation channel can yield a four-quark final state through $\chi_1\chi_1\to Z_BZ_B\to4q$, whereas IceCube does not provide dedicated limits for this topology \cite{IceCubeSolar2025}. Any bound therefore requires a recast of available two-body final states, such as $b\bar b$, and is not included in the primary global exclusion.

\subsection{Physical regimes of the combined scan}
\label{subsec:combinedvalidation}

Two mediator thresholds are useful for organizing the phenomenology of the
combined scan.
The first occurs at $M_{Z_B}=m_\chi$, below which the annihilation channel $\chi_1\chi_1\to Z_BZ_B$ becomes kinematically accessible. This channel is included in the relic-density
calculation together with the other relevant annihilation and coannihilation
processes. Its opening changes the relative importance of the different
freeze-out channels and is also relevant for the Solar phenomenology, since it
provides a ground-state annihilation mode that does not require a real
$\chi_2$ in the initial state.

The second threshold occurs at $M_{Z_B}\simeq2m_\chi$, where the mediator decay $Z_B\to\chi_1\chi_2$
becomes kinematically allowed. This threshold mainly affects the collider
phenomenology: the additional dark-sector contribution increases the total
mediator width and reduces the visible dijet branching fraction. The collider
constraint is therefore evaluated using the physical width and branching
fractions, with the broad-resonance information included where available.

The intermediate region $m_\chi\leq M_{Z_B}<2m_\chi$
is characterized by both $Z_BZ_B$ production at the nonrelativistic annihilation threshold and the
two-body dark decay of the mediator being closed. At the reference point, freeze-out is dominated by quark coannihilation and the mediator is predominantly visible in dijet searches. Elsewhere in this interval, scalar or mixed bosonic channels can also be open. The reference
benchmark lies in this region.

For $M_{Z_B}<m_\chi$, the opening of the $Z_BZ_B$ channel changes the
freeze-out composition and can also have important consequences for Solar
annihilation. For $M_{Z_B}>2m_\chi$, the relic-density calculation remains
unchanged in its treatment, while the collider interpretation must account for
the reduced dijet branching fraction and the larger mediator width.

These thresholds separate physically distinct regions of the same scan. Convergence of the implemented relic calculation is assessed independently from omitted processes and the validity of each experimental recast. The Solar constraints are discussed separately
below, since their interpretation requires an additional recast for the
relevant multi-quark final states.

\subsection{Solar capture as a diagnostic}
\label{subsec:combinedsolar}

The Solar analysis is not included in the primary exclusion of the parameter
space. Our baseline viable region is defined by the simultaneous requirements
from LZ, the relic density, collider searches, and theory consistency. Solar
capture is instead used as an additional diagnostic of this surviving region.

As discussed in Sec.~\ref{subsec:solar}, the Solar phenomenology depends
sensitively on the available present-day annihilation channels and on whether
capture--annihilation equilibrium is reached. We therefore evaluate the
equilibrium parameter $\xi=t_\odot\sqrt{C_\odot C_A}$
for the surviving points and use the Solar calculation primarily to identify
the regions where an appreciable neutrino signal from the Sun could be
expected.

In the region where $M_{Z_B}<m_\chi$,
the annihilation channel $\chi_1\chi_1\to Z_BZ_B$
is kinematically open and can significantly affect the Solar annihilation
rate. The corresponding visible final state is predominantly a cascade into
quarks, $\chi_1\chi_1\to Z_BZ_B\to4q$.
However, IceCube does not provide a dedicated Solar-DM limit for this
four-quark topology \cite{IceCubeSolar2025}. Any constraint therefore requires a recast of available
two-body channels, such as $b\bar b$, together with assumptions about the
resulting neutrino spectrum.

For this reason, we do not use the Solar/IceCube result as a hard exclusion
of the parameter space. All primary results are shown without imposing the
Solar constraint, while the Solar calculation is presented separately as a
complementary diagnostic of the surviving configurations and as an indication
of the regions where a dedicated neutrino analysis of the relevant multi-quark
final state would be most useful.

\subsection{Profiling over the inelastic splitting}
\label{subsec:deltaprofile}

The reference point has often been illustrated with $\delta_\chi=300\keV$, but the inelastic splitting is not fixed in the full analysis. For each pair $(m_\chi,M_{Z_B})$, we vary $\delta_\chi$ and determine whether there exists a common coupling $g_B$ that reproduces the LZ high-energy event, satisfies the low-energy recoil constraint, gives the required relic abundance, and remains compatible with the collider bounds. The direct-detection conditions are evaluated first, and the resulting candidate points are subsequently tested against the remaining requirements.

Increasing $\delta_\chi$ raises the minimum incoming velocity in Eq.~\eqref{eq:vmininelastic}, progressively restricting the endothermic transition to the high-velocity tail of the halo. The splitting therefore controls both the normalization required to explain the LZ candidate and the relative event rates in the low- and high-energy windows. However, our recoil calculation also includes diagonal elastic channels, which remain accessible when the endothermic rate becomes small. Consequently, the behavior near the endothermic kinematic endpoint must be interpreted using the complete recoil spectrum rather than the transition contribution alone.

Specifically, we include the off-diagonal vector transitions $\chi_1N\to\chi_2N$ and $\chi_2N\to\chi_1N$, weighted by the corresponding halo populations, together with diagonal axial-vector elastic scattering and scalar-mediated spin-independent elastic scattering. The excited-state abundance is evaluated using the depletion prescription described above and is negligible in the relevant depleted-state benchmark. The diagonal axial contribution is evaluated using the coherent, velocity-suppressed $\mathcal O_8$ response; this implementation does not include the complete nuclear orbital and spin responses. The scalar contribution includes interference between the two scalar mediators. Transition and elastic rates are added incoherently because they correspond to different final DM states, while interference within each scattering amplitude is retained where implemented.

The first restriction concerns the kinematic accessibility of the endothermic interpretation. The SHM used to generate the figures has $v_0=238\kms$, $v_{\rm esc}=544\kms$, and $v_E=250.2\kms$, giving $v_{\max}=v_{\rm esc}+v_E=794.2\kms$. For a xenon isotope of nuclear mass $m_A$ and a recoil energy $E_R$, the largest accessible endothermic splitting is, in units with $c=1$,
\begin{equation}
\delta_\chi^{\rm max}(m_\chi,A,E_R)
=
v_{\max}\sqrt{2m_AE_R}
-
\frac{m_AE_R}{\mu_{\chi A}}.
\label{eq:deltamaxER}
\end{equation}
We define the signal-window kinematic ceiling by
\begin{equation}
\delta_{\chi,\max}^{\rm SHM}(m_\chi)
=\max_{\substack{A\\200\keV\leq E_R\leq300\keV}}
\delta_\chi^{\rm max}(m_\chi,A,E_R).
\label{eq:deltamaxSHM}
\end{equation}
Above this ceiling, no endothermic transition can produce a recoil in the adopted signal window within the SHM. Elastic scattering remains possible, but such points fall outside the endothermic interpretation considered in this scan. Moreover, the ceiling specifies where kinematic support disappears; it does not guarantee an appreciable transition rate close to that boundary.

At each trial splitting below this ceiling, we integrate the complete recoil spectrum with the detector acceptance specified in Sec.~\ref{subsec:LZ} and Eq.~\eqref{eq:NexpLZ}. In the $200$--$300\keV$ signal window, we require the expected count to lie within the exact central $68.27\%$ Poisson interval for one observed event,
\begin{equation}
\mu_{\rm sig}^{\rm low}
\simeq0.17275
\leq
N_{\rm sig}(g_B,\delta_\chi)
\leq
3.29953
\simeq\mu_{\rm sig}^{\rm high}.
\label{eq:NsigPoissondelta}
\end{equation}
Numerically solving the count equations defines the coupling band
\begin{equation}
g_{B,{\rm LZ}}^{\rm low}(\delta_\chi)
\leq
g_B
\leq
g_{B,{\rm LZ}}^{\rm high}(\delta_\chi).
\label{eq:gBLZdelta}
\end{equation}
The dominant baryonic transition contribution scales as $g_B^4$ at fixed mediator mass. Nevertheless, the complete count is not assumed to obey an exact fourth-power scaling: the ordinary-$Z$ transition amplitude, scalar exchange, and their interference terms generate additional coupling dependence. The band boundaries are therefore obtained from the implemented full count rather than by applying a universal $g_B^4$ rescaling.

The low-energy condition is evaluated using the same spectrum and detector response:
\begin{equation}
N_{\rm low}
=
\mathcal E
\int_{0}^{200\keV}
dE_R\,\epsilon(E_R)\,\frac{dR}{dE_R}.
\label{eq:Nlowdelta}
\end{equation}
For this simplified count construction, we treat the interval as a zero-count control region, $n_{\rm low}=0$, without an additional background component. The integration includes the detector acceptance, so the nominal lower endpoint at zero recoil does not imply sensitivity down to zero energy. This construction is a diagnostic count constraint rather than a reconstruction of the complete LZ experimental likelihood.

For zero observed events, the one-sided $95\%$ CL upper limit on the Poisson mean is
\begin{equation}
\mu_{\rm low}^{95}
=
-\ln(0.05)
\simeq2.996.
\label{eq:mulow95}
\end{equation}
We impose
\begin{equation}
N_{\rm low}(g_B,\delta_\chi)
\leq
\mu_{\rm low}^{95},
\label{eq:Nlow95condition}
\end{equation}
and determine the corresponding coupling limit numerically,
\begin{equation}
g_B
\leq
g_{B,{\rm low}}^{95}(\delta_\chi).
\label{eq:gBlow95}
\end{equation}
The direct-detection conditions overlap only when
\begin{equation}
g_{B,{\rm LZ}}^{\rm low}(\delta_\chi)
\leq
\min\left[
g_{B,{\rm LZ}}^{\rm high}(\delta_\chi),
g_{B,{\rm low}}^{95}(\delta_\chi)
\right].
\label{eq:LZoverlapdelta}
\end{equation}
This overlap condition is checked at every trial splitting.

We define the lower splitting boundary by the first intersection of the low-energy upper-limit curve with the lower edge of the signal-region Poisson band,
\begin{equation}
g_{B,{\rm low}}^{95}(\delta_{\chi,\min})
=
g_{B,{\rm LZ}}^{\rm low}(\delta_{\chi,\min}).
\label{eq:deltamincondition}
\end{equation}
Below this crossing, the coupling required even at the lower edge of the signal band overproduces low-energy events. This definition differs from using the central normalization $N_{\rm sig}=1$ to locate the crossing: the latter is more restrictive and gives approximately $306.7\keV$ for the illustrated benchmark, whereas the lower-band definition gives approximately $184.8\keV$.

Importantly, the first crossing and the SHM ceiling do not by themselves delimit an entirely allowed interval. Elastic scattering can cause the low-energy limit and the lower signal-band edge to cross again at larger splitting. We therefore define the direct-detection-compatible set by
\begin{widetext}
\begin{equation}
\mathcal D_{\rm LZ}
=
\left\{
\delta_\chi:
\delta_{\chi,\min}\leq\delta_\chi<
\delta_{\chi,\max}^{\rm SHM},
\,\,
g_{B,{\rm LZ}}^{\rm low}(\delta_\chi)
\leq
\min\left[
g_{B,{\rm LZ}}^{\rm high}(\delta_\chi),
g_{B,{\rm low}}^{95}(\delta_\chi)
\right]
\right\}.
\label{eq:deltarangefinal}
\end{equation}
\end{widetext}
Additional relic-density, collider, and theoretical requirements further restrict the candidate points within this set.

\begin{figure}[htbp]
\centering
\includegraphics[width=0.98\columnwidth]{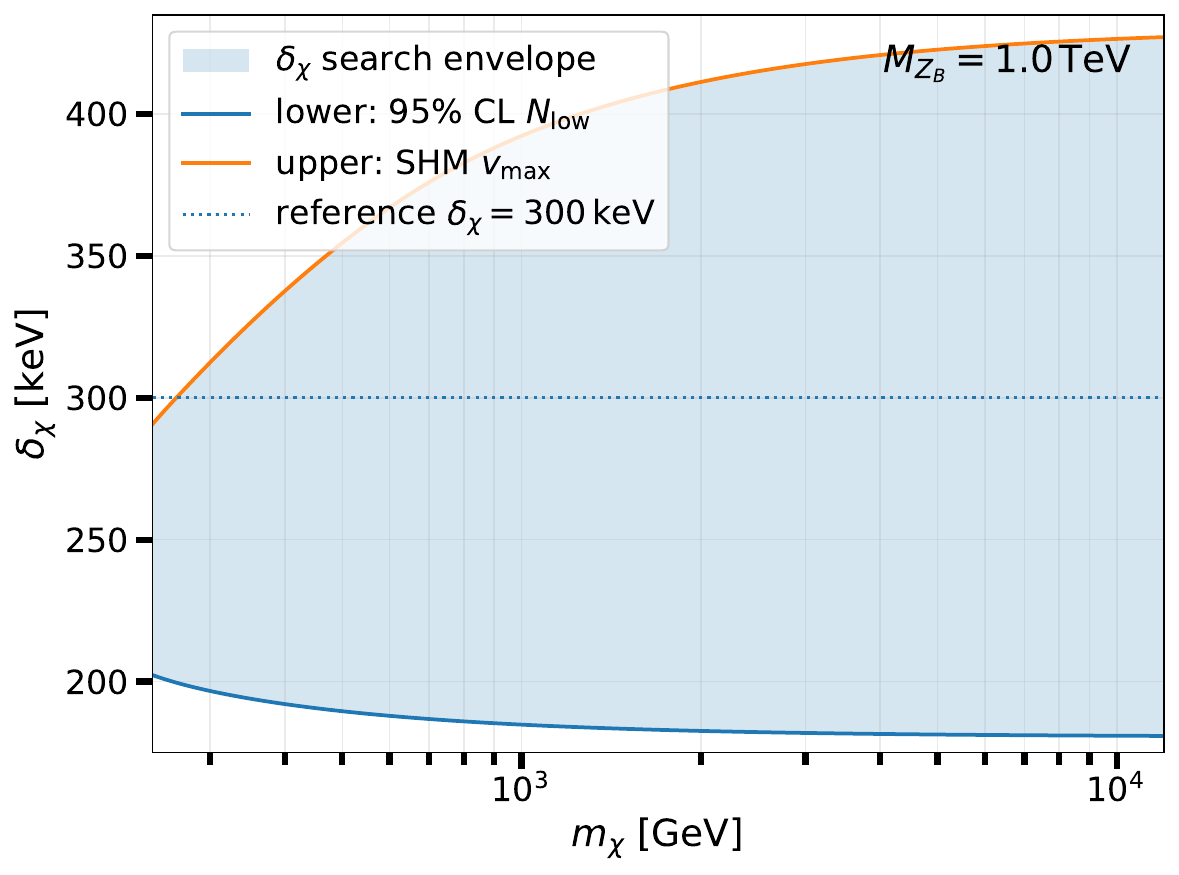}
\caption{Splitting search envelope as a function of $m_\chi$ for fixed $M_{Z_B}=1\TeV$. The solid blue curve gives the first intersection of the $0$--$200\keV$ one-sided $95\%$ CL coupling limit with the lower edge of the $200$--$300\keV$ central $68.27\%$ Poisson band. The solid orange curve is the SHM kinematic ceiling for the endothermic transition, evaluated with $v_{\max}=794.2\kms$. The shading gives the initial $\delta_\chi$ search envelope; each point must also satisfy the full count-overlap condition in Eq.~\eqref{eq:deltarangefinal}. The horizontal dotted line marks $\delta_\chi=300\keV$.}
\label{fig:deltaband}
\end{figure}

Fig.~\ref{fig:deltaband} displays the first-crossing lower boundary and the SHM kinematic ceiling as functions of $m_\chi$, with the mediator mass fixed to $M_{Z_B}=1\TeV$. The shaded region is the initial search envelope between these two boundaries. It should not be interpreted as a fully accepted direct-detection region, because the plotted upper curve is the endothermic kinematic ceiling rather than the upper boundary obtained from the full overlap condition. The ceiling rises with $m_\chi$ and approaches its heavy-DM limit as the DM--nucleus reduced mass saturates. The lower curve instead follows the comparison of the accepted low- and high-energy spectra. The horizontal reference line shows where $\delta_\chi=300\keV$ lies within this envelope; its presence there does not establish compatibility with all constraints.

In the contact-interaction implementation used here, the direct-detection coupling enters through $g_B/M_{Z_B}$, with the remaining benchmark parameters fixed. Changing $M_{Z_B}$ therefore rescales the coupling curves without changing their intersections in $\delta_\chi$. This explains why the splitting-envelope figure can be illustrated at a single mediator mass. This property applies to the adopted recoil calculation and does not imply that relic-density or collider constraints are independent of $M_{Z_B}$.

\begin{figure}[htbp]
\centering
\includegraphics[width=0.98\columnwidth]{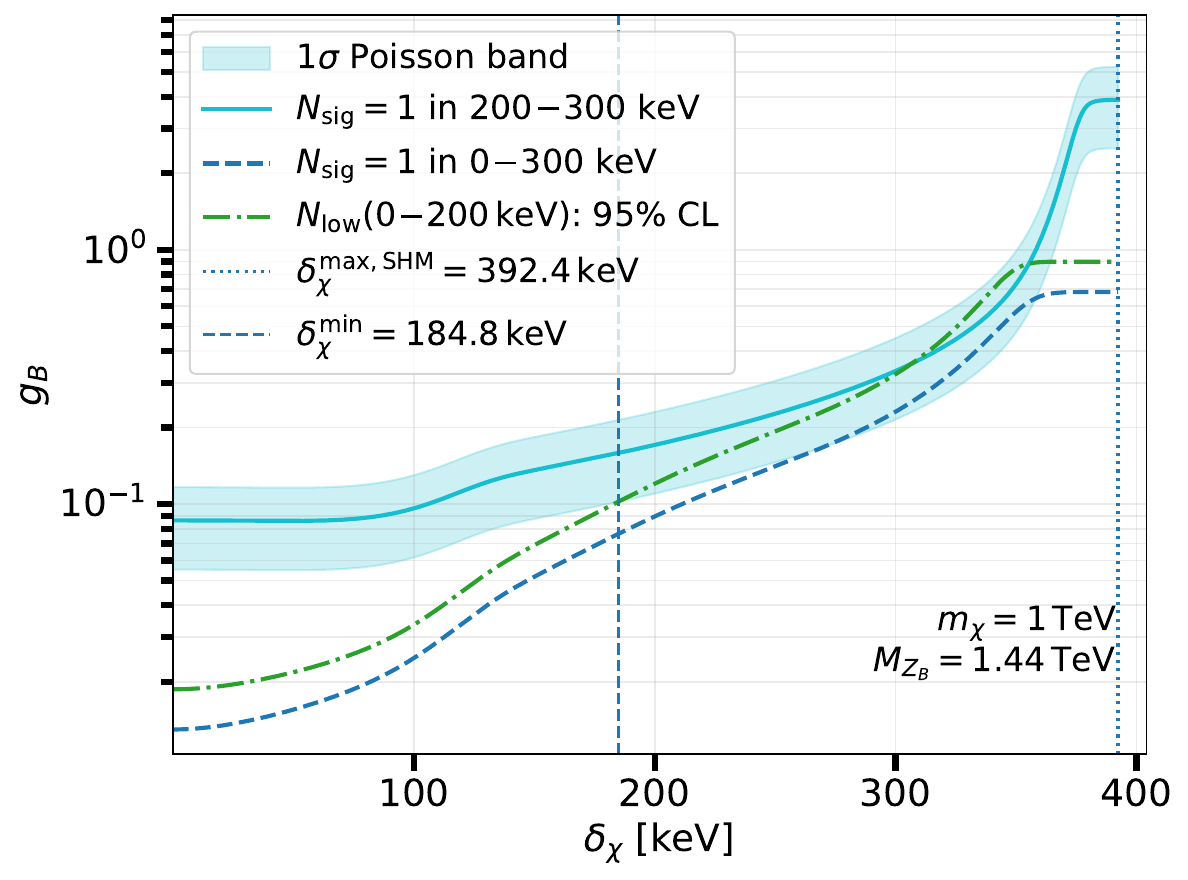}
\caption{Coupling as a function of $\delta_\chi$ for $m_\chi=1\TeV$ and $M_{Z_B}=1.44\TeV$, using the complete implemented transition and elastic recoil spectrum. The cyan band shows the central $68.27\%$ Poisson interval for one event in $200$--$300\keV$, with the solid cyan curve giving $N_{\rm sig}=1$. The blue dashed curve gives one event in $0$--$300\keV$ and is displayed only for comparison. The green dash-dotted curve is the one-sided $95\%$ CL upper limit from the zero-count $0$--$200\keV$ control region. Acceptance requires overlap of the cyan band with the region below the green curve. The vertical dashed line marks the first crossing at approximately $184.8\keV$, and the dotted line marks the endothermic SHM ceiling at approximately $392.4\keV$. Elastic scattering produces the large-splitting plateaus and a second crossing near $364.4\keV$, above which the signal band is incompatible with the low-energy limit.}
\label{fig:deltamchi}
\end{figure}

Fig.~\ref{fig:deltamchi} shows the coupling curves explicitly for $m_\chi=1\TeV$ and $M_{Z_B}=1.44\TeV$. The cyan band represents the signal-region Poisson interval, and its solid curve gives the normalization $N_{\rm sig}=1$. The green dash-dotted curve is the low-energy upper limit. At a given splitting, the accepted coupling range is the part of the cyan band below the green curve. Their first intersection gives $\delta_{\chi,\min}\simeq184.8\keV$, while the vertical dotted line marks the endothermic SHM ceiling at $\delta_{\chi,\max}^{\rm SHM}\simeq392.4\keV$.

The large-splitting flattening in Fig.~\ref{fig:deltamchi} has a different origin from the rapidly rising inelastic-only normalization. For the transition alone, the accepted rate tends to zero near the SHM ceiling, so maintaining a fixed event count requires a rapidly increasing coupling. In the implemented recoil approximation, diagonal elastic scattering supplies a nonzero count even when the transition disappears. The curves consequently approach finite couplings, dominated here by the diagonal axial contribution: approximately $g_B=3.89$ for one event in $200$--$300\keV$, $g_B=0.682$ for one event in $0$--$300\keV$, and $g_B=0.898$ for the low-energy upper limit.

These plateaus do not provide an additional viable explanation of the candidate. At the signal-region plateau, the coupling that yields one event in $200$--$300\keV$ predicts approximately $1.05\times10^3$ events below $200\keV$, far above the adopted limit. The lower edge of the cyan band crosses the green curve a second time at approximately $\delta_\chi=364.4\keV$. For this benchmark, the implemented direct-detection count conditions therefore overlap approximately over $184.8\keV\lesssim\delta_\chi\lesssim364.4\keV$, before imposing the other constraints, rather than over the entire interval extending to $392.4\keV$. The dashed full-window normalization is shown only for comparison and is not used to define acceptance.

\subsection{Combined limits and surviving fixed-mass slices}
\label{subsec:fixedmasses_combined}

We now combine the LZ, relic-density, and collider requirements in a way that
makes explicit the role of the profiled splitting interval discussed in
Sec.~\ref{subsec:deltaprofile}.  For each fixed DM mass $m_\chi$ and mediator
mass $M_{Z_B}$, the splitting is first allowed to vary over the interval
\begin{equation}
\delta_{\chi,\min}(m_\chi)
\le \delta_\chi \le
\delta_{\chi,\max}^{\rm SHM}(m_\chi),
\end{equation}
where the lower boundary is defined by the intersection of the $0$--$200\keV$
$95\%$ CL upper limit with the lower edge of the exact $68.27\%$ Poisson band
for the single $200$--$300\keV$ LZ event, while the upper boundary is the SHM
kinematic ceiling.

At each trial splitting we determine the central $68.27\%$ signal-count coupling band and retain only its overlap with the low-energy upper limit. The union of these accepted bands over $\mathcal D_{\rm LZ}$ in Eq.~\eqref{eq:deltarangefinal} defines the profiled LZ count-compatible region in the $(M_{Z_B},g_B)$ plane.  This LZ region is then intersected with the thermal relic
condition $\Omega_\chi h^2=0.12\pm10\%$ and with the observed dijet upper
limit.  The surviving subset therefore corresponds to points for which one can
simultaneously choose a splitting $\delta_\chi\ge\delta_{\chi,\min}$ and a
coupling $g_B$ such that the model (i) reproduces the high-energy LZ event,
(ii) does not overproduce low-energy recoils, (iii) yields the observed relic
density within the adopted tolerance, and (iv) remains below the present
ATLAS/CMS dijet bound.

\begin{figure*}[tbp]
\centering
\includegraphics[width=0.49\textwidth]{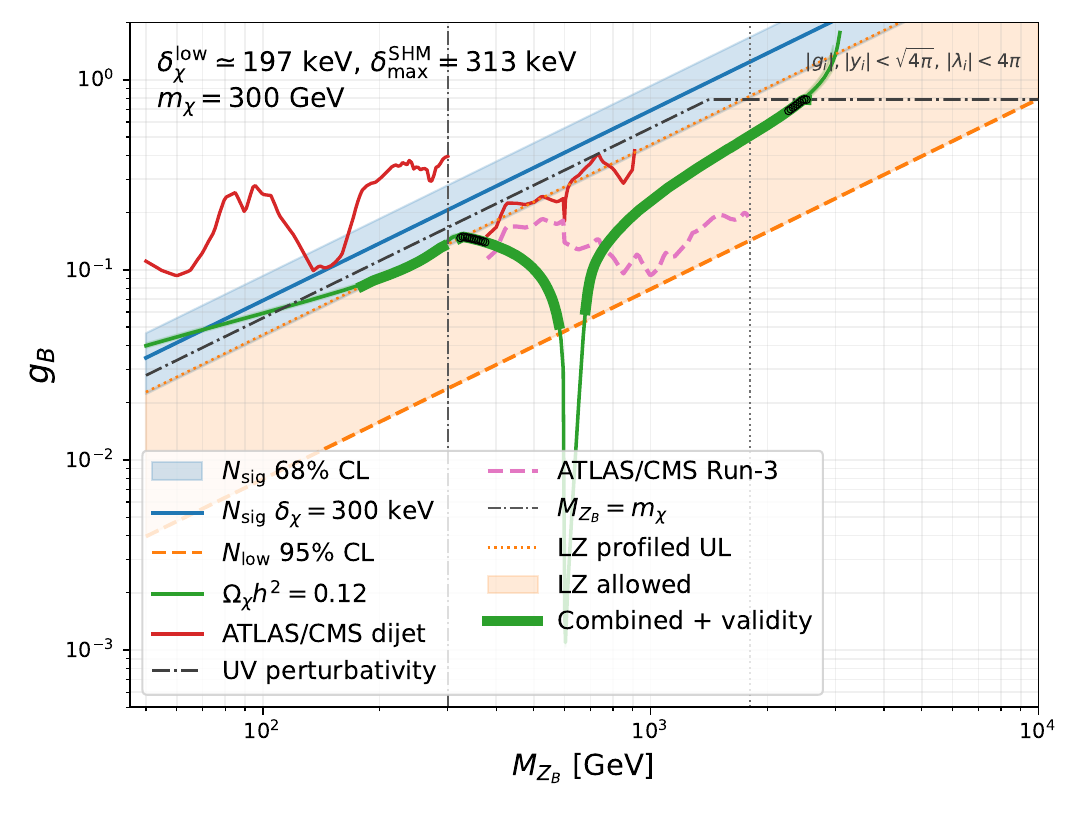}
\includegraphics[width=0.49\textwidth]{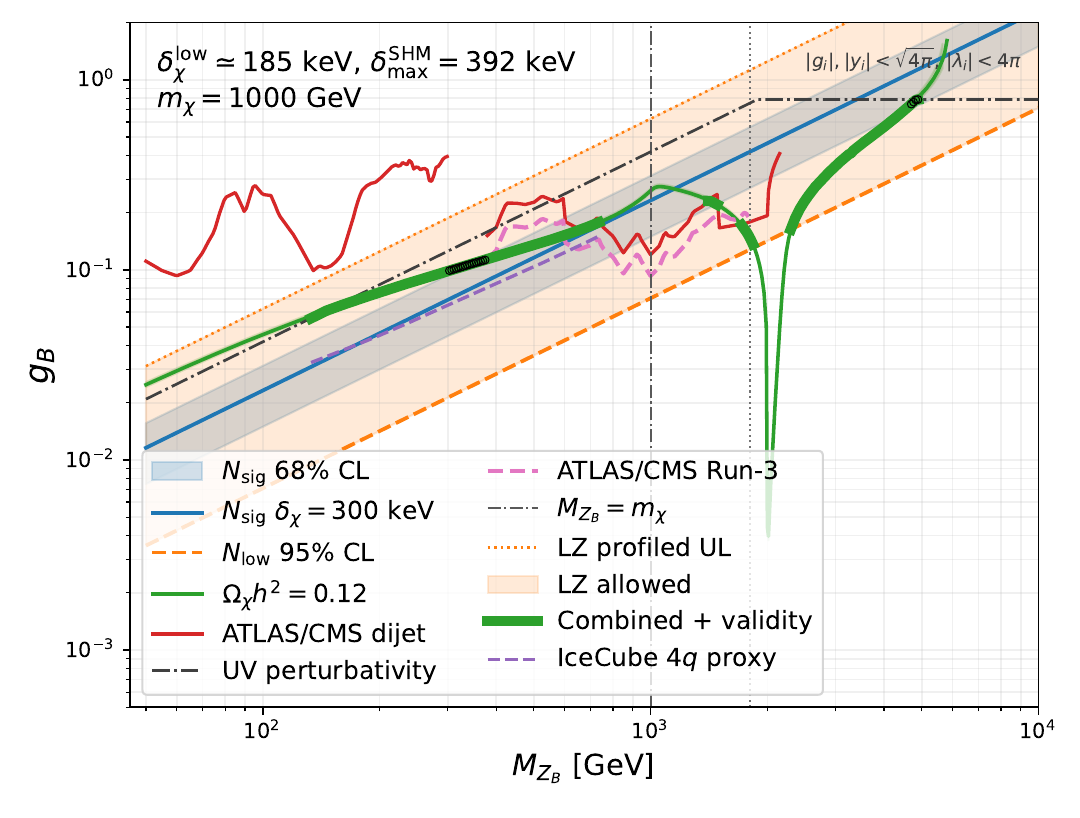}
\includegraphics[width=0.49\textwidth]{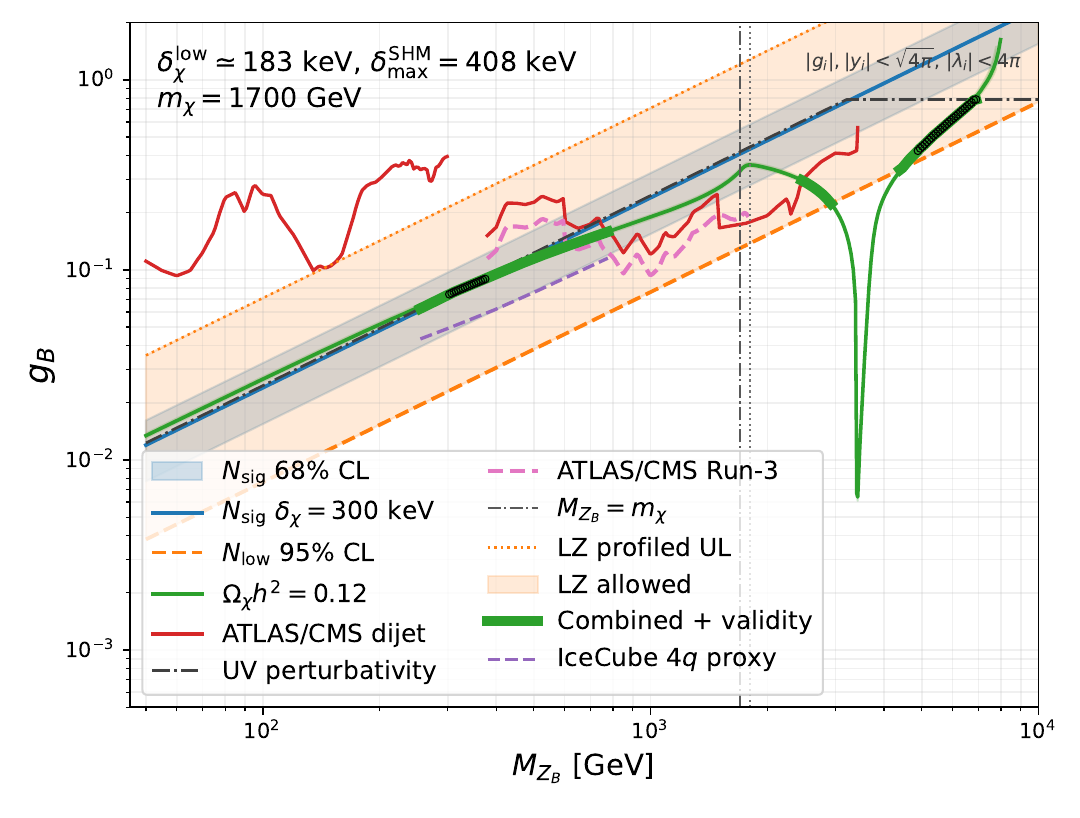}
\includegraphics[width=0.49\textwidth]{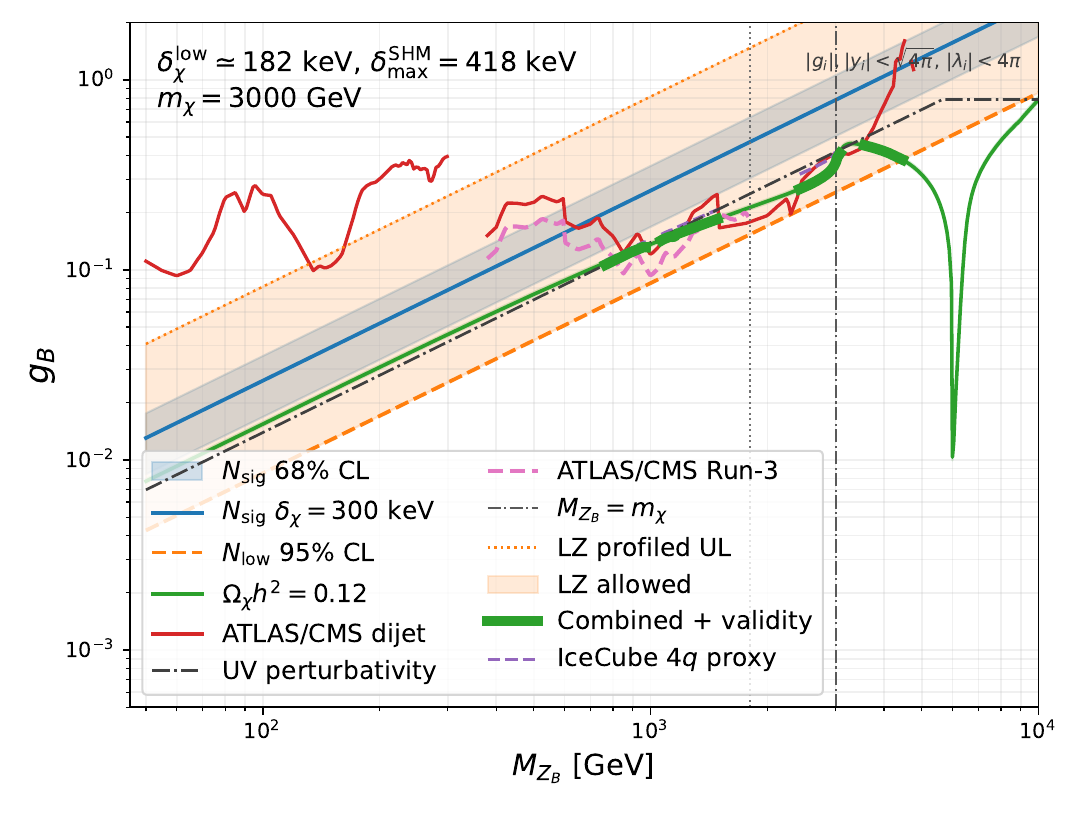}
\caption{Constraints in the $M_{Z_B}$--$g_B$ plane for $m_\chi=300$, $1000$, $1700$, and $3000\GeV$, from top left to bottom right. Solid blue gives the central one-event normalization at $\delta_\chi=300\keV$, with the adopted $68\%$ signal-count band. Orange shows the LZ envelope after profiling over the annotated splitting range and imposing the $200$--$300\keV$ signal and $0$--$200\keV$ control-count requirements; dashed and dotted orange curves mark its lower and upper edges. Thin green and its band give the thermal relic target $\Omega_\chi h^2=0.12$ and tolerance. Thick green identifies sampled LZ--relic--UV solutions passing the assessed dijet requirements where available; open black markers flag candidates lacking assessed dijet coverage. Red shows observed ATLAS/CMS dijet upper limits, pink dashed curves the luminosity-scaled Run-3 sensitivity, and purple dashed curves the diagnostic IceCube Solar-capture limits from the adopted four-quark proxy. IceCube does not enter the thick-green selection. Gray dash-dotted gives the UV perturbativity ceiling, without a mediator-width cut. Vertical guides mark $M_{Z_B}=m_\chi$ and the $1.8\TeV$ dijet search-range boundary.}
\label{fig:fixedmasses}
\end{figure*}

Fig.~\ref{fig:fixedmasses} shows the reference $\delta_\chi=300\keV$ signal-count band and its central blue curve. The orange region is the projected LZ coupling envelope after profiling the splitting and applying both count cuts. The thin green line and band show the central relic target and its $10\%$ tolerance. Thick green segments use an explicitly accepted witness within that tolerance; they need not follow the central relic curve. The red line is the observed dijet envelope, the pink dashed line is the Run-3 luminosity target, and the gray dash-dotted line is the UV perturbativity ceiling. Open black markers identify witnesses without assessed dijet coverage. The vertical line at $1.8\TeV$ is the inherited current search-range marker, not the dark threshold.

Several points follow immediately from the figure.  First, the existence of a
combined solution is not unique to the $m_\chi=1\TeV$ benchmark.  A surviving
branch is present also for lighter and heavier DM masses, demonstrating that
the overlap between LZ, relic density, and dijet bounds persists over a broad
mass range.  Second, the viable region is always much narrower than the full
orange LZ-allowed domain.  This shows that the relic-density and collider
requirements are genuinely restrictive and do not merely trim a very broad
LZ-compatible parameter space.

The $m_\chi=300\GeV$ panel lies relatively close to the kinematic turn-on of
the endothermic process.  As a consequence, the LZ-allowed region is more
sensitive to the high-velocity tail of the halo and the surviving branch is
comparatively compressed.  By contrast, the $m_\chi=1\TeV$ panel provides the
cleanest benchmark.  The LZ-compatible region is broad enough to accommodate a
thermal solution, while the collider curve remains close enough that the model
predicts a visible baryonic resonance near present sensitivity.  This is the
same qualitative feature already emphasized by the reference solution: the
point that explains LZ also lies close to the present dijet reach.  In this
sense the collider prediction is not an independent afterthought, but a direct
consequence of matching the LZ event and the relic abundance with the same
coupling.

The heavier slices, $m_\chi=1700\GeV$ and $3\TeV$, show that this geometric
interplay continues into the multi-TeV regime.  The surviving branch moves
toward larger couplings and larger mediator masses, but it does not disappear.
The threshold $M_{Z_B}=2m_\chi$ also affects these slices.  To the right of this threshold the invisible decay
$Z_B\to\chi_1\chi_2$ opens, modifying both the mediator width and the dijet
branching fraction.  The persistence of a viable branch on both sides of this
line shows that the combined solution is not an artifact of a single kinematic
configuration, although the most robust and phenomenologically clean region
remains the TeV-scale benchmark around $m_\chi\sim1\TeV$.

We also show the IceCube limits as a diagnostic of the sensitivity of Solar-capture searches. Within the assumptions of the adopted recast, IceCube can probe part of the parameter space that remains compatible with the LZ count selections, thermal relic abundance, observed dijet limits, and UV perturbativity criterion. A dedicated IceCube analysis of four-quark final states would allow a more direct assessment of this complementary sensitivity.

Overall, Fig.~\ref{fig:fixedmasses} shows that the combined constraints do not
select an isolated point, but rather a structured branch in parameter space.
The LZ profiling over $\delta_\chi$ enlarges the region relative to a fixed
$\delta_\chi=300\keV$ treatment, yet the simultaneous relic-density and collider
requirements still isolate a much more restricted subset.  This surviving
subset is the phenomenologically relevant target for the global interpretation
of the model.

\subsection{Global $m_\chi$--$M_{Z_B}$--$g_B$ scan}
\label{subsec:globalscan}

\begin{figure*}
\centering
\includegraphics[width=1.29\columnwidth]{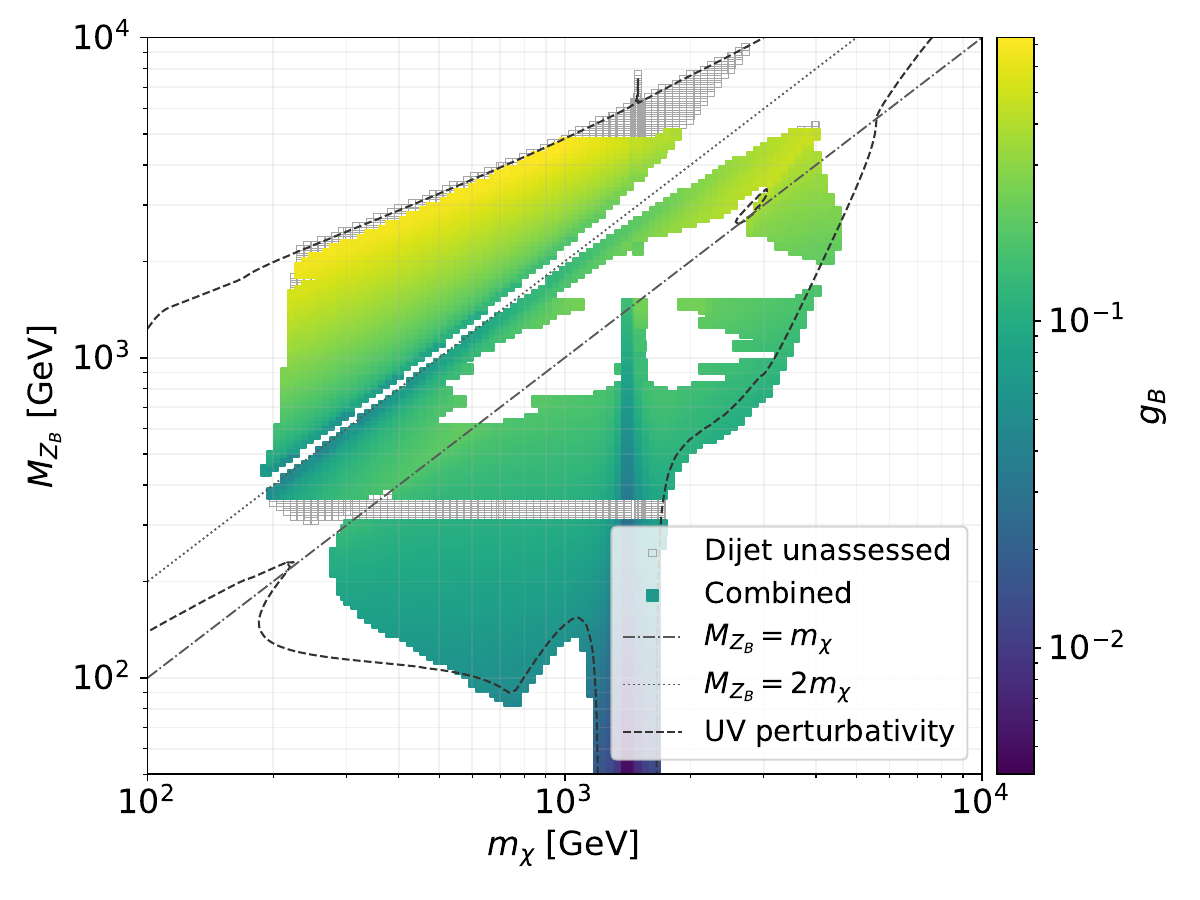}
\caption{Global projection of the LZ, relic-density, dijet, and UV constraints onto the $m_\chi$--$M_{Z_B}$ plane, profiling over the inelastic splitting. Filled points admit an explicitly tested coupling and splitting satisfying all primary requirements, with color indicating the representative $g_B$. Open gray points satisfy the other requirements but lack assessed dijet coverage. Dash-dotted and dotted lines mark $M_{Z_B}=m_\chi$ and $M_{Z_B}=2m_\chi$, respectively; dashed curves show the UV perturbativity boundary evaluated at the relic-density target. The diagnostic IceCube limits are not imposed.}
\label{fig:global}
\end{figure*}

Fig.~\ref{fig:global} projects the three-dimensional scan onto the $m_\chi$--$M_{Z_B}$ plane. Filled colored points have a tested coupling and splitting satisfying the primary count, relic, collider, and adopted numerical-domain requirements. Open gray points have an LZ--relic--domain intersection but lack assessed dijet coverage at that coupling. They are conditional candidates, not collider-allowed points. Color encodes the representative $g_B$. No Solar capture bounds are imposed.

The overlap forms a continuous pattern, rather than an isolated grid point. The reference intermediate-mediator solution has quark-dominated freeze-out and a narrow visible mediator. The open-vector region needs a fuller ground-state and Solar analysis, while the invisible region additionally motivates missing-energy searches. At larger masses, scalar thresholds and possible heavy-state effects must be examined.

The sampled global overlap spans $m_\chi\simeq0.19$--$4.5\TeV$, $M_{Z_B}\simeq0.05$--$5.0\TeV$, and $g_B\simeq0.004$--$0.74$. These rounded ranges are projections of a correlated region supported by tested coupling and splitting configurations, rather than independent bounds on three free parameters, and their extrema are limited by the scan domain. The selection profiles over the inelastic splitting and includes the adopted LZ count interval, relic abundance, UV perturbativity criterion, and observed dijet recast, requiring assessed dijet coverage.

Points outside the assessed collider coverage extend to approximately $M_{Z_B}=9.4\TeV$ and $g_B=0.79$ in the sampled LZ--relic--UV candidate set. They remain unassessed conditional candidates and are not included in the collider-compatible ranges above. In particular, the absence of a width veto does not turn the absence of an applicable search template into collider acceptance.

The parameter space compatible with the adopted LZ count interval, thermal relic abundance, UV perturbativity criterion, and assessed observed dijet limits can be tested through complementary DD, collider, and Solar-capture searches. Its sampled projections span $m_\chi\simeq0.19$--$4.5\TeV$, $M_{Z_B}\simeq0.05$--$5.0\TeV$, and $g_B\simeq0.004$--$0.74$, with substantial correlations among these parameters. For a fixed model configuration and unchanged recoil acceptance, the expected xenon signal grows linearly with exposure: a tenfold increase in effective exposure would turn an expectation of one event into ten events. In the background-free limit, the corresponding coupling sensitivity improves as $g_B^{\rm lim}\propto\mathcal E^{-1/4}$, so exposure increases by factors of ten and one hundred improve the coupling reach by factors of approximately $1.8$ and $3.2$, respectively. Additional LZ data and an independent XENONnT measurement could therefore test the predicted high-energy recoil spectrum and its suppressed low-energy counterpart, provided their analyses retain sensitivity to the relevant recoil energies. At the LHC, our luminosity-scaled Run-3 estimates indicate sensitivity to $g_B\simeq0.20$--$0.23$ for $M_{Z_B}\simeq1.4$--$1.5\TeV$, allowing ATLAS and CMS to probe the surviving near-TeV mediator branch across different DM masses and splittings. IceCube provides complementary sensitivity on the open-mediator branch, $m_\chi>M_{Z_B}$, where annihilation into two mediators produces four-quark final states: the diagnostic recast already reaches part of the otherwise allowed region, while a dedicated analysis is needed to establish a reliable exclusion reach. These measurements can test different portions of the correlated allowed region, although the present estimates do not establish complete coverage of its full extent.

\section{Conclusions}
\label{sec:conclusions}

Can the $248\keV$ nuclear-recoil candidate reported by LZ be explained by DM while remaining consistent with other observations? Answering this question requires more than fitting a recoil rate. In a UV-complete theory, the interactions producing the xenon signal also determine the relic abundance, collider signatures, excited-state lifetime, and Solar capture. These observables must be tested together using the same masses and couplings.

We have carried out a comprehensive investigation of this possibility in a scalar-minimal, anomaly-free $U(1)_B$ theory. Our analysis connects the gauge and fermion structure to the recoil spectrum, finite-width thermal freeze-out, collider constraints, and ID prospects. We also examine radiative kinetic mixing, excited-state depletion and regeneration, the captured Solar population, and the sensitivity to the high-speed halo tail. This combined treatment identifies both viable configurations and the assumptions that require further experimental or theoretical refinement.

The model provides a physically motivated origin of endothermic scattering. Baryonic symmetry breaking generates the mediator and new-fermion masses, while a small Majorana interaction splits the neutral Dirac fermion into two nearby states. Its vector current becomes off diagonal, producing the transition $\chi_1A\to\chi_2A$, and a residual discrete gauge parity stabilizes $\chi_1$. The required sub-MeV splitting is technically natural because setting its generating coupling to zero restores a fermion-number symmetry. Its small size is therefore radiatively stable, although explaining the coupling's numerical value requires an additional UV mechanism.

An extended, correlated region survives the adopted LZ count requirements, relic-density target, UV perturbativity criteria, and assessed observed dijet constraints. Within the scanned domain, its projections span $m_\chi\simeq0.19$--$4.5\TeV$, $M_{Z_B}\simeq0.05$--$5.0\TeV$, and $g_B\simeq0.004$--$0.74$. These are projections of compatible configurations, rather than independent allowed intervals, and depend on the adopted halo, detector response, and UV spectrum. The overlap reflects the uncertainty of a single-event interpretation: thermal solutions can satisfy the adopted count interval even where the central one-event normalization is excluded by dijet searches.

A distinctive result is that this UV-complete endothermic model predicts recoils at both high and low energies. The coherent vector transition produces the high-energy signal, while diagonal axial and scalar interactions generate elastic scattering below the endothermic threshold. The axial contribution is suppressed by its nonrelativistic velocity and momentum dependence; the scalar contribution includes the portal and direct Higgs couplings, together with their interference. These smaller components provide a correlated low-energy prediction rather than an independently adjustable signal. They also motivate examining whether the lower-energy event classified by LZ as an accidental coincidence could be compatible with this spectrum, although establishing such an interpretation requires a dedicated detector-level likelihood analysis. Additional LZ data and XENONnT measurements covering both recoil regimes can test this spectral connection.

The small electroweak admixture needed in the neutral sector also generates neutrino and radiative decays of $\chi_2$. The calculated radiative lifetime is approximately $3.1$ days at the reference thermal solution, removing primordial excited states before recombination. This suppresses exothermic DD and present-day halo coannihilation while leaving the dominant endothermic interaction essentially unchanged. Continuous production inside the Sun nevertheless requires a separate captured-population calculation.

The surviving region offers complementary targets for ATLAS, CMS, and IceCube. Our luminosity-scaled Run-3 estimates reach $g_B\simeq0.20$--$0.23$ near $M_{Z_B}\simeq1.4$--$1.5\TeV$, testing the corresponding mediator branch across different DM masses and splittings. Solar capture provides a complementary route, particularly when annihilation into two mediators is open and their decays produce four-quark final states. The diagnostic IceCube recast indicates sensitivity to configurations surviving the primary constraints. A dedicated four-quark analysis, incorporating elemental capture, captured-state evolution, thermalization, and neutrino transport, is needed to determine its quantitative reach. Together, these searches could test a substantial portion of the surviving region, although complete coverage is not established by the present projections.

The central outcome is a consistent and experimentally testable connection between the LZ high-energy candidate, a residual low-energy recoil spectrum, and the interactions governing cosmology and mediator production. Further xenon data, ATLAS and CMS Run-3 analyses, and a dedicated IceCube Solar search can jointly determine whether this connection supports a dark matter interpretation of the LZ event.

\begin{acknowledgments}
MDM acknowledges support from the research grant {\sl TAsP (Theoretical Astroparticle Physics)} funded by Istituto Nazionale di Fisica Nucleare (INFN).
\end{acknowledgments}

\appendix

\section{Anomaly cancellation and the baryonic charge assignment}
\label{app:anomalies}

This appendix derives the baryonic charges used in Sec.~\ref{subsec:gaugefield} and Tab.~\ref{tab:fields}. The derivation has two steps: anomaly cancellation fixes the difference $B_1-B_2$, while gauge invariance of the Majorana interaction fixes the individual charges. We compute all anomaly coefficients using left-handed Weyl fields, replacing each right-handed field by its charge conjugate.

\subsection{Cancellation of gauge anomalies}

In the SM, gauged baryon number produces two nonzero anomaly coefficients: $[SU(2)_L]^2U(1)_B$ and $[U(1)_Y]^2U(1)_B$. The first receives a contribution from the left-handed quark doublets,
\begin{equation}
\mathcal A_{22B}^{\rm SM}
=
3_{\rm gen}\times3_{\rm c}\times\frac12\times\frac13
=
\frac32.
\label{eq:A22BSM}
\end{equation}
Here the factor $1/2$ is the Dynkin index of an $SU(2)_L$ doublet. In the new fermion sector, only $\Psi_L$ and $\Psi_R$ contribute to this anomaly. Their baryonic charges are $B_1$ and $B_2$, respectively, giving
\begin{equation}
\mathcal A_{22B}^{\rm new}
=
\frac12\left[B_1-B_2\right].
\label{eq:A22Bnew}
\end{equation}
The minus sign follows from converting $\Psi_R$ into a left-handed charge-conjugate field. Cancellation requires
\begin{equation}
B_1-B_2=-3.
\label{eq:anomalycondition}
\end{equation}

The second nonzero SM coefficient is $\mathcal A_{YYB}^{\rm SM}=-3/2$. For the charge pattern in Tab.~\ref{tab:fields}, the doublet $\Psi$ has hypercharge $-1/2$, the charged singlet $\eta$ has hypercharge $-1$, and the neutral singlet $\chi$ has zero hypercharge. The baryonic charges of the singlets are assigned in the opposite chiral order to those of the doublet. Consequently,
\begin{equation}
\mathcal A_{YYB}^{\rm new}
=
2\left[\frac14\right]\left[B_1-B_2\right]
-\left[B_1-B_2\right]
=
-\frac12\left[B_1-B_2\right].
\end{equation}
The first term is the contribution of the two doublet components, and the second is that of the charged singlet. Eq.~\eqref{eq:anomalycondition} therefore cancels both electroweak mixed anomalies.

The remaining new-sector coefficients vanish without imposing another condition:
\begin{align}
\mathcal A_{YBB}^{\rm new}
&=
-\left[B_1^2-B_2^2\right]
+\left[B_1^2-B_2^2\right]
=0,
\nonumber\\
\mathcal A_{BBB}^{\rm new}
&=
2\left[B_1^3-B_2^3\right]
-\left[B_1^3-B_2^3\right]
-\left[B_1^3-B_2^3\right]
=0,
\nonumber\\
\mathcal A_{{\rm grav}B}^{\rm new}
&=
2\left[B_1-B_2\right]
-\left[B_1-B_2\right]
-\left[B_1-B_2\right]
=0.
\end{align}
In the last two expressions, the three terms come from $\Psi$, $\eta$, and $\chi$, respectively. The factor of two counts the two components of $\Psi$. The corresponding SM coefficients vanish, as does the SM $[SU(3)_c]^2U(1)_B$ anomaly; the new fermions are color singlets. Their SM-vectorlike representations also preserve cancellation of the pure SM gauge anomalies, and the two additional Weyl doublets preserve the absence of the global $SU(2)$ anomaly. The new sector is nevertheless chiral under $U(1)_B$, which allows it to cancel the nonzero SM electroweak mixed anomalies.

\subsection{Fixing the individual baryonic charges}

Anomaly cancellation fixes $B_1-B_2$, but does not determine $B_1$ and $B_2$ separately. The additional condition follows from the scalar and Majorana interactions specified in Sec.~\ref{subsec:pseudodirac}.

For $B\left[S_B\right]=-3$, the Dirac operator $S_B\overline\chi_R\chi_L$ is gauge invariant when $B\left[S_B\right]-B_1+B_2=0$. This reproduces Eq.~\eqref{eq:anomalycondition} and introduces no independent restriction. By contrast, the one-sided Majorana operator $S_B^\dagger\overline{\chi_L^c}\chi_L$ requires
\begin{equation}
-B\left[S_B\right]+2B_2=0.
\end{equation}
It therefore fixes
\begin{equation}
B_2=-\frac32,
\,\,
B_1=B_2-3=-\frac92,
\end{equation}
in agreement with Eq.~\eqref{eq:Bcharges}.

Thus, within the scalar-minimal realization adopted here, the charge assignment follows jointly from anomaly cancellation and the requirement that the same baryonic scalar generate the Dirac masses and the left-handed Majorana mass responsible for the inelastic splitting. These are the charges used to derive the baryonic currents and mass operators in the main text.

\section{Radiatively generated kinetic mixing}
\label{app:kinmix}

The kinetic-mixing assumption in Sec.~\ref{subsec:lagrangian} is specified here. Even if hypercharge--baryon kinetic mixing is set to zero at a particular ultraviolet scale, fields charged under both $U(1)_Y$ and $U(1)_B$ regenerate it radiatively.  At one loop,
\begin{equation}
\frac{d\epsilon_{YB}}{d\ln\mu}=
-\frac{2g_Yg_B}{16\pi^2}b_{YB}+\mathcal O(\epsilon_{YB}),
\label{eq:kinRGE}
\end{equation}
where
\begin{equation}
b_{YB}=\frac23\sum_{\rm Weyl}YB
+\frac13\sum_{\rm complex\ scalars}YB.
\label{eq:bYB}
\end{equation}
Above all anomalon thresholds, for the field content of Sec.~\ref{sec:model}, the scalar contribution vanishes and
\begin{equation}
\sum_{\rm Weyl}YB=2-2(B_1+B_2)=14,
\end{equation}
so that
\begin{equation}
b_{YB}=\frac{28}{3}.
\label{eq:bYBvalue}
\end{equation}
The charge trace differs from an anomaly trace: conjugating a right-handed field changes both $Y$ and $B$, leaving $YB$ unchanged. The SM quarks contribute $2$, while the doublet and charged singlet each contribute $-(B_1+B_2)=6$; the neutral singlet and scalars contribute zero. Anomaly cancellation therefore does not prevent kinetic mixing from running.

If all charged fields were retained down to the mediator scale, imposing $\epsilon_{YB}(10\TeV)=0$ gives approximately
\begin{equation}
\epsilon_{YB}(M_{Z_B})\simeq
-\frac{g_Yg_B}{8\pi^2}\frac{28}{3}
\ln\left(\frac{M_{Z_B}}{10\TeV}\right).
\label{eq:epssolution}
\end{equation}
At $M_{Z_B}=1.44\TeV$ and $g_B\simeq0.23$, this unthresholded estimate gives $|\epsilon_{YB}|\simeq1.9\times10^{-2}$. Below the charged-anomalon thresholds, the SM quarks instead give $b_{YB}=4/3$ when all quarks are active. Matching at $M_\eta=2\TeV$ and $M_\Psi=1.5\TeV$, and neglecting the small charged mixing, gives $|\epsilon_{YB}|\simeq1.7\times10^{-2}$ for the same $10\TeV$ boundary condition. After diagonalizing the gauge kinetic terms, the baryonic vector obtains an induced coupling of order $\epsilon_{YB}g_Y$ to the hypercharge current.  This produces small couplings to charged leptons, a small $Z$--$Z_B$ mixing after electroweak symmetry breaking, and additional decay channels of the excited state.

These estimates depend on the specified ultraviolet boundary condition. In the numerical analysis of the main text, we instead impose zero renormalized kinetic mixing at the low-energy reference scale. This is a parameter choice, not a symmetry protection, and the illustrative running estimates above are not inserted into the scan. A nonzero low-energy value would require a consistent revision of the recoil, relic-density, mediator-width, and collider calculations.

We set the kinetic mixing to zero at the mediator mass scale and neglect its loop-induced effects in the numerical analysis. This is a chosen reference condition, since fields carrying both hypercharge and baryon number can regenerate the mixing even in an anomaly-free theory.

\section{Present-day abundance of the excited state}
\label{app:excitedstate}

This appendix provides the population and decay calculations used in
Secs.~\ref{subsec:pseudodirac},~\ref{subsec:indirect}, and~\ref{subsec:solar}.
The endothermic interpretation requires a Galactic halo dominated by
$\chi_1$: a residual $\chi_2$ population would generate exothermic
scattering, $\chi_2A\to\chi_1A$, and restore the unsuppressed mixed-pair
annihilation channel $\chi_1\chi_2\to q\bar q$. We therefore examine
primordial depletion, late-time regeneration, and the distinct population
maintained by Solar capture.

We define
\begin{equation}
f_2(t)=\frac{n_2(t)}{n_1(t)+n_2(t)},
\,\,
f_1(t)=1-f_2(t),
\label{eq:f2def}
\end{equation}
where $n_i$ are the cosmological number densities. At thermal freeze-out,
$\delta_\chi\ll T_f$, so the two states have approximately equal
abundances. Their subsequent relative population is controlled by
state-changing scatterings and intrinsic decay. For the reference
splitting $\delta_\chi=300\keV<2m_e$, the decay
$\chi_2\to\chi_1e^+e^-$ is forbidden, making neutrino and radiative
transitions particularly important.
The representative decay mechanisms discussed in this appendix are illustrated in Fig.~\ref{fig:chi2_decay_diagrams}, including the electroweak neutrino decay, the one-loop radiative transition, and the kinetic-mixing-induced channels.

\subsection{Exothermic direct detection and the depletion requirement}

The minimum speeds for upward and downward scattering are
\begin{equation}
v_{\rm min}^{\rm up/down}(E_R)
=
\frac{1}{\sqrt{2m_AE_R}}
\left|
\frac{m_AE_R}{\mu_{\chi A}}\pm\delta_\chi
\right|.
\label{eq:vminupdownapp}
\end{equation}
For downward scattering, the minimum vanishes at
\begin{equation}
E_R^{\rm down}
=
\frac{\mu_{\chi A}}{m_A}\delta_\chi.
\label{eq:ERdown}
\end{equation}
For $m_\chi=1\TeV$, xenon, and $\delta_\chi=300\keV$, this gives
$E_R^{\rm down}\simeq270\keV$. Consequently, even slowly moving excited
particles can generate recoils near the LZ candidate.

Using the same SHM, xenon isotopic mixture, Helm form factor, and
detector response as the main analysis, let $\widehat R_{\rm up}$ and
$\widehat R_{\rm down}$ denote the accepted rates for the same reference
local density assigned entirely to the relevant initial state. In the
$200$--$300\keV$ signal window,
\begin{equation}
\frac{\widehat R_{\rm down}}{\widehat R_{\rm up}}
\simeq5.30\times10^2.
\label{eq:RdownRupunit}
\end{equation}
The corresponding ratios are $8.20\times10^2$ over $3$--$400\keV$ and
$9.08\times10^2$ in the $0$--$200\keV$ control window. These ratios
depend on the recoil interval; the small ordinary-$Z$ contribution at
the benchmark changes them only slightly.

For local fractions $f_1=1-f_2$ and $f_2$, the signal-window ratio is
\begin{equation}
\frac{R_{\rm down}}{R_{\rm up}}
=
\frac{f_2\widehat R_{\rm down}}
{(1-f_2)\widehat R_{\rm up}}
\simeq5.30\times10^2\frac{f_2}{1-f_2}.
\label{eq:Rratiof2}
\end{equation}
Requiring the exothermic contribution not to exceed the endothermic
one gives the diagnostic target
\begin{equation}
f_2(t_0)\lesssim
f_2^{\rm DD}
\equiv
\frac{1}{1+5.30\times10^2}
\simeq1.88\times10^{-3}.
\label{eq:f2DD}
\end{equation}
This is not an LZ confidence-level limit: a statistical bound requires
a joint fit of the upward and downward recoil spectra. It nevertheless
quantifies the depletion needed for a predominantly endothermic
interpretation.

Number-conserving conversions, such as
$\chi_2\chi_2\leftrightarrow\chi_1\chi_1$ and
$\chi_2f\leftrightarrow\chi_1f$, can reduce the excited fraction before
they decouple. If chemical equilibrium between the states persists,
$n_2/n_1\simeq\exp[-\delta_\chi/T_\chi]$, so reaching
Eq.~\eqref{eq:f2DD} through conversions alone requires
$T_\chi\lesssim47.8\keV$. Whether equilibrium lasts this long depends
on the collision rates and dark kinetic history; after kinetic
decoupling, $T_\chi$ need not equal the plasma temperature, and
quark-level rates cannot be extrapolated through the QCD transition
without hadronic matching. We therefore do not use the equilibrium
factor as proof of depletion. The decay calculation below instead
allows an initial fraction $f_2=1/2$ once inverse production is
negligible. Both conversion and decay preserve the total number of
dark particles, unlike annihilation.

\subsection{Electroweak-mixing-induced decay into neutrinos}
\label{app:EWchi2decay}

The singlet--doublet mixing introduced in
Sec.~\ref{subsec:anomalons} gives the light Dirac mass eigenstate $N_L$
an ordinary-$Z$ interaction,
\begin{equation}
\mathcal L_Z\supset
\frac{g_Z}{2}Z_\mu\overline{N_L}\gamma^\mu
\left[
\sin^2\theta_LP_L+\sin^2\theta_RP_R
\right]N_L,
\label{eq:lightZcurrentApp}
\end{equation}
where $g_Z=g/c_W$. Its vector and axial coefficients are
\begin{equation}
\begin{aligned}
g_V^Z&=
\frac{g_Z}{4}
\left[\sin^2\theta_L+\sin^2\theta_R\right],
\\
g_A^Z&=
\frac{g_Z}{4}
\left[\sin^2\theta_R-\sin^2\theta_L\right].
\end{aligned}
\label{eq:gVAgeneral}
\end{equation}

With the positive-mass Majorana convention of
Sec.~\ref{subsec:pseudodirac},
\begin{equation}
N_L=\frac{\chi_2-i\chi_1}{\sqrt2},
\,\,
N_L^c=\frac{\chi_2+i\chi_1}{\sqrt2}.
\label{eq:NLmajoranadecomposition}
\end{equation}
The vector and axial currents consequently become
\begin{align}
\overline{N_L}\gamma^\mu N_L
&=
i\overline{\chi_1}\gamma^\mu\chi_2,
\label{eq:vectorcurrentmajorana}
\\
\overline{N_L}\gamma^\mu\gamma^5N_L
&=
\frac12
\left[
\overline{\chi_1}\gamma^\mu\gamma^5\chi_1+
\overline{\chi_2}\gamma^\mu\gamma^5\chi_2
\right].
\label{eq:axialcurrentmajorana}
\end{align}
Thus the vector current changes the state, whereas the axial current
is diagonal, up to corrections of order $\delta_\chi/m_\chi$. This is
the same distinction underlying the inelastic vector and elastic
axial contributions to DD.

The transition coupling relevant for the neutrino decay is
\begin{equation}
g_{Z12}
=
\frac{g_Z}{4}
\left[\sin^2\theta_L+\sin^2\theta_R\right].
\label{eq:gZ12general}
\end{equation}
For $\theta_L=\theta_R=\theta_N$, the axial ordinary-$Z$ coefficient
vanishes at leading order and
$g_{Z12}=(g_Z/2)\sin^2\theta_N$. Defining
$J_{12}^\mu=i\overline{\chi_1}\gamma^\mu\chi_2$, the relevant interactions
are
\begin{equation}
\mathcal L\supset
g_{Z12}Z_\mu J_{12}^\mu+
\frac{g_Z}{2}Z_\mu
\sum_{\alpha=e,\mu,\tau}
\overline{\nu_\alpha}\gamma^\mu P_L\nu_\alpha.
\label{eq:ZdecayLag}
\end{equation}
Since the decay momentum is much smaller than $m_Z$, integrating out
the $Z$ gives
\begin{equation}
\mathcal L_{\rm eff}^{\nu,{\rm EW}}
=
-G_\nu^{\rm EW}J_{12}^\mu
\sum_{\alpha=e,\mu,\tau}
\overline{\nu_\alpha}\gamma_\mu P_L\nu_\alpha,
\label{eq:LeffEWnu}
\end{equation}
with positive coefficient
\begin{equation}
G_\nu^{\rm EW}
=
\frac{g_{Z12}g_Z}{2m_Z^2}
=
\frac{g_Z^2}{8m_Z^2}
\left[\sin^2\theta_L+\sin^2\theta_R\right].
\label{eq:GEWgeneral}
\end{equation}
For equal angles,
\begin{equation}
G_\nu^{\rm EW}
=
\frac{g_Z^2}{4m_Z^2}\sin^2\theta_N.
\label{eq:GEWequal}
\end{equation}

For a constant transition coefficient $G_\nu$, the exact three-body
phase-space expression, including three neutrino flavors, is
\begin{align}
\Gamma_{\nu\bar\nu}
={}&
\frac{3}{384\pi^3m_{\chi_2}^3}
\int_0^{\delta_\chi^2}\!ds\,
|G_\nu|^2
[\delta_\chi^2-s]^{3/2}
\nonumber\\
&\times
\left[(m_{\chi_1}+m_{\chi_2})^2-s\right]^{1/2}
\left[(m_{\chi_1}+m_{\chi_2})^2+2s\right].
\label{eq:exactNuIntegral}
\end{align}
For $\delta_\chi\ll m_\chi$, this reduces to
\begin{equation}
\Gamma_{\chi_2\to\chi_1\nu\bar\nu}^{\rm EW}
=
\frac{[G_\nu^{\rm EW}]^2\delta_\chi^5}{40\pi^3}
\left[
1+\mathcal O\left[\frac{\delta_\chi}{m_\chi}\right]
\right].
\label{eq:EWnuwidth}
\end{equation}
The width therefore scales as
$[\sin^2\theta_L+\sin^2\theta_R]^2\delta_\chi^5$,
or $\theta_N^4\delta_\chi^5$ for small equal angles.

At $\theta_N=0.005$ and $\delta_\chi=300\keV$,
$g_{Z12}\simeq9.29\times10^{-6}$ and
$G_\nu^{\rm EW}\simeq4.13\times10^{-10}\GeV^{-2}$, giving
\begin{equation}
\begin{aligned}
\Gamma_{\chi_2\to\chi_1\nu\bar\nu}^{\rm EW}
&\simeq3.34\times10^{-40}\GeV,
\\
\tau_{\nu\bar\nu}^{\rm EW}
&\simeq6.24\times10^7\ {\rm yr}.
\end{aligned}
\label{eq:EWbenchmarkLifetime}
\end{equation}
For small equal angles, the lifetime scales as
\begin{equation}
\tau_{\nu\bar\nu}^{\rm EW}
\simeq6.24\times10^7\ {\rm yr}
\left[\frac{0.005}{\theta_N}\right]^4
\left[\frac{300\keV}{\delta_\chi}\right]^5.
\label{eq:EWlifetimeScaling}
\end{equation}

After conversion freeze-out and once inverse decays are negligible,
the decay-only primordial fraction follows
\begin{equation}
f_2^{\rm prim}(t)
=
f_2(t_i)
\exp\left[-\frac{t-t_i}{\tau_2}\right].
\label{eq:f2decay}
\end{equation}
Taking $f_2(t_i)=1/2$ and $t_0\simeq1.38\times10^{10}$ yr, the
neutrino channel alone gives
\begin{equation}
f_2^{\rm prim}(t_0)
\simeq
\frac12
\exp\left[
-\frac{1.38\times10^{10}}{6.24\times10^7}
\right]
\sim4\times10^{-97}.
\label{eq:f2todayEW}
\end{equation}
The corresponding primordial exothermic-to-endothermic rate ratio is
of order $2\times10^{-94}$.

More generally, Eq.~\eqref{eq:f2DD} requires
$\tau_2<t_0/\ln[(1/2)/f_2^{\rm DD}]
\simeq2.47\times10^9$ yr. If depletion proceeds only through the
electroweak neutrino channel, this corresponds to
\begin{equation}
\theta_N\gtrsim1.99\times10^{-3}
\label{eq:thetaMinDD}
\end{equation}
at $\delta_\chi=300\keV$. The reference mixing therefore suffices to
remove the primordial excited population by today, even before
including the faster radiative channel.

\subsection{Neutrino decay from kinetic mixing}
\label{app:mixing}

The electroweak transition above must be distinguished from pure
hypercharge--baryon kinetic mixing. In the exact-singlet limit,
$\theta_L=\theta_R=0$, there is no direct ordinary-$Z$ coupling of
the dark fermion. Kinetic mixing can still connect its baryonic current
to neutrinos, but the complete neutral-gauge propagator makes this
interaction momentum suppressed.

The kinetic terms are
\begin{equation}
\mathcal L_{\rm kin}\supset
-\frac14B_{\mu\nu}B^{\mu\nu}
-\frac14Z_{B\mu\nu}Z_B^{\mu\nu}
-\frac{\epsilon_{YB}}{2}B_{\mu\nu}Z_B^{\mu\nu}.
\label{eq:kineticMixingLagApp}
\end{equation}
After electroweak symmetry breaking,
$B_\mu=c_WA_{0\mu}-s_WZ_{0\mu}$, so the operator mixes $Z_B$ with
both neutral SM gauge fields. To first order in $\epsilon_{YB}$,
the transverse inverse propagator in the basis $(A_0,Z_0,Z_B)$ is
\begin{equation}
K(s)=
\begin{pmatrix}
s&0&\epsilon_{YB}c_Ws\\
0&s-m_Z^2&-\epsilon_{YB}s_Ws\\
\epsilon_{YB}c_Ws&-\epsilon_{YB}s_Ws&s-M_{Z_B}^2
\end{pmatrix},
\,\,
s=q^2.
\label{eq:neutralPropagatorK}
\end{equation}
The off-diagonal entries contain $q^2$, reflecting the derivative
structure of kinetic mixing. Because neutrinos have no electric
charge, their transition amplitude uses the mixed $Z_0$--$Z_B$
propagator,
\begin{equation}
[K^{-1}(s)]_{Z_0Z_B}
\simeq
\frac{\epsilon_{YB}s_Ws}
{(s-m_Z^2)(s-M_{Z_B}^2)}.
\label{eq:mixedPropagator}
\end{equation}
Multiplication by the baryonic transition coupling $3g_B$ and the
neutrino coupling $g_Z/2$ gives
\begin{equation}
G_\nu^{\rm kin}(s)
=
\frac{3g_B\epsilon_{YB}g_Y}{2}
\frac{s}{(s-M_{Z_B}^2)(s-m_Z^2)}.
\label{eq:GkineticExact}
\end{equation}
At low momentum,
\begin{equation}
G_\nu^{\rm kin}(s)\simeq C_\nu s,
\,\,
C_\nu=
\frac{3g_B\epsilon_{YB}g_Y}
{2M_{Z_B}^2m_Z^2}.
\label{eq:GkineticLowE}
\end{equation}
There is therefore no constant neutrino contact operator from pure
kinetic mixing. In a basis with canonical kinetic terms, the same
result follows from cancellation between the correlated $Z$ and
$Z_B$ exchanges. Keeping only an induced $Z_B\nu\bar\nu$ vertex would
miss this cancellation. It does not apply to the genuine
doublet-induced coupling in Eq.~\eqref{eq:GEWgeneral}; an additional
independent mass-mixing interaction would also require a separate
analysis.

Replacing $G_\nu$ by $C_\nu s$ in the phase-space integral gives
\begin{equation}
\Gamma_{\chi_2\to\chi_1\nu\bar\nu}^{\rm kin}
=
\frac{C_\nu^2\delta_\chi^9}{315\pi^3},
\label{eq:kineticNuWidthCorrect}
\end{equation}
in the heavy-DM limit. Relative to a constant transition operator,
the extra momentum factor in the amplitude supplies four additional
powers of $\delta_\chi$ in the width:
\begin{equation}
\Gamma_{\nu\bar\nu}^{\rm kin}
\propto
\epsilon_{YB}^2g_B^2
\frac{\delta_\chi^9}{M_{Z_B}^4m_Z^4}.
\label{eq:kineticScaling}
\end{equation}
For $g_B=0.22782$, $M_{Z_B}=1.44\TeV$,
$|\epsilon_{YB}|=9.5\times10^{-3}$ as an independent illustrative input, and
$\delta_\chi=300\keV$, one obtains
$|C_\nu|\simeq6.7\times10^{-14}\GeV^{-4}$ and
\begin{equation}
\begin{aligned}
\Gamma_{\chi_2\to\chi_1\nu\bar\nu}^{\rm kin}
&\simeq9.1\times10^{-63}\GeV,
\\
\tau_{\nu\bar\nu}^{\rm kin}
&\simeq2.3\times10^{30}\ {\rm yr}.
\end{aligned}
\label{eq:kineticNuLifetime}
\end{equation}
This illustrative loop-size mixing cannot deplete the excited state.
A related kinetic-mixing decay analysis is given in
Ref.~\cite{Gemmell:2026yaw}.

Below the electron-pair threshold, kinetic mixing also permits
$\chi_2\to\chi_1+3\gamma$ through an electron box. Charge conjugation
forbids the corresponding two-photon channel through an electron
triangle. In the limit $\delta_\chi\ll m_e$, the Euler--Heisenberg
interaction gives~\cite{CarrilloGonzalezToro2022}
\begin{equation}
\Gamma_{3\gamma}^{\rm kin}
\propto
\frac{g_B^2\epsilon_{YB}^2\alpha^4}
{M_{Z_B}^4m_e^8}
\delta_\chi^{13}.
\label{eq:threeGammaScaling}
\end{equation}
At $300\keV$, however, $\delta_\chi/m_e\simeq0.59$, so this asymptotic
expression is not a precision lifetime calculation. We do not use
it to establish depletion; the electroweak radiative transition
calculated below already provides a much faster mechanism.

\subsection{One-loop radiative decay $\chi_2\to\chi_1\gamma$}
\label{app:radiativedecay}

The same singlet--doublet mixing generates an electromagnetic
transition through the charged anomalons and the $W$ boson. This
interaction follows from the renormalizable model rather than from
an independently chosen dipole operator.

In the physical charged-fermion basis,
\begin{equation}
\mathcal L_W=
\frac{g}{\sqrt2}W^-_\mu
\sum_{a=1,2}
\overline{X_a^-}\gamma^\mu
\left[c_{aL}P_L+c_{aR}P_R\right]N_L
+{\rm h.c.},
\label{eq:WchargedCurrentDipole}
\end{equation}
where
\begin{equation}
c_{aL}
=
-\sin\theta_L\,
\overline{(U_-^L)_{1a}},
\,\,
c_{aR}
=
-\sin\theta_R\,
\overline{(U_-^R)_{1a}}.
\label{eq:chargedCurrentCoeffs}
\end{equation}
Here the overbar denotes complex conjugation, and $U_-^{L,R}$ are
the rotations defined in Eq.~\eqref{eq:chargeddiag}.

The complete one-loop amplitude includes photon attachments to the
charged fermion and $W$, together with charged-Goldstone and mixed
$W$--Goldstone diagrams~\cite{Lavoura2003}. For an on-shell photon,
write the Dirac electromagnetic vertex as
\begin{equation}
\Gamma^\mu_\gamma
=
i\sigma^{\mu\nu}q_\nu
\left[\mu_D+i d_D\gamma^5\right].
\label{eq:DiracDipoleVertex}
\end{equation}
Resolving $N_L$ into the Majorana pair converts this dipole into an
off-diagonal transition. We define
$\mu_{12}=\sqrt{|\mu_M|^2+|\mu_E|^2}$ through the physical photon
vertex, avoiding an ambiguity from Majorana bilinear normalization.
For the real, CP-conserving benchmark, one transition structure
suffices and
\begin{equation}
\mu_{12}=|\mu_D|.
\label{eq:mu12muD}
\end{equation}

For the adopted real symmetric charged and neutral sector, define
$w_j=(g/\sqrt2)c_{jL}=(g/\sqrt2)c_{jR}$. The gauge-complete result,
to leading order in $\delta_\chi/m_\chi$, reduces to
\begin{align}
|\mu_{12}|={}&
\frac{e}{16\pi^2}
\left|
\sum_{j=1}^2w_j^2
\int_0^1\!dt\,
\frac{P_j(t)}{D_j(t)}
\right|,
\label{eq:matchedDipole}
\\
D_j(t)={}&
(1-t)m_W^2+t m_{X_j}^2-t(1-t)m_\chi^2,
\nonumber\\
P_j(t)={}&
2(1-t)
\left[2(m_{X_j}-m_\chi)+m_\chi t\right]
\nonumber\\
&+
\frac{t(m_{X_j}-m_\chi)^2}{m_W^2}
\left[m_{X_j}+m_\chi-m_\chi t\right].
\nonumber
\end{align}
The charged-Goldstone couplings are fixed by the Ward relation and
the same mass matrices, with magnitude proportional to
$w_j(m_\chi-m_{X_j})/m_W$ in this symmetric limit. They introduce no
additional parameter. The mass-gap terms in $P_j$ retain this
longitudinal-$W$ physics, and the external mass must be kept in
$D_j$ for the TeV spectrum. Since $m_{X_j}>m_\chi$ in the adopted
spectrum, the denominator is positive throughout the integration
interval. Neglecting the small splitting inside the loop produces
only corrections of order $\delta_\chi/m_\chi$.

In the small charged-mixing limit, the $\Psi^-$-like eigenstate
dominates and $m_{X_1}\simeq M_\Psi$. For equal neutral angles, the
result can equivalently be written as
\begin{equation}
\mu_{12}^{W}
=
\frac{eg^2}{32\pi^2}
\frac{m_\chi}{m_W^2}
\sin^2\theta_N\,
\mathcal F_W
\left[
\frac{m_\chi^2}{m_W^2},
\frac{M_\Psi^2}{m_W^2}
\right].
\label{eq:mu12LoopFunction}
\end{equation}
At $m_\chi=1\TeV$, $M_\Psi=1.5\TeV$, and $\theta_N=0.005$,
\begin{equation}
\mathcal F_W\simeq0.337,
\,\,
\mu_{12}\simeq5.34\times10^{-10}\GeV^{-1},
\label{eq:mu12Benchmark}
\end{equation}
or approximately $1.8\times10^{-12}\mu_B$.

The full charged-sector completion with
$y_1=y_3=y_2=y_4=0.01437$ gives
$m_{X_1}\simeq1499.9875\GeV$,
$m_{X_2}\simeq2000.0125\GeV$, and
$\mu_{12}=5.34190\times10^{-10}\GeV^{-1}$.
Thus the second charged state changes the leading result negligibly
for this spectrum; this need not hold for larger charged mixing.
The supplied package checks the matching against unreduced
two-dimensional loop integrals, adaptive quadrature, the Goldstone
Ward relation, the SM neutrino-moment limit, and an independent
spinor evaluation of the decay width.

The exact two-body width is
\begin{equation}
\Gamma_{\chi_2\to\chi_1\gamma}
=
\frac{\mu_{12}^2
[m_{\chi_2}^2-m_{\chi_1}^2]^3}
{8\pi m_{\chi_2}^3},
\label{eq:transitionDipoleWidth}
\end{equation}
which reduces to
\begin{equation}
\Gamma_{\chi_2\to\chi_1\gamma}
\simeq
\frac{\mu_{12}^2\delta_\chi^3}{\pi}
\label{eq:transitionDipoleWidthApprox}
\end{equation}
in the pseudo-Dirac limit. At the reference splitting,
\begin{equation}
\begin{aligned}
\Gamma_{\chi_2\to\chi_1\gamma}
&\simeq2.45\times10^{-30}\GeV,
\\
\tau_\gamma
&\simeq2.68\times10^5\ {\rm s}
\simeq3.11\ {\rm days}.
\end{aligned}
\label{eq:radiativeLifetimeBenchmark}
\end{equation}
At fixed reference masses and for small equal angles,
\begin{equation}
\tau_\gamma
\simeq3.11\ {\rm days}
\left[\frac{0.005}{\theta_N}\right]^4
\left[\frac{300\keV}{\delta_\chi}\right]^3,
\label{eq:radiativeLifetimeScaling}
\end{equation}
neglecting small changes of the charged spectrum and loop function.
The radiative width is approximately $7\times10^9$ times the
electroweak neutrino width, so it determines the physical benchmark
lifetime and removes the primordial excited population well before
recombination. The representative decay mechanisms are shown in
Fig.~\ref{fig:chi2_decay_diagrams}.

\tikzset{
  mixing insertion/.style={
    circle,
    draw,
    minimum size=7pt,
    inner sep=0pt,
    path picture={
      \draw[line width=0.8pt]
        (path picture bounding box.south west) --
        (path picture bounding box.north east);
      \draw[line width=0.8pt]
        (path picture bounding box.north west) --
        (path picture bounding box.south east);
    }
  }
}

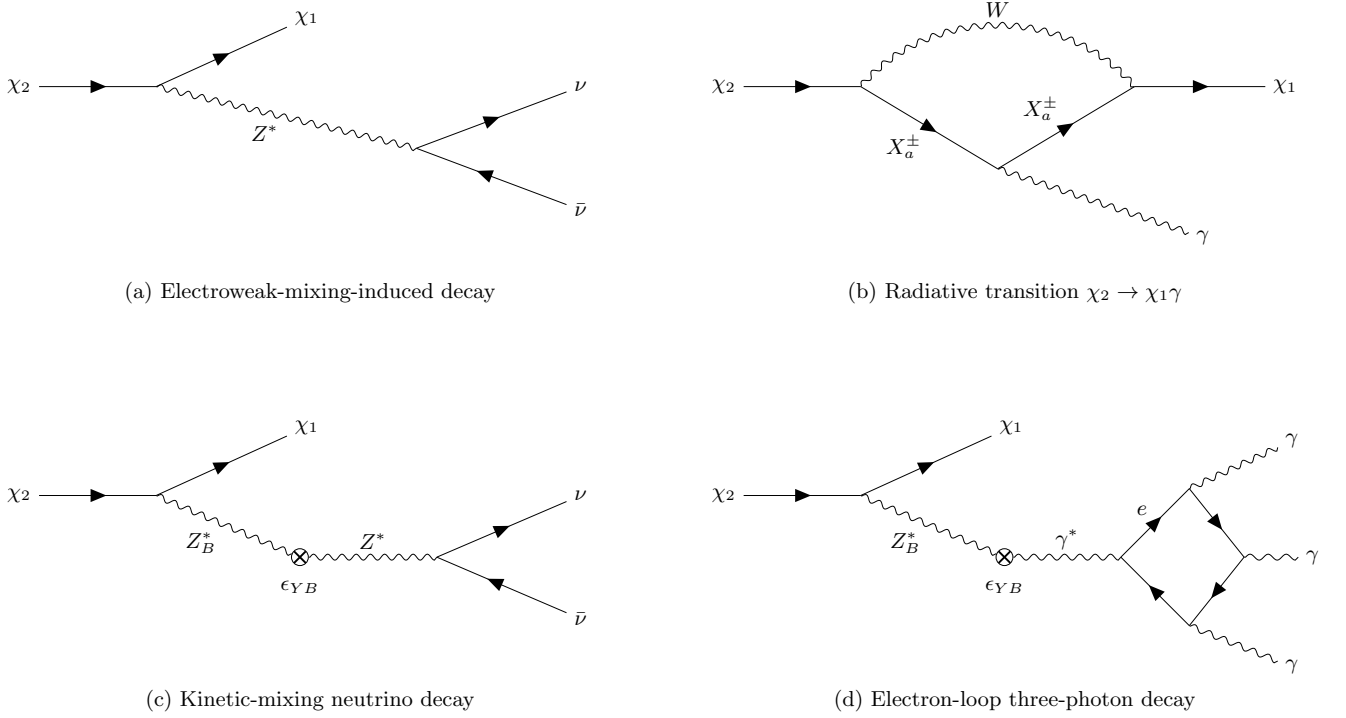
\begin{figure*}[t]
\centering

\begin{minipage}{0.48\textwidth}
\centering
\resizebox{\linewidth}{!}{%
\begin{tikzpicture}
\useasboundingbox (-0.5,-3.4) rectangle (9.0,2.2);
\begin{feynman}
\vertex (i) at (0,0) {$\chi_2$};
\vertex (v1) at (2.0,0);
\vertex (f) at (4.2,1.0) {$\chi_1$};
\vertex (v2) at (5.8,-0.9);
\vertex (n1) at (8.2,0.0) {$\nu$};
\vertex (n2) at (8.2,-1.8) {$\bar\nu$};

\diagram*{
  (i) -- [fermion] (v1),
  (v1) -- [fermion] (f),
  (v1) -- [boson, edge label'={$Z^\ast$}] (v2),
  (v2) -- [fermion] (n1),
  (v2) -- [anti fermion] (n2),
};
\end{feynman}
\node at (4.25,-3.0)
  {\small (a) Electroweak-mixing-induced decay};
\end{tikzpicture}%
}
\end{minipage}\hfill%
\begin{minipage}{0.48\textwidth}
\centering
\resizebox{\linewidth}{!}{%
\begin{tikzpicture}
\useasboundingbox (-0.5,-3.4) rectangle (9.0,2.2);
\begin{feynman}
\vertex (i) at (0,0) {$\chi_2$};
\vertex (v1) at (2.0,0);
\vertex (v3) at (4.0,-1.2);
\vertex (v4) at (6.0,0);
\vertex (f) at (8.2,0) {$\chi_1$};
\vertex (g) at (7.0,-2.2) {$\gamma$};

\diagram*{
  (i) -- [fermion] (v1),
  (v1) -- [boson, bend left=50, edge label={$W$}] (v4),
  (v1) -- [fermion, edge label'={$X_a^\pm$}] (v3),
  (v3) -- [fermion, edge label={$X_a^\pm$}] (v4),
  (v4) -- [fermion] (f),
  (v3) -- [photon] (g),
};
\end{feynman}
\node at (4.25,-3.0)
  {\small (b) Radiative transition $\chi_2\to\chi_1\gamma$};
\end{tikzpicture}%
}
\end{minipage}

\vspace{0.3cm}

\begin{minipage}{0.48\textwidth}
\centering
\resizebox{\linewidth}{!}{%
\begin{tikzpicture}
\useasboundingbox (-0.5,-3.4) rectangle (9.0,2.2);
\begin{feynman}
\vertex (i) at (0,0) {$\chi_2$};
\vertex (v1) at (2.0,0);
\vertex (f) at (4.2,1.0) {$\chi_1$};
\vertex (mix) [mixing insertion] at (4.1,-0.9) {};
\vertex (v2) at (6.1,-0.9);
\vertex (n1) at (8.2,0.0) {$\nu$};
\vertex (n2) at (8.2,-1.8) {$\bar\nu$};

\diagram*{
  (i) -- [fermion] (v1),
  (v1) -- [fermion] (f),
  (v1) -- [boson, edge label'={$Z_B^\ast$}] (mix),
  (mix) -- [boson, edge label={$Z^\ast$}] (v2),
  (v2) -- [fermion] (n1),
  (v2) -- [anti fermion] (n2),
};
\end{feynman}
\node[anchor=north, inner sep=1pt]
  at ([yshift=-5pt]mix.south)
  {\small $\epsilon_{YB}$};
\node at (4.25,-3.0)
  {\small (c) Kinetic-mixing neutrino decay};
\end{tikzpicture}%
}
\end{minipage}\hfill%
\begin{minipage}{0.48\textwidth}
\centering
\resizebox{\linewidth}{!}{%
\begin{tikzpicture}
\useasboundingbox (-0.5,-3.4) rectangle (9.0,2.2);
\begin{feynman}
\vertex (i) at (0,0) {$\chi_2$};
\vertex (v1) at (2.0,0);
\vertex (f) at (4.2,1.0) {$\chi_1$};
\vertex (mix) [mixing insertion] at (4.1,-0.9) {};
\vertex (e1) at (5.8,-0.9);
\vertex (e2) at (6.8,0.1);
\vertex (e3) at (7.6,-0.9);
\vertex (e4) at (6.8,-1.9);
\vertex (g1) at (8.3,0.8) {$\gamma$};
\vertex (g2) at (8.6,-0.9) {$\gamma$};
\vertex (g3) at (8.3,-2.5) {$\gamma$};

\diagram*{
  (i) -- [fermion] (v1),
  (v1) -- [fermion] (f),
  (v1) -- [boson, edge label'={$Z_B^\ast$}] (mix),
  (mix) -- [photon, edge label={$\gamma^\ast$}] (e1),
  (e1) -- [fermion, edge label={$e$}] (e2),
  (e2) -- [fermion] (e3),
  (e3) -- [fermion] (e4),
  (e4) -- [fermion] (e1),
  (e2) -- [photon] (g1),
  (e3) -- [photon] (g2),
  (e4) -- [photon] (g3),
};
\end{feynman}
\node[anchor=north, inner sep=1pt]
  at ([yshift=-5pt]mix.south)
  {\small $\epsilon_{YB}$};
\node at (4.25,-3.0)
  {\small (d) Electron-loop three-photon decay};
\end{tikzpicture}%
}
\end{minipage}

\caption{Representative excited-state decays. Panel (a) shows the
ordinary-$Z$ neutrino channel induced by singlet--doublet mixing.
Panel (b) shows the charged-anomalon and $W$ contribution to the
radiative transition; photon attachments to all charged lines and the
Goldstone diagrams are required for the complete amplitude. Panel (c)
illustrates the momentum-suppressed neutrino channel from kinetic
mixing, and panel (d) the electron-loop three-photon channel below
the electron-pair threshold.}
\label{fig:chi2_decay_diagrams}
\end{figure*}

The transition moment also contributes to inelastic DD. Its leading
NR nuclear charge response is
\begin{equation}
\frac{d\sigma_\gamma^{\rm charge}}{dE_R}
=
\frac{\alpha Z^2|\mu_{12}|^2F_A^2(E_R)}{E_R}
\left[
1-\frac{v_{\rm min}^2}{v^2}
+\frac{m_AE_R}{2m_\chi^2v^2}
\right],
\label{eq:dipoleChargeResponse}
\end{equation}
with energies expressed in consistent units and velocities in units
of $c$. The last term is the charge Darwin contribution. Folding
this response with the production xenon mixture and acceptance
gives $1.35\times10^{-12}$ signal-window events and
$4.82\times10^{-12}$ control-window events at the reference point.
A pointwise charge-response bound, combined with
Cauchy--Schwarz for interference, limits the fractional correction
to below $7.2\times10^{-4}$ throughout the selected primary
witnesses. This supports retaining the tree-level recoil scan within
this response scope, without reclassifying points at its numerical
boundaries. Magnetic nuclear responses and off-shell charge-radius
or anapole terms are not included in this bound.

\subsection{Abundance at recombination}

Present-day depletion alone does not establish depletion at
recombination. If only the electroweak neutrino channel were present,
$t_{\rm rec}\simeq3.8\times10^5$ yr would give
\begin{equation}
f_2^{\rm prim}(t_{\rm rec})
\simeq
\frac12
\exp\left[
-\frac{3.8\times10^5}{6.24\times10^7}
\right]
\simeq0.497.
\label{eq:f2recombination}
\end{equation}
The mixed-pair $s$-wave annihilation channel would then retain an
effective population-weighted rate
\begin{equation}
2f_1(t_{\rm rec})f_2(t_{\rm rec})
\langle\sigma v\rangle_{12}
\simeq
\frac12\langle\sigma v\rangle_{12}.
\label{eq:CMBcoann}
\end{equation}
Using the DD-inspired target at recombination, neutrino decay alone
would require $\theta_N\gtrsim2.8\times10^{-2}$, substantially larger
than the reference mixing.

The neutrino-only calculation illustrates why depletion by today does not necessarily imply depletion before recombination. In the full UV model, however, the much faster radiative decay is also present: its lifetime ranges from $7.4\times10^4$ to $5.1\times10^6\s$ across the sampled parameter points satisfying the primary constraints. It therefore removes the primordial $\chi_2$ population before recombination and suppresses annihilation involving a $\chi_1\chi_2$ pair. 

At the reference radiative lifetime, a radiation-dominated background
gives $T_\gamma\simeq2.22\keV$ and $z\simeq9.4\times10^6$. Even with
half of the initial DM in $\chi_2$, the released energy is only
$7.7\times10^{-11}$ of the photon energy density at that epoch.
The emitted photon has energy approximately $300\keV$, below the
deuterium photodisintegration threshold. These estimates indicate
a negligible energy release at the benchmark, although they are not
a full BBN or spectral-distortion likelihood.

For comparison, the conservative benchmark estimate obtained
without radiative depletion, $p_{\rm ann}\lesssim2.5\times10^{-29}\cm^3\s^{-1}\GeV^{-1}$ for $f_{\rm eff}\leq1$, is already
below the Planck bound
$3.2\times10^{-28}\cm^3\s^{-1}\GeV^{-1}$~\cite{Planck2018}.
Depletion suppresses the mixed-pair contribution further; it does
not suppress all ground-state annihilation channels. Open bosonic
channels require their own CMB response.


\subsection{Late-time regeneration}
\label{app:regeneration}

The exponentially depleted primordial component is not the complete
local population if an environment continues to excite $\chi_1$.
Schematically, decay and conversion contribute
\begin{align}
\dot n_2+3Hn_2={}&
-\Gamma_{2\to1}n_2+
\Gamma_{1\to2}^{\rm inv}n_1
\nonumber\\
&-
\langle\sigma_{22\to11}v\rangle n_2^2+
\langle\sigma_{11\to22}v\rangle n_1^2
\nonumber\\
&-
\Gamma_{2f\to1f}n_2+
\Gamma_{1f\to2f}n_1+S_2,
\label{eq:fulln2schematic}
\end{align}
where $S_2$ denotes additional sources. This expression identifies
the population-changing terms rather than a complete annihilation
and transport system.

In the smooth Galactic halo, self-excitation
$\chi_1\chi_1\to\chi_2\chi_2$ requires
\begin{equation}
v_{\rm rel}
\gtrsim
\sqrt{\frac{8\delta_\chi}{m_\chi}}
\simeq464\ {\rm km\,s^{-1}}
\label{eq:selfexcitationthreshold}
\end{equation}
at the reference point. A contact estimate gives
$\sigma_{\chi\chi}^{\rm up}
\sim(3g_B)^4m_\chi^2/(4\pi M_{Z_B}^4)
\sim2\times10^{-36}\cm^2$.
With $\rho_\chi=0.4\GeV\,{\rm cm}^{-3}$ and a characteristic
relative speed of $300\kms$, the excitation time is of order
$10^{24}$ yr even before imposing the threshold. Scattering on
Galactic gas is also inefficient.

This dimensional estimate is not a full mixed-Majorana scattering
calculation. The supplied local-source audit instead uses an enlarged
tree-level production envelope, obtaining $f_2<8.6\times10^{-9}$
over the selected witnesses in the specified smooth-halo environment,
and checks the resulting reverse-transition recoil contribution.
Rapid decay therefore permits a small source-supported fraction,
but not a phenomenologically important one in this environment.
The estimate does not apply unchanged to dense subhalos or compact
objects.

The Sun is different because endothermic capture continuously
injects $\chi_2$. Its population must therefore be evolved using
the physical radiative lifetime, $\tau_\gamma\simeq3.11$ days,
together with scattering and annihilation. The intrinsic decay
photons and neutrinos have sub-MeV energies and do not generate
the high-energy IceCube signal; decay matters by changing the
captured state composition.

\subsection{Captured two-state population}
\label{app:solarpopulation}

Let $\mathcal N_i$ denote the captured populations, $C_i$ the capture
rates into each state, $\kappa_{ij}$ the scattering-conversion rates,
and $\mathcal A_{ij}=\langle\sigma_{ij}v\rangle/V_{\rm eff}$ the
one-zone annihilation coefficients. Neglecting evaporation in the
mass range considered,
\begin{align}
\dot{\mathcal N}_1={}&
C_1+(\Gamma_2+\kappa_{21})\mathcal N_2
-\kappa_{12}\mathcal N_1
\nonumber\\
&-
\mathcal A_{11}\mathcal N_1^2
-\mathcal A_{12}\mathcal N_1\mathcal N_2,
\nonumber\\
\dot{\mathcal N}_2={}&
C_2+\kappa_{12}\mathcal N_1
-(\Gamma_2+\kappa_{21})\mathcal N_2
\nonumber\\
&-
\mathcal A_{22}\mathcal N_2^2
-\mathcal A_{12}\mathcal N_1\mathcal N_2.
\label{eq:solarTwoState}
\end{align}
Here $\Gamma_2$ includes the radiative width. The annihilation event
rate is
\begin{equation}
\Gamma_A=
\frac12\mathcal A_{11}\mathcal N_1^2+
\mathcal A_{12}\mathcal N_1\mathcal N_2+
\frac12\mathcal A_{22}\mathcal N_2^2.
\label{eq:solarTwoStateAnnihilation}
\end{equation}
The identical-state factors count annihilation events; summing the
population equations gives
$\dot{\mathcal N}_1+\dot{\mathcal N}_2=C_1+C_2-2\Gamma_A$,
as required by number balance. A small excited fraction must still
be retained when evaluating the mixed-pair term.

For a population test using the thermalized volume and capture
normalization of Sec.~\ref{subsec:solar}, we take initially empty
populations, capture into state 2, and neglect thermal upward
conversion because $\delta_\chi/T_\odot\simeq230$ at the reference
point. In the capture-plus-decay limit,
$\mathcal N_2\simeq C_2\tau_\gamma$ while $\mathcal N_1$ accumulates
or annihilates. Including diagonal annihilation and the leading
mixed-pair quark channel with the physical mediator width, the
stiff ODE solution gives
\begin{equation}
\begin{aligned}
\mathcal N_1&=2.07\times10^{37},
\,\,
\mathcal N_2=3.86\times10^{25},
\\
f_2^\odot&=1.86\times10^{-12},
\,\,
\Gamma_A=1.91\times10^{13}\s^{-1}.
\end{aligned}
\label{eq:solarPopulationBenchmark}
\end{equation}
The total annihilation rate is approximately $1.1\%$ above the
one-population estimate. Thus continuous injection leaves a nonzero
excited population, but its mixed-pair contribution is subdominant
under these assumptions.

An illustrative open-channel point with $m_\chi=1\TeV$,
$M_{Z_B}=500\GeV$, and $g_B=0.10$ gives
$f_2^\odot=1.9\times10^{-11}$ and reproduces the one-population
equilibrium rate. This tests a different population limit rather
than providing a fitted thermal benchmark; the analytic
capture-plus-decay solution supplies an independent ODE check.

These tests support the rapid-decay population approximation within
the adopted one-zone treatment. They do not establish orbital
thermalization or exclude nonthermal upward scattering of newly
captured particles. A precision Solar constraint additionally
requires element-dependent capture and scattering, orbital
evolution, all relevant annihilation channels, and neutrino yields
and transport for the actual final states. The Solar curves in the
main text therefore remain diagnostic.

\section{Subleading direct-detection interactions}
\label{app:subleadingDD}

The coherent endothermic transition $\chi_1N\to\chi_2N$ mediated by $Z_B$ dominates the DD signal discussed in Secs.~\ref{subsec:LZ} and~\ref{subsec:deltaprofile}. The UV completion also predicts a mixing-induced SM $Z$ contribution, diagonal axial $Z_B$ exchange, and elastic scalar exchange. This appendix derives their coefficients and recoil spectra, explains their different suppressions, and specifies which amplitudes interfere.

\subsection{Reference $Z_B$-mediated endothermic interaction}

For $\theta_L=\theta_R=\theta_N$, the effective vector charge of the light Dirac state is $B_V^{\rm eff}=-3$, independently of the neutral mixing angle. Using the Hermitian transition current $J_{12}^\mu$ defined in Sec.~\ref{subsec:current}, the interaction is
\begin{equation}
\mathcal L\supset-3g_BZ_{B\mu}J_{12}^\mu.
\end{equation}
Both nucleons have baryon number one. Integrating out $Z_B$ therefore gives
\begin{equation}
\mathcal L_{\rm eff}^{B}
=
G_N^BJ_{12}^\mu\left[\overline N\gamma_\mu N\right],
\,\,
G_N^B=\frac{3g_B^2}{M_{Z_B}^2}.
\label{eq:GBappendix}
\end{equation}
The coefficient is isoscalar, $G_p^B=G_n^B=G_N^B$, so the nuclear amplitude scales as $AG_N^B$. For $g_B=0.22782$ and $M_{Z_B}=1.44\TeV$,
\begin{equation}
G_N^B\simeq7.51\times10^{-8}\GeV^{-2}.
\label{eq:GBbenchmarkappendix}
\end{equation}

The leading nonrelativistic operator is $\mathcal O_1=\mathbf 1_\chi\mathbf 1_N$, acting off diagonally in the DM-state index. Its contact amplitude has no explicit velocity or momentum suppression, and the differential cross section scales as $d\sigma_B^A/dE_R\propto A^2|G_N^B|^2F_A^2/v^2$ within the allowed recoil interval. The absence of low-energy recoils instead follows from the endothermic minimum speed in Eq.~\eqref{eq:vmininelastic}: its splitting contribution grows as $\delta_\chi/\sqrt{E_R}$, eventually exceeding the largest halo speed as $E_R$ decreases.

\subsection{Electroweak $Z$ exchange induced by neutral-fermion mixing}

The singlet $\chi$ has no SM gauge interaction, whereas the neutral doublet component $\Psi^0$ has $T_3=1/2$ and $Q=0$. Its $Z$ coupling is
\begin{equation}
\mathcal L_Z^\Psi
=
\frac{g}{2c_W}Z_\mu
\left[
\overline{\Psi_L^0}\gamma^\mu\Psi_L^0
+
\overline{\Psi_R^0}\gamma^\mu\Psi_R^0
\right].
\label{eq:PsiZcoupling}
\end{equation}
The light neutral Dirac state contains the doublet admixtures
\begin{align}
N_{LL}&=-\sin\theta_L\,\Psi_L^0+\cos\theta_L\,\chi_L,
\nonumber\\
N_{LR}&=-\sin\theta_R\,\Psi_R^0+\cos\theta_R\,\chi_R.
\end{align}
Projecting onto this state gives
\begin{equation}
\mathcal L_Z^{N_L}
=
Z_\mu\overline N_L\gamma^\mu
\left[g_V^{\chi Z}+g_A^{\chi Z}\gamma^5\right]N_L,
\label{eq:lightZcoupling}
\end{equation}
where
\begin{align}
g_V^{\chi Z}
&=
\frac{g}{4c_W}
\left[\sin^2\theta_L+\sin^2\theta_R\right],
\label{eq:gVchiZ}\\
g_A^{\chi Z}
&=
\frac{g}{4c_W}
\left[\sin^2\theta_R-\sin^2\theta_L\right].
\label{eq:gAchiZ}
\end{align}
For equal angles, $g_A^{\chi Z}=0$ and $g_V^{\chi Z}=g\sin^2\theta_N/\left[2c_W\right]$. After the Majorana splitting, the vector interaction becomes the off-diagonal transition
\begin{equation}
\mathcal L_Z^{12}
=
i g_V^{\chi Z}Z_\mu
\overline{\chi_1}\gamma^\mu\chi_2.
\label{eq:Ztransition}
\end{equation}
It therefore contributes to the same endothermic final state as baryonic vector exchange.

The nucleon interaction is
\begin{equation}
\mathcal L_Z^N
=
\frac{g}{2c_W}Z_\mu\overline N\gamma^\mu
\left[g_V^N-g_A^N\gamma^5\right]N,
\end{equation}
with vector charges $g_V^p=1/2-2s_W^2$ and $g_V^n=-1/2$. In the convention where $G_N^B$ is positive, the coherent weak coefficient is
\begin{equation}
G_Z^N
=
-\frac{g^2}{8c_W^2m_Z^2}
\left[\sin^2\theta_L+\sin^2\theta_R\right]g_V^N.
\label{eq:GZN}
\end{equation}
For equal angles,
\begin{equation}
G_p^Z
=
-\frac{\sin^2\theta_N}{v^2}
\left[\frac12-2s_W^2\right],
\,\,
G_n^Z
=
\frac{\sin^2\theta_N}{2v^2}.
\label{eq:GZNEqual}
\end{equation}
The nuclear coefficients are consequently
\begin{equation}
G_Z^A=ZG_p^Z+\left[A-Z\right]G_n^Z,
\,\,
G_B^A=AG_N^B.
\label{eq:GZAGBA}
\end{equation}
The full proton and neutron coefficients used in the coherent calculation are $C_p=G_N^B+G_p^Z$ and $C_n=G_N^B+G_n^Z$.

For xenon, the neutron weak charge dominates. The amplitude ratio
\begin{equation}
r_Z^A\equiv\left|\frac{G_Z^A}{G_B^A}\right|
\label{eq:rZA}
\end{equation}
is approximately
\begin{equation}
r_Z^{\rm Xe}\simeq1.5\times10^{-3}
\label{eq:rZXebenchmark}
\end{equation}
for $A\simeq131$, $Z=54$, and the reference parameters $\theta_N=0.005$, $g_B=0.22782$, and $M_{Z_B}=1.44\TeV$. With the signed coefficients above, the interference is constructive and changes the reference vector rate by approximately $0.3\%$. The weak-only rate scales as $\sin^4\theta_N$, whereas its interference with the baryonic amplitude scales as $\sin^2\theta_N$. The cross term must therefore be retained even when the separate weak curve is very small. Its numerical importance elsewhere in the scan depends on the baryonic coefficient.

The displayed weak spectrum includes the coherent vector-nucleon response. The spin and velocity-dependent inelastic terms from the axial nucleon current are not included in this comparison. If $\theta_L\neq\theta_R$, a diagonal DM axial coupling also appears through Eq.~\eqref{eq:gAchiZ}, generating elastic $\mathcal O_8$, $\mathcal O_9$, and spin-dependent $\mathcal O_4$ contributions. These additional diagonal electroweak terms vanish at leading order in the equal-angle benchmark.

\subsection{Elastic scattering from the diagonal axial $Z_B$ current}

The diagonal baryonic axial interaction is present even in the singlet limit. For equal neutral mixing angles, $B_A^{\rm eff}=-3\cos\left[2\theta_N\right]/2$, and the lighter Majorana state has the interaction
\begin{equation}
\mathcal L_{\chi_1\chi_1Z_B}^{A}
=
-\frac{3g_B}{4}\cos\left[2\theta_N\right]
Z_{B\mu}\overline{\chi_1}\gamma^\mu\gamma^5\chi_1.
\label{eq:axialZBchi1appendix}
\end{equation}
The factor $1/2$ associated with the diagonal Majorana bilinear is canceled by the identical-field contractions in the physical vertex. Accordingly, we write
\begin{equation}
\mathcal L_{\rm eff}^{AV}
=
-\frac{G_{A,N}^B}{2}
\left[\overline{\chi_1}\gamma^\mu\gamma^5\chi_1\right]
\left[\overline N\gamma_\mu N\right],
\label{eq:AVeffective}
\end{equation}
where the physical amplitude coefficient is
\begin{equation}
G_{A,N}^B
=
-\frac{3g_B^2}{2M_{Z_B}^2}\cos\left[2\theta_N\right]
=
-\frac{\cos\left[2\theta_N\right]}{2}G_N^B.
\label{eq:GAVB}
\end{equation}
At the reference point, $G_{A,N}^B\simeq-3.75\times10^{-8}\GeV^{-2}$. Applying another factor of $1/2$ to this physical coefficient would underestimate the axial rate by a factor of four.

The suppression arises in the nonrelativistic operators~\cite{FitzpatrickEFTDD,AnandNREFT},
\begin{equation}
\mathcal O_8=\mathbf S_\chi\cdot\mathbf v^\perp,
\,\,
\mathcal O_9=
i\mathbf S_\chi\cdot
\left[\mathbf S_N\times\frac{\mathbf q}{m_N}\right].
\label{eq:O8O9}
\end{equation}
Here $\mathbf S_\chi$ and $\mathbf S_N$ are the DM and nucleon spin operators. Their coefficients follow from the same relativistic interaction, with signs determined consistently with the momentum-transfer convention.

The $\mathcal O_8$ amplitude contains the transverse velocity and couples to the coherent nuclear charge response $M$. Its coherent rate is suppressed by a factor of order $v^2\sim10^{-6}$ relative to an elastic $\mathcal O_1$ interaction with the same relativistic coefficient. By contrast, $\mathcal O_9$ contains $q/m_N$ and probes nuclear spin: its rate carries a momentum factor $q^2/m_N^2$ and lacks the full coherent $A^2$ enhancement. These are different suppressions; $\mathcal O_9$ is not subject to a universal additional $v^2$ factor.

For the elastic nuclear center-of-mass response, $q^2=2m_AE_R$ and $v_A^{\perp\,2}=v^2-q^2/\left[4\mu_{\chi A}^2\right]=v^2-\left[v_{\min,A}^{\rm el}\right]^2$, with velocities in units of $c$. The coherent axial cross section therefore scales as
\begin{equation}
\frac{d\sigma_{AV,M}^A}{dE_R}
\propto
\frac{A^2|G_{A,N}^B|^2F_A^2}{v^2}
\left[v^2-\frac{q^2}{4\mu_{\chi A}^2}\right].
\label{eq:axialcoherentScaling}
\end{equation}
The velocity factor vanishes at the minimum incident speed for a given recoil. Its halo average must be evaluated explicitly rather than replaced by a constant suppression multiplying an SI spectrum.

Although suppressed, axial scattering is elastic and remains accessible below the endothermic threshold. It therefore generates the residual low-energy baryonic tail in Sec.~\ref{subsec:LZ}. This does not contradict its small contribution near the LZ candidate: the elastic and endothermic processes have different kinematic support.

The figure and scan retain only the coherent $\mathcal O_8$ charge response, approximated by a Helm form factor and folded with the transverse-velocity integral in Eq.~\eqref{eq:DDhalomoments}. The $\mathcal O_8$ orbital response, $\mathcal O_9$ spin response, and their interference are omitted. A full xenon NREFT calculation is needed to quantify their effect; their smallness cannot be inferred from the coherent velocity suppression alone. The displayed elastic rates and the resulting selection boundaries are conditional on this approximation, whose physical uncertainty is not measured by numerical convergence.

\subsection{Scalar-mediated elastic SI scattering}

The light neutral state couples to both the baryonic scalar $s_0$ and, through fermion mixing, the SM Higgs component $h_0$. Writing the diagonal Majorana scalar interaction as $\mathcal L\supset-g_{s_0\chi\chi}s_0\overline{\chi_1}\chi_1/2$, the physical couplings for the symmetric mass prescription are
$g_{h_0\chi\chi}=-2\Delta_\Psi\sin^2\theta_N/v$ and
$g_{s_0\chi\chi}=\left[m_\chi+2\Delta_\Psi\sin^2\theta_N\right]/v_B$,
where $\Delta_\Psi=M_\Psi-m_\chi$ and negligible Majorana corrections are omitted.

Rotating with Eq.~\eqref{eq:scalarrotation} gives the nucleon coefficient
\begin{align}
G_S^N=\frac{f_Nm_N}{v}\bigg[&
\frac{\left[g_{h_0\chi\chi}c_B+g_{s_0\chi\chi}s_B\right]c_B}{m_{h_1}^2}
\nonumber\\
&-
\frac{\left[-g_{h_0\chi\chi}s_B+g_{s_0\chi\chi}c_B\right]s_B}{m_{h_2}^2}
\bigg],
\label{eq:scalarcoefficient}
\end{align}
where $c_B=\cos\theta_B$ and $s_B=\sin\theta_B$. Separating its two origins,
\begin{align}
G_{S,\rm portal}^N
&=
\frac{f_Nm_N}{v}\,g_{s_0\chi\chi}s_Bc_B
\left[\frac{1}{m_{h_1}^2}-\frac{1}{m_{h_2}^2}\right],
\nonumber\\
G_{S,\rm Higgs}^N
&=
\frac{f_Nm_N}{v}\,g_{h_0\chi\chi}
\left[\frac{c_B^2}{m_{h_1}^2}+\frac{s_B^2}{m_{h_2}^2}\right],
\nonumber\\
G_S^N&=G_{S,\rm portal}^N+G_{S,\rm Higgs}^N.
\label{eq:scalarAmplitudeParts}
\end{align}
Both scalar propagators and their relative signs are included. In particular, $\theta_B\to0$ removes the portal contribution but leaves the direct-Higgs term when $\theta_N\neq0$.

Scalar exchange generates elastic $\mathcal O_1$ scattering, with $d\sigma_S^A/dE_R\propto A^2|G_S^N|^2F_A^2/v^2$. There is no additional nonrelativistic velocity or momentum suppression. Its small rate instead follows from the mixing-dependent couplings and their possible cancellation.

For $f_N=0.30$, $\sin\theta_B=+3\times10^{-4}$, and the reference fermion spectrum,
$G_{S,\rm portal}^N\simeq1.04\times10^{-11}\GeV^{-2}$ and
$G_{S,\rm Higgs}^N\simeq-7.42\times10^{-12}\GeV^{-2}$.
Their destructive interference gives
\begin{equation}
|G_S^N|\simeq2.96\times10^{-12}\GeV^{-2},
\label{eq:GSbenchmark}
\end{equation}
corresponding to $|G_S^N/G_N^B|\simeq3.94\times10^{-5}$ and a scalar rate approximately $0.081$ times the portal-only rate. The cancellation depends on the signed scalar angle, neutral mixing, and fermion mass gap; it is not a universal property of the scan. At the reference point, scalar scattering is negligible near the LZ candidate and smaller than the axial contribution at low recoil energy.

\subsection{Recoil spectra and interference between amplitudes}

The relevant halo integrals are
\begin{align}
\eta\left[w\right]
&=
\int_{v\geq w}d^3v\,
\frac{f_{\rm lab}\left[\mathbf v\right]}{v},
\nonumber\\
\eta_\perp\left[w\right]
&=
\int_{v\geq w}d^3v\,
\frac{f_{\rm lab}\left[\mathbf v\right]}{v}
\left[v^2-w^2\right].
\label{eq:DDhalomoments}
\end{align}
The distribution is normalized to unity and velocities are expressed in units of $c$. Endothermic vector scattering uses $\eta$ at $v_{\min,A}^{\rm up}$ from Eq.~\eqref{eq:vmininelastic}; elastic scalar and coherent axial scattering use $\eta$ and $\eta_\perp$, respectively, at $v_{\min,A}^{\rm el}$.

\begin{figure}[htbp]
\centering
\includegraphics[width=0.98\columnwidth]{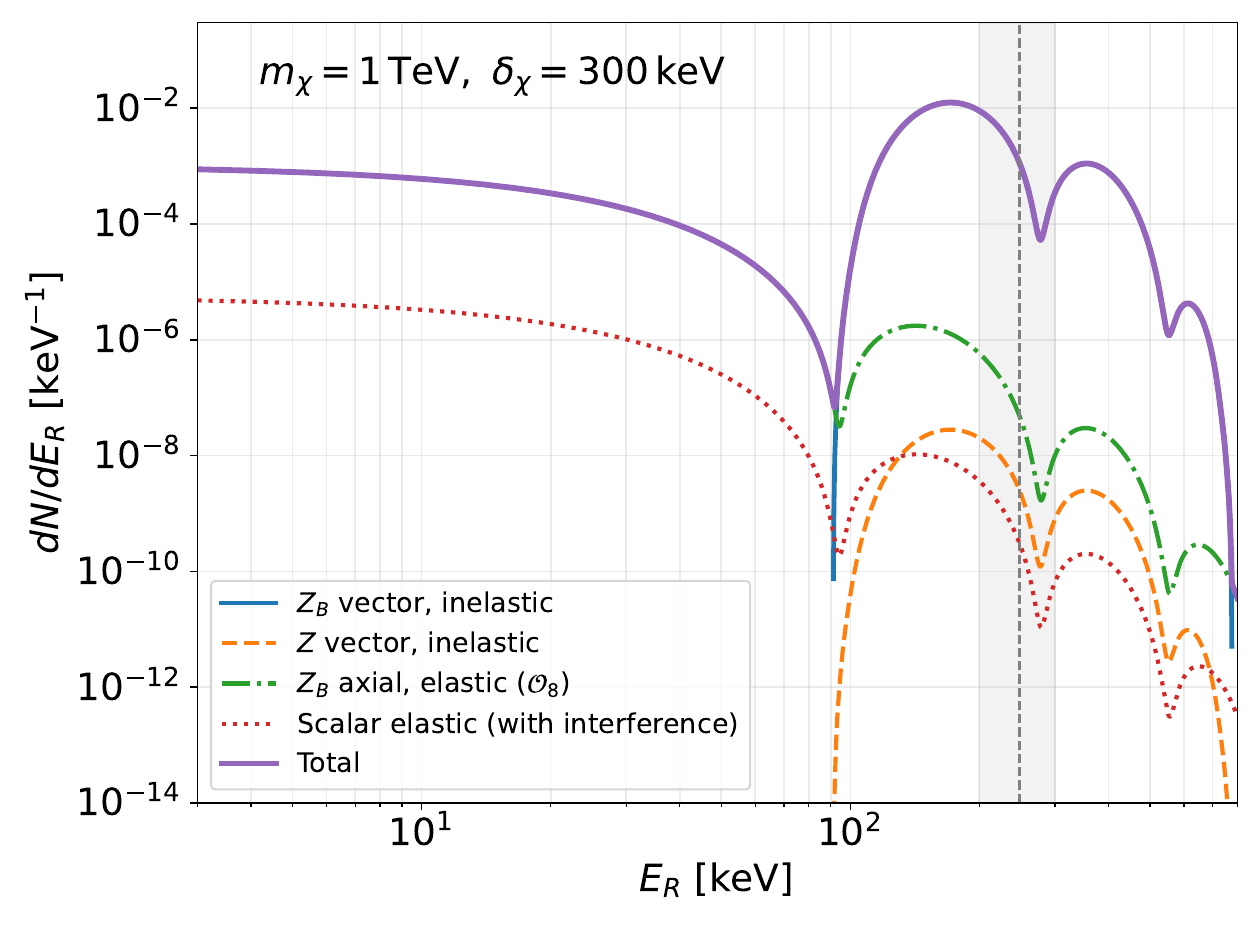}
\caption{Recoil spectra at the reference benchmark, with the physical coefficients and fixed normalization used in the scan. Blue and orange show the separate squared baryonic and weak endothermic vector amplitudes; their constructive interference is included in the purple total. Green shows elastic axial $Z_B$ exchange in the coherent $\mathcal O_8$ approximation. Red shows elastic scalar exchange, including destructive portal--Higgs interference. The elastic spectra remain accessible below the endothermic threshold. The dashed line marks $248\keV$, and the shaded interval is $200$--$300\keV$. Detector acceptance is omitted.}
\label{fig:ddcomponents}
\end{figure}

For the depleted halo used in Fig.~\ref{fig:ddcomponents}, the individual spectra before detector acceptance are
\begin{align}
\frac{dN_B}{dE_R}
&=
\mathcal K\sum_A a_AF_A^2
|AG_N^B|^2\eta\left[v_{\min,A}^{\rm up}\right],
\nonumber\\
\frac{dN_Z}{dE_R}
&=
\mathcal K\sum_A a_AF_A^2
|G_Z^A|^2\eta\left[v_{\min,A}^{\rm up}\right],
\nonumber\\
\frac{dN_{BZ}}{dE_R}
&=
\mathcal K\sum_A a_AF_A^2
2\operatorname{Re}\left[AG_N^B\overline{G_Z^A}\right]
\eta\left[v_{\min,A}^{\rm up}\right],
\nonumber\\
\frac{dN_{AV,M}}{dE_R}
&=
\mathcal K\sum_A a_AF_A^2
A^2|G_{A,N}^B|^2
\eta_\perp\left[v_{\min,A}^{\rm el}\right],
\nonumber\\
\frac{dN_S}{dE_R}
&=
\mathcal K\sum_A a_AF_A^2
A^2|G_S^N|^2\eta\left[v_{\min,A}^{\rm el}\right].
\label{eq:DDcomponentSpectra}
\end{align}
Here $a_A\equiv\xi_A$ is the isotope mass fraction defined in Eq.~\eqref{eq:rate}, $F_A$ is its Helm form factor, and $\mathcal K$ contains the common exposure calibration, DM flux factor $1/m_\chi$, and unit conversions. The normalization is shared by all processes and is not adjusted to force the current model to predict one event.
We combine the xenon isotope contributions using mass fractions and evaluate the elastic axial spectrum in the leading coherent $\mathcal O_8$ approximation; the impact of the omitted orbital, $\mathcal O_9$ spin, and associated interference responses on low-energy counts is negligible.

Two interference effects are retained. The $Z_B$ and $Z$ vector amplitudes produce the same endothermic final state and must be added before squaring; their cross term is $dN_{BZ}/dE_R$. Likewise, the portal and direct-Higgs amplitudes produce the same elastic scalar final state, giving
\begin{align}
|G_S^N|^2 \!=\! |G_{S,\rm portal}^N|^2 \!+\! |G_{S,\rm Higgs}^N|^2 \!+\! 2\operatorname{Re}\left[G_{S,\rm portal}^N\overline{G_{S,\rm Higgs}^N}\right].
\label{eq:scalarRateInterference}
\end{align}
Both cross terms are signed contributions, not separate event populations. The scalar cross term is negative at the reference point and is already included in $dN_S/dE_R$.

The inelastic and elastic amplitudes do not interfere because their outgoing DM states are orthogonal. The leading elastic scalar--axial interference also vanishes for unpolarized DM: $\mathcal O_1$ is spin independent, whereas $\mathcal O_8$ and $\mathcal O_9$ are linear in $\mathbf S_\chi$ and have a vanishing spin trace. The implemented total is therefore
\begin{equation}
\frac{dN_{\rm DD}}{dE_R}
=
\frac{dN_B}{dE_R}
+\frac{dN_Z}{dE_R}
+\frac{dN_{BZ}}{dE_R}
+\frac{dN_{AV,M}}{dE_R}
+\frac{dN_S}{dE_R}.
\label{eq:DDhierarchy}
\end{equation}
The amplitudes are combined separately for each isotope before summing its recoil contribution. The omitted nuclear axial spin--orbital interference is distinct from the vector-mediator and scalar interference retained here.

Fig.~\ref{fig:ddcomponents} illustrates these spectra for $m_\chi=1\TeV$, $\delta_\chi=300\keV$, $M_{Z_B}=1.44\TeV$, $g_B=0.22782$, and $\theta_N=0.005$. The SHM parameters are $v_0=238\kms$, $v_{\rm esc}=544\kms$, and $v_E=250.2\kms$. Detector acceptance is omitted to display the physical recoil shapes.

At $E_R=248\keV$, the baryonic spectrum is approximately $1.13\times10^{-3}\keV^{-1}$. The weak-only spectrum is $2.52\times10^{-9}\keV^{-1}$, while the constructive vector interference contributes $3.38\times10^{-6}\keV^{-1}$, approximately $0.30\%$ of the baryonic contribution. Thus the leading weak correction is the interference, rather than the separate weak squared amplitude.

At $E_R=20\keV$, endothermic scattering has no halo support. The coherent axial and scalar spectra are approximately $3.35\times10^{-4}\keV^{-1}$ and $1.87\times10^{-6}\keV^{-1}$, respectively. The scalar result combines portal-only and direct-Higgs-only spectra of $2.30\times10^{-5}\keV^{-1}$ and $1.17\times10^{-5}\keV^{-1}$ with an interference contribution of $-3.28\times10^{-5}\keV^{-1}$. The negative cross term reduces the positive physical scalar rate.

Integrated over $1$--$35\keV$ before acceptance, the axial and scalar components give approximately $0.015$ and $8.16\times10^{-5}$ events. The reference point therefore predicts a small elastic tail, rather than an order-one low-energy event. After applying the adopted acceptance, the total counts are $0.218$ in the $200$--$300\keV$ signal window and $0.732$ in the control window below $200\keV$, as used in Sec.~\ref{subsec:LZ}.

The reference high-energy signal is therefore dominated by baryonic vector exchange, while axial and scalar scattering produce a residual elastic tail below the endothermic threshold. The raw spectra illustrate this hierarchy; statistical selections use the same normalization with detector acceptance applied over the specified recoil windows.

The general DD calculation includes both off-diagonal transitions, $\chi_1N\to\chi_2N$ and $\chi_2N\to\chi_1N$, weighted by their halo populations, together with scalar elastic scattering and the coherent axial approximation described above. The matched electromagnetic transition and its leading nuclear charge response are discussed in App.~\ref{app:radiativedecay}, where the stated interference bound establishes its subleading contribution for the selected configurations within that response prescription.

\section{Possible ultraviolet origin of the pseudo-Dirac mass hierarchy}
\label{app:uvsplitting}

The phenomenology in the main text requires a small Majorana mass compared with the Dirac mass,
\begin{equation}
\frac{m_M}{M_D}
=
\frac{y_\chi}{\lambda_\chi}
\sim{\cal O}\left[10^{-7}\right].
\label{eq:uvhierarchy}
\end{equation}
For the reference point, $m_M/M_D\simeq300\keV/1\TeV\simeq3\times10^{-7}$, up to the small neutral-mixing correction. Such a hierarchy is characteristic of pseudo-Dirac DM~\cite{DeSimoneSanzSato2010,CarrilloGonzalezToro2022}. As discussed in Sec.~\ref{subsec:pseudodirac}, it is technically natural because $m_M\to0$ restores a fermion-number symmetry~\cite{tHooftNaturalness}.

Technical naturalness protects a small Majorana mass but does not determine its magnitude. We therefore describe two possible UV origins of the hierarchy. These are illustrative extensions, not additional ingredients of the phenomenological scan. Since $S_B^\dagger\overline{\chi_L^c}\chi_L$ is already gauge invariant, either interpretation requires an additional symmetry or selection rule that suppresses the direct renormalizable Majorana interaction while allowing the Dirac mass and the electroweak Yukawa interactions.

\subsubsection{Higher-dimensional origin}

Introduce a scalar spurion $\Phi$ that is neutral under the SM and $U(1)_B$ gauge groups but charged under the additional symmetry. If the renormalizable Majorana interaction is forbidden, the leading allowed operator can be
\begin{equation}
-\mathcal L_{\rm UV}
\supset
\frac{c_\chi}{2}
\left[\frac{\Phi}{\Lambda}\right]^n
S_B^\dagger\overline{\chi_L^c}\chi_L
+{\rm h.c.},
\label{eq:higherdimchi}
\end{equation}
where $\Lambda$ is the heavy threshold and $c_\chi$ is of order unity. This suppression by symmetry-breaking insertions is familiar from Froggatt--Nielsen constructions~\cite{FroggattNielsen1979} and has also been considered in pseudo-Dirac dark sectors~\cite{CarrilloGonzalezToro2022}.

For $\langle\Phi\rangle=f/\sqrt2$, the effective Yukawa coupling is
\begin{equation}
y_\chi^{\rm eff}
=
c_\chi
\left[\frac{f}{\sqrt2\Lambda}\right]^n.
\label{eq:yeffUV}
\end{equation}
Using $m_M=y_\chi^{\rm eff}v_B/\sqrt2$ and $M_D=\lambda_\chi v_B/\sqrt2$ gives
\begin{equation}
\frac{m_M}{M_D}
=
\frac{c_\chi}{\lambda_\chi}
\left[\frac{f}{\sqrt2\Lambda}\right]^n.
\label{eq:masshierarchyUV}
\end{equation}
The small Majorana coupling can thus encode a hierarchy of scales rather than a fundamental dimensionless parameter of order $10^{-7}$.

For $c_\chi\simeq1$ and $f\simeq v_B\simeq2.1\TeV$, the reference value $y_\chi^{\rm eff}\simeq2\times10^{-7}$ corresponds to
\begin{align}
\Lambda&\simeq7\times10^9\GeV,
&&n=1,
\nonumber\\
\Lambda&\simeq3\times10^6\GeV,
&&n=2.
\end{align}
The operator has dimension $4+n$, and the effective expansion requires $f/\Lambda\ll1$. Provided the symmetry-breaking insertions shown in Eq.~\eqref{eq:higherdimchi} are the leading allowed ones, radiative corrections respect the same selection rule and cannot generate an unsuppressed Majorana term.

Because $\Phi$ is neutral under $U(1)_B$, its vacuum expectation value does not contribute to $M_{Z_B}=3g_Bv_B$ or alter the baryonic residual parity. A specific completion must nevertheless identify the heavy mediators and check the other operators generated by matching. Here their additional effects are assumed to decouple, leaving the Majorana entry whose projection onto the light neutral state gives $\delta_\chi\simeq m_M\cos^2\theta_L$, as in the main text.

\subsubsection{Seesaw-like interpretation}

Alternatively, a small Majorana entry can arise from mixing with a much heavier neutral fermion, in analogy with the ordinary seesaw mechanism~\cite{Minkowski1977,GellMannRamondSlansky1979,MohapatraSenjanovic1980}. In the symmetry-broken theory, consider the schematic mass matrix
\begin{equation}
\mathcal M_{\rm UV}
=
\begin{pmatrix}
0 & M_D & m_{\chi N}\\
M_D & 0 & 0\\
m_{\chi N} & 0 & M_N
\end{pmatrix},
\label{eq:darkseesawmatrix}
\end{equation}
in the basis $\left[\chi_L,\chi_R^c,N\right]$, with $M_N\gg M_D,m_{\chi N}$. The mixing entry may originate from additional scalar or higher-dimensional interactions.

Integrating out $N$, or taking the Schur complement of its mass block, generates a left-handed Majorana entry with magnitude
\begin{equation}
m_M\simeq\frac{m_{\chi N}^2}{M_N}.
\label{eq:darkseesaw}
\end{equation}
The induced entry has a negative sign in the displayed real mass convention; this sign can be absorbed by rephasing the light fields. The Dirac entry remains $M_D$, so the light states form a pseudo-Dirac pair with splitting $\delta_\chi\simeq\cos^2\theta_L\,m_{\chi N}^2/M_N$.

Numerically,
\begin{equation}
m_{\chi N}
\simeq
\frac{55\GeV}{|\cos\theta_L|}
\left[\frac{\delta_\chi}{300\keV}\right]^{1/2}
\left[\frac{M_N}{10^7\GeV}\right]^{1/2}.
\label{eq:darkseesawnumeric}
\end{equation}
For the small reference mixing angle, the projection factor is effectively unity. This example shows how a sub-MeV splitting can coexist with a TeV-scale Dirac mass through a heavy mass hierarchy. Corrections associated with the heavy-state admixture are suppressed by that hierarchy and are distinct from the singlet--doublet mixing of the minimal model.

Eq.~\eqref{eq:darkseesawmatrix} is a symmetry-broken illustration, not a complete gauge-invariant extension. Since $\chi_L$ carries baryon number, mixing with a gauge-neutral Majorana state requires charged scalar or spurion insertions. A baryon-charged heavy state instead requires a gauge-consistent origin for its Majorana mass. The completion must also suppress the direct renormalizable Majorana operator. Any additional charged vacuum expectation value must be checked for its contribution to the mediator mass and its effect on the stabilizing parity of Sec.~\ref{subsec:residual}.

Both interpretations can reproduce the same low-energy hierarchy: the spurion construction suppresses the Yukawa coupling by powers of $f/\Lambda$, while the seesaw generates a Majorana mass proportional to $m_{\chi N}^2/M_N$. Their connection to the phenomenology requires the additional sector to decouple while preserving the assumed baryonic mass relations, residual parity, and low-energy interactions. The scan itself continues to use the minimal field content defined in the main text.

\section{Large Magellanic Cloud effects in the high-velocity tail}
\label{app:lmc_tail}

The endothermic interpretation of the LZ candidate is particularly sensitive to the high-speed DM population, as discussed in Secs.~\ref{subsec:LZ} and~\ref{subsec:deltaprofile}. For splittings of several hundred keV, the minimum incident speed approaches the endpoint of the local velocity distribution, so changes in its tail can strongly affect the recoil spectrum. The Large Magellanic Cloud (LMC) can modify this tail both by perturbing the native MW halo and by contributing its own high-speed particles~\cite{BeslaPeterGaravitoLMC,DonaldsonPetersenPenarrubiaLMC,GaravitoCamargoWake,GaravitoCamargoBFE,SmithOrlikLMC}. Related effects include the halo wake and the reflex motion of the MW~\cite{PetersenPenarrubiaReflex,CunninghamLMCResponse,GomezLMCReflex}.

For the dominant vector transition, the velocity dependence enters through
\begin{equation}
\eta\left[v_{\min},t\right]
=
\int_{|\mathbf v|>v_{\min}}d^3v\,
\frac{f_{\rm lab}\left[\mathbf v,t\right]}{v}.
\label{eq:eta_lmc_appendix}
\end{equation}
We adopt the present-day cosmological MW--LMC model and halo-integral profiles of Ref.~\cite{SmithOrlikLMC} to illustrate their effect on the xenon recoil prediction.

\subsubsection{SHM, native-MW, and MW+LMC distributions}
\label{app:lmc_definitions}

The comparison distinguishes three distributions. The SHM is an isotropic, truncated Maxwellian and contains neither the dynamical response to the LMC nor LMC-origin particles. The reference comparison in Fig.~11 of Ref.~\cite{SmithOrlikLMC} uses $v_0=220\kms$ and $v_{\rm esc}=544\kms$, whereas the main analysis here uses $v_0=238\kms$, $v_{\rm esc}=544\kms$, and $v_E=250.2\kms$.

The profile labeled MW (Pres.) contains only native-MW particles in the present-day simulation, but those particles have already responded to the LMC infall. It is therefore not an unperturbed MW halo. The difference between this profile and the SHM includes both the properties of the cosmological MW analogue and its dynamical response to the LMC.

The MW+LMC (Pres.) profile includes native-MW and LMC-origin particles from the same snapshot. Its difference from MW (Pres.) isolates the additional LMC-origin population within that simulation. Although its local fraction is small, this population can contribute substantially to the extreme high-speed tail.

\begin{figure}[t]
\centering
\includegraphics[width=0.99\linewidth]{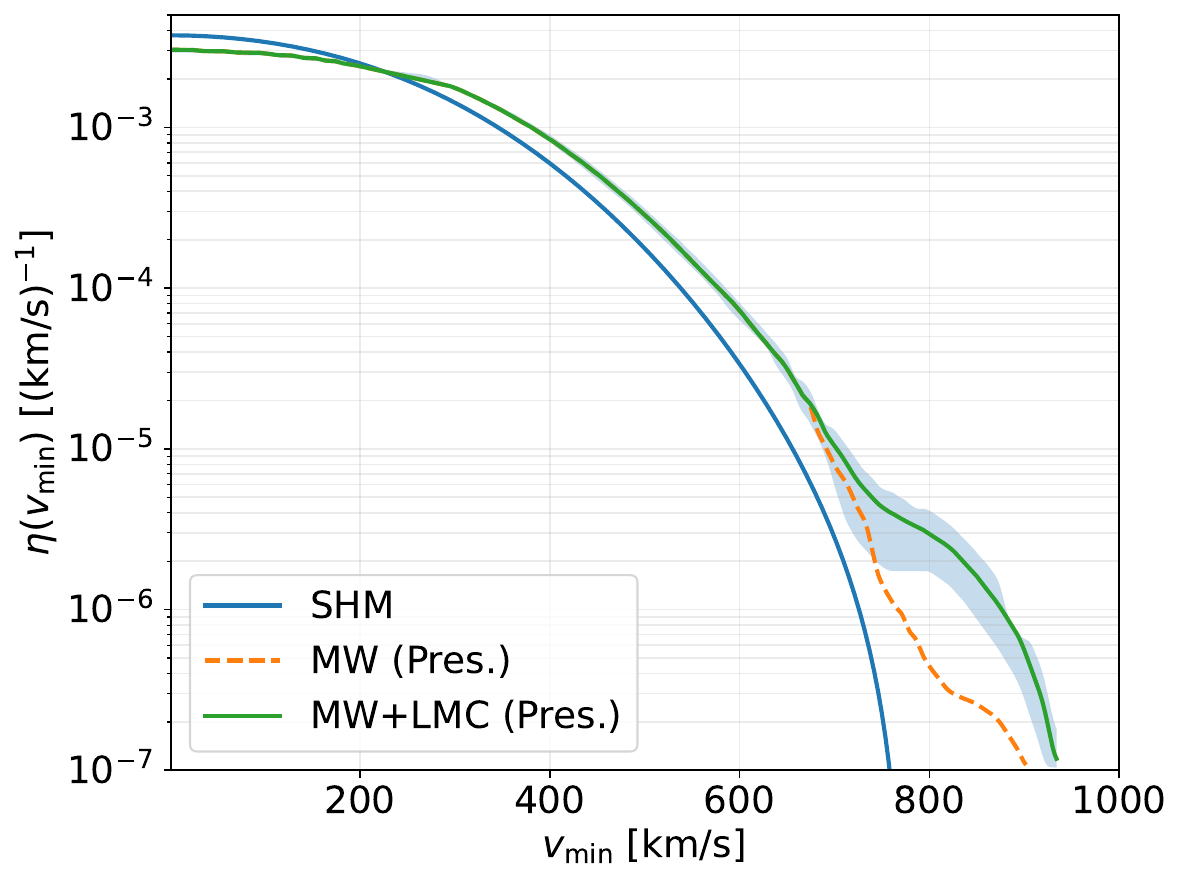}
\caption{Inverse-speed profiles adopted from the model of Ref.~\cite{SmithOrlikLMC}. The SHM follows the reference definition used in that work. The dashed profile contains present-day native-MW particles, including their dynamical response to the LMC; the solid profile also includes LMC-origin particles. The band shows the reported $1\sigma$ MW+LMC uncertainty.}
\label{fig:lmc_full_eta}
\end{figure}

As shown in Fig.~\ref{fig:lmc_full_eta}, the differences become substantial above approximately $600$--$700\kms$, where the SHM integral decreases rapidly. Both the response of native-MW particles and the additional LMC population can therefore affect endothermic scattering with $\delta_\chi\sim300\keV$. The adopted model also supports speeds beyond the SHM endpoint.

\subsubsection{Impact on the xenon recoil spectrum}
\label{app:lmc_recoil}

We propagate these profiles through the endothermic xenon recoil calculation, using $m_\chi=1\TeV$, natural isotope abundances, the Helm form factor, and splittings $\delta_\chi=300\keV$ and $350\keV$. For this comparison, the SHM has $v_0=220\kms$, $v_{\rm esc}=544\kms$, and $v_E=232\kms$, consistently with the reference comparison.

A common interaction normalization is chosen such that the SHM spectrum at $\delta_\chi=300\keV$ gives one event in $215$--$270\keV$. Detector acceptance is omitted. This illustrative normalization is separate from the absolute benchmark normalization and accepted $200$--$300\keV$ counts used in the main text; the comparison is not an additional LZ fit.

\begin{figure}[t]
\centering
\includegraphics[width=0.98\linewidth]{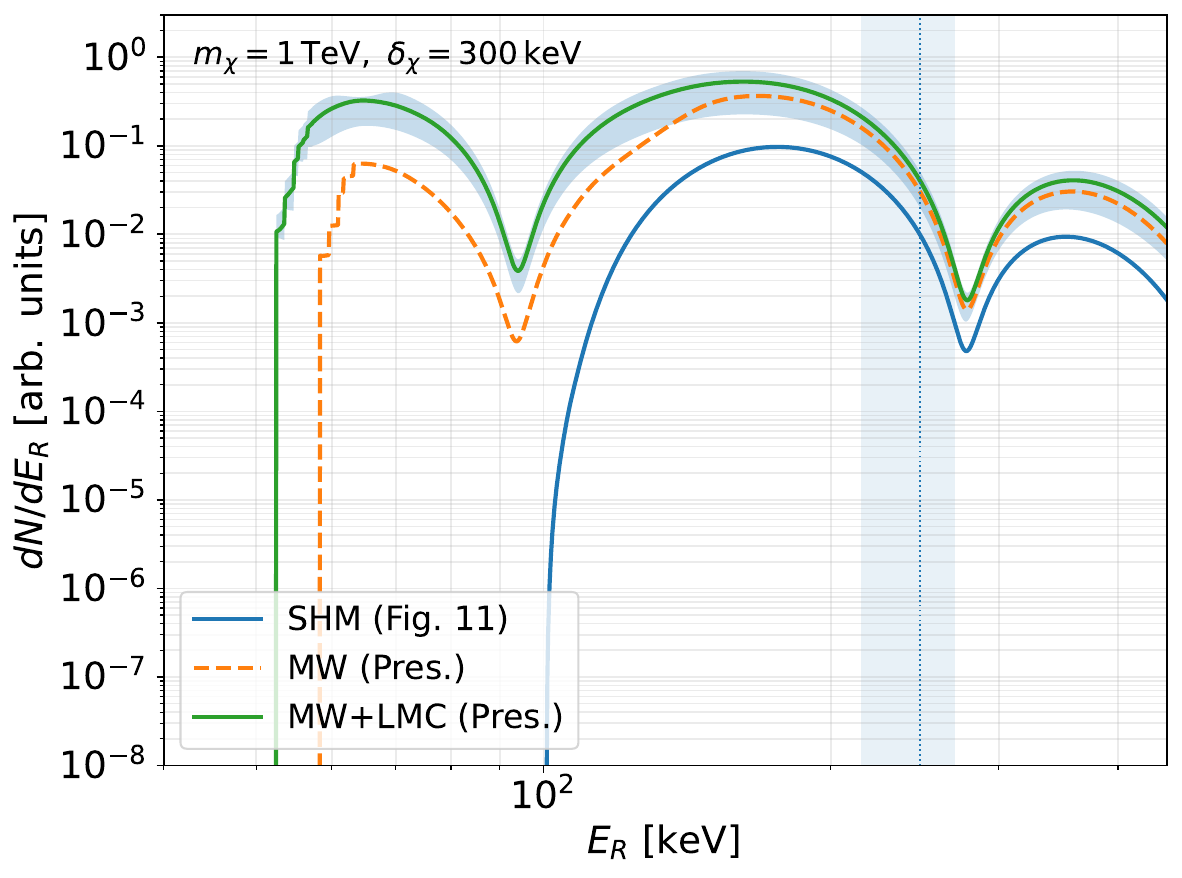}
\includegraphics[width=0.98\linewidth]{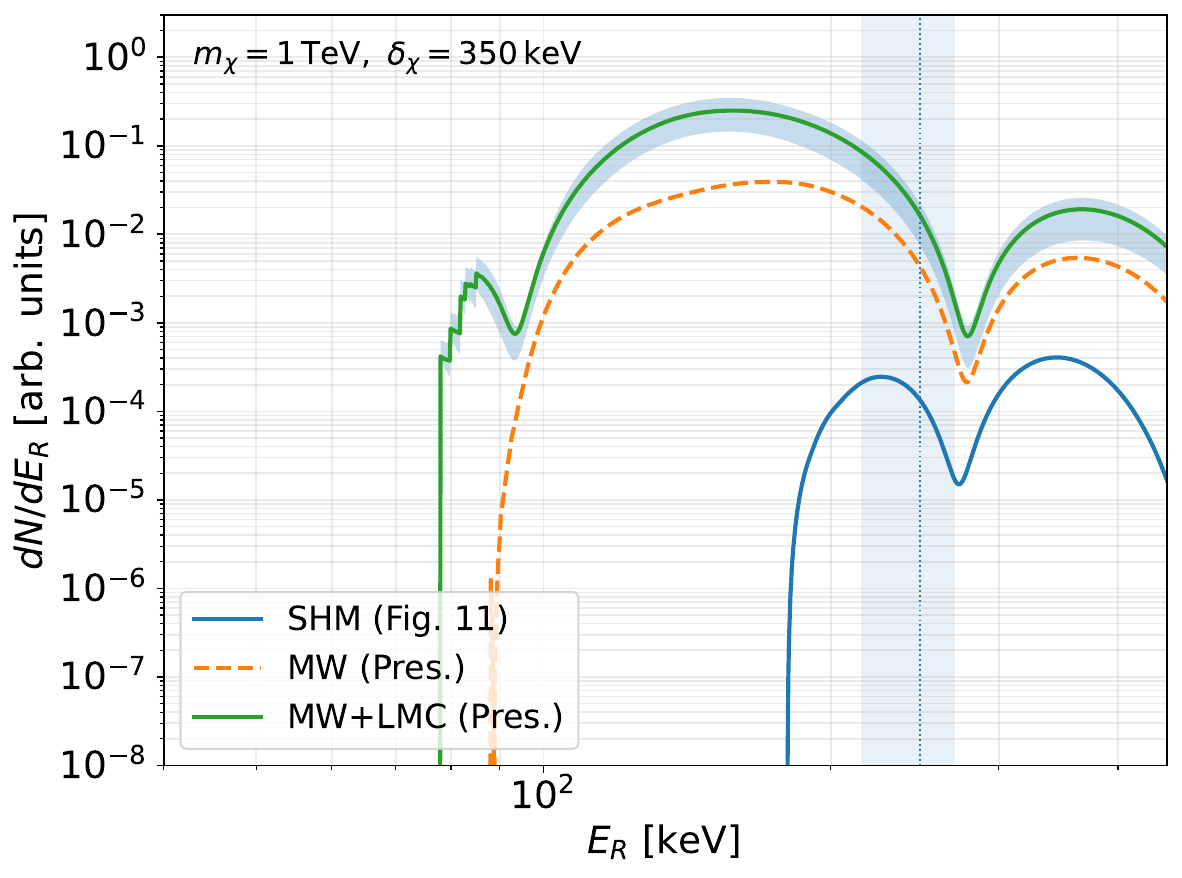}
\caption{Xenon recoil spectra for $m_\chi=1\TeV$, using the SHM and the present-day native-MW and MW+LMC profiles of Ref.~\cite{SmithOrlikLMC}. The upper and lower panels show $\delta_\chi=300\keV$ and $350\keV$, respectively, with the same interaction normalization and no detector acceptance. The bands propagate the reported $1\sigma$ MW+LMC profile uncertainty. The dotted line marks $248\keV$, and the shaded interval is $215$--$270\keV$. The larger splitting probes a more extreme portion of the high-speed tail.}
\label{fig:lmc_recoil}
\end{figure}

For $\delta_\chi=300\keV$, the yield in $215$--$270\keV$ increases from $1.00$ for the SHM to approximately $3.11$ for MW (Pres.) and $4.13$ for MW+LMC (Pres.). The propagated MW+LMC band spans approximately $2.03$--$5.23$ events. The enhancement also changes the spectral shape, including the relative heights of the Helm-form-factor lobes and the population of lower-energy recoils.

At $\delta_\chi=350\keV$, the same interaction normalization gives approximately $8.39\times10^{-3}$ events for the SHM, $0.419$ for MW (Pres.), and $1.66$ for MW+LMC (Pres.), with a MW+LMC band of $0.783$--$2.23$. The central MW+LMC yield is therefore approximately two hundred times the SHM yield. This stronger dependence arises because increasing the splitting pushes the required speeds further into the tail. Related sensitivity has been emphasized in studies of the Higgsino interpretation~\cite{FanReeceSeaOfFog,OHareHighVelocity,McCabeSeasonal}.

The halo changes the relative recoil-lobe heights, while the isotope-specific form-factor zeros remain fixed by nuclear structure. The uncertainty band propagates only the reported uncertainty of the selected MW+LMC profile; it does not cover the full variation among Galactic models, simulation realizations, Solar-motion assumptions, or detector responses.

\subsubsection{Implications for the parameter-space results}
\label{app:lmc_impact_results}

The enhanced high-speed population can substantially reduce the coupling required to reproduce a fixed recoil yield. For dominant baryonic vector exchange at fixed $m_\chi$, $\delta_\chi$, and $M_{Z_B}$, the yield in a specified recoil window scales as $g_B^4$. If $\mathcal B$ denotes the ratio of the window-integrated yields evaluated with two halo profiles at the same coupling, the corresponding count-normalization coupling scales as $g_B\to g_B\mathcal B^{-1/4}$. This relation applies to the integrated count even when the enhancement is energy dependent, but does not establish compatibility with the full spectrum.

For $\delta_\chi=300\keV$, the yields quoted above imply
$g_B^{\rm MW}/g_B^{\rm SHM}\simeq3.11^{-1/4}\simeq0.75$ and
$g_B^{\rm MW+LMC}/g_B^{\rm SHM}\simeq4.13^{-1/4}\simeq0.70$.
Thus, reproducing the same count in $215$--$270\keV$ requires approximately $25\%$ and $30\%$ smaller couplings, respectively. The propagated MW+LMC profile band corresponds to $g_B^{\rm MW+LMC}/g_B^{\rm SHM}\simeq0.66$--$0.84$. Comparing the two present-day simulation profiles directly, adding LMC-origin particles reduces the required coupling by approximately $7\%$ relative to the native-MW profile.

The effect is stronger for $\delta_\chi=350\keV$. Relative to the SHM yield of $8.39\times10^{-3}$, the MW and MW+LMC yields correspond to enhancement factors of approximately $50$ and $198$. The required couplings are therefore approximately $0.38$ and $0.27$ times their SHM values, reductions of about $62\%$ and $73\%$. The MW+LMC band gives a coupling ratio of approximately $0.25$--$0.32$. Adding LMC-origin particles to the present-day native-MW profile alone reduces the coupling by approximately $29\%$. The larger changes relative to the SHM include both the properties and dynamical response of the simulated MW halo and the additional LMC population.

These estimates quantify the change in the coupling inferred from one recoil window, rather than a shift of the combined allowed region. The enhanced tail also changes the spectral shape and the control- and sideband yields, while a change in $g_B$ affects the relic abundance and collider predictions. Moreover, the illustrative SHM parameters differ from those of the main analysis. Updating the contours therefore requires a new DD fit with the adopted MW+LMC profile, followed by combination with the relic-density and collider constraints of Sec.~\ref{sec:results}.

\renewcommand{\bibfont}{\small\setlength{\baselineskip}{10pt}}
\bibliographystyle{apsrev4-2}
\bibliography{paper}
\end{document}